\documentstyle[harvard,12pt]{article}
\citationstyle{agsm}
\let\citen=\citeasnoun

\let\ssection=\section
\let\ssubsection=\subsection
\renewcommand{\section}{\setcounter{equation}{0}\ssection}
\renewcommand{\subsection}{\setcounter{equation}{0}\ssubsection}

\newcommand{\be}{\begin{enumerate}}
\newcommand{\ee}{\end{enumerate}}
\newcommand{\bi}{\begin{itemize}}
\newcommand{\ei}{\end{itemize}}
\newcommand{\bd}{\begin{description}}
\newcommand{\ed}{\end{description}}
\newcommand{\beq}{\begin{equation}}
\newcommand{\eeq}{\end{equation}}
\newcommand{\beqa}{\begin{eqnarray}}
\newcommand{\eeqa}{\end{eqnarray}}

\renewcommand{\a}{\alpha}                   
\renewcommand{\b}{\beta}                    
\newcommand{\de}{\delta}
\newcommand{\eq}[1]{(\ref{#1})}
\newcommand{\eqs}[2]{(\ref{#1}--\ref{#2})}
\newcommand{\f}{\phi}

\newcommand{\k}{\kappa}
\newcommand{\m}{\mu}

\newcommand{\r}{\rho}
\newcommand{\s}{\sigma}

\newcommand{\w}{\omega}

\newcommand{\bra}[1]{\la #1 |}
\newcommand{\brac}[2]{\la #1 |#2 \ra}
\newcommand{\cf}{{\em cf\/}}
\newcommand{\dg}{|g|}
\newcommand{\eg}{{\em e.g.,\ }}
\newcommand{\etc}{{\em etc}}
\newcommand{\g}{\gamma}                    
\newcommand{\half}{\frac1 2}
\newcommand{\ie}{{\em i.e.,\ }}
\newcommand{\ket}[1]{| #1 \ra}
\newcommand{\la}{\langle}
\newcommand{\map}{\rightarrow}
\newcommand{\naive}{na\"{\i}ve\ }
\newcommand{\op}[1]{\widehat{#1}}
\newcommand{\pa}{\partial}

\newcommand{\ra}{\rangle}
\newcommand{\rref}{{\rm ref}}
\newcommand{\tr}{{\rm tr}}

\newcommand{\DM}{{\rm Diff}(\M)}
\newcommand{\DS}{{\rm Diff}(\Si)}
\newcommand{\E}{{\rm Emb}(\Si,\M)}

\newcommand{\Em}{{\cal E}}                  
\newcommand{\Fo}{{\cal F}}                  
\newcommand{\For}{{\cal F}^{\rm ref}}
\renewcommand{\H}{{\cal H}}                 
\newcommand{\Hp}{{\cal H}_\perp}
\newcommand{\Ha}{{\cal H}_a}
\newcommand{\M}{{\cal M}}
\newcommand{\N}{\vec{N}}
\newcommand{\Np}{N}
\newcommand{\Om}{\Omega}
\renewcommand{\P}{{\cal P}}                 
\newcommand{\Prob}{{\rm Prob}}
\newcommand{\R}{{\rm I\! R}}                
\newcommand{\RS}{{\rm Riem}(\Sigma)}

\newcommand{\Si}{\Sigma}

\newcommand{\SiR}{\Sigma\times\R}
\newcommand{\T}{{\cal T}}
\newcommand{\X}{{\cal X}}
\newcommand{\Z}{{\cal Z}}

\begin{document}


\begin{titlepage}
\hspace{4truein} Imperial/TP/91-92/25

\begin{center}
                {\LARGE\bf Canonical Quantum Gravity \\[0.2cm]
                            and the Problem of Time
\footnote{Lectures presented at the NATO Advanced Study Institute
``Recent Problems in Mathematical Physics'', Salamanca, June 15--27,
1992.}
\footnote{Research supported in part by SERC grant GR/G60918.}}
\end{center}
\vspace{2 truecm}
\begin{center}
        C.J.~Isham \\[0.5cm]
        Blackett Laboratory\\
        Imperial College\\
        South Kensington\\
        London SW7 2BZ\\
        United Kingdom
\end{center}

\begin{center} August 1992\end{center}

\begin{abstract}
The aim of this paper is to provide a general introduction to the
problem of time in quantum gravity. This problem originates in the
fundamental conflict between the way the concept of `time' is used in
quantum theory, and the role it plays in a diffeomorphism-invariant
theory like general relativity. Schemes for resolving this problem
can be sub-divided into three main categories: (I) approaches in
which time is identified before quantising; (II) approaches in which
time is identified after quantising; and (III) approaches in which
time plays no fundamental role at all. Ten different specific schemes
are discussed in this paper which also contain an introduction to
the relevant parts of the canonical decomposition of general
relativity.
\end{abstract}
\end{titlepage}

\tableofcontents\clearpage


\section{INTRODUCTION}
\label{Sec:Intro}
\subsection{Preamble}
These notes are based on a course of lectures given at the NATO
Advanced Summer Institute ``Recent Problems in Mathematical
Physics'', Salamanca, June 15--27, 1992. The notes reflect part of an
extensive investigation with Karel Kucha\v{r} into the problem of
time in quantum gravity. An excellent recent review is
\citen{Kuc92a}, to which the present article is complementary to some
extent. In particular, my presentation is slanted towards the more
conceptual aspects of the problem and, as this a set of lecture notes
(rather than a review paper proper), I have also included a fairly
substantial technical introduction to the canonical theory of general
relativity. However, there is inevitably a strong overlap with many
portions of Kucha\v{r}'s paper, and I am grateful to him for
permission to include this material plus a number of ideas that have
emerged in our joint discussions. Therefore, the credit for any good
features in the present account should be shared between us; the
credit for the mistakes I claim for myself alone.

\subsection{Preliminary Remarks}
The problem of `time' is one of the deepest issues that must be
addressed in the search for a coherent theory of quantum gravity.
The major conceptual problems with which it is closely connected
include:
\bi
\item the status of the concept of {\em probability\/} and the
      extent to which it is {\em conserved\/};

\item the status of the associated concepts of {\em causality\/}
      and {\em unitarity};

\item the time-honoured debate about whether quantum gravity
      should be approached via a {\em canonical\/}, or a {\em
      covariant\/}, quantisation scheme;

\item the extent to which {\em spacetime\/} is a meaningful
      concept;

\item the extent to which classical {\em geometrical\/} concepts
      can, or should, be maintained in the quantum theory;

\item the way in which our {\em classical world\/} emerged from
      some primordial quantum event at the big-bang;

\item the whole question of the {\em interpretation\/} of quantum
      theory and, in particular, the domain of applicability of the
      conventional Copenhagen view.
\ei

   The prime source of the problem of time in quantum gravity is the
invariance of classical general relativity under the group $\DM$ of
diffeomorphisms of the spacetime manifold $\M$. This stands against the
simple Newtonian picture of a fixed time parameter, and tends to
produce quantisation schemes that apparently lack any fundamental
notion of time at all. From this perspective, the heart of the
problem is contained in the following questions:
\be
\item How should the notion of time be re-introduced into the
      quantum theory of gravity?

\item In particular, should attempts to identify time be made at the
      classical level, \ie before quantisation, or should the theory be
      quantised first?

\item   Can `time' still be regarded as a fundamental concept in a
quantum theory of gravity, or is its status purely phenomenological?
If the concept of time is not fundamental, should it be replaced by
something that is: for example, the idea of a {\em history\/} of a
system, or {\em process\/}, or an {\em ordering structure\/} that is
more general than that afforded by the conventional idea of time?

\item   If `time' is only an approximate concept, how reliable is the
rest of the quantum-mechanical formalism in those regimes where the
normal notion of time is not applicable? In particular, how closely
tied to the concept of time is the idea of probability? This is
especially relevant in those approaches to quantum gravity in which
the notion of time emerges only {\em after\/} the theory has been
quantised.
\ee
In addition to these questions---which apply to quantum gravity in
general---there is also the partly independent issue of the
applicability of the concept of time (and, indeed, of quantum theory
in general) in the context of quantum cosmology. Of particular
relevance here are questions of (i) the status of the Copenhagen
interpretation of quantum theory (with its emphasis on the role of
measurements); and (ii) the way in which our present classical
universe, including perhaps the notion of time, emerged from the
quantum origination event.

    A key ingredient in all these questions is the realisation that
the notion of time used in conventional quantum theory is grounded
firmly in Newtonian physics. Newtonian time is a fixed structure,
external to the system: a concept that is manifestly incompatible
with diffeomorphism-invariance and also with the idea of constructing
a quantum theory of a truly closed system (such as the universe
itself). Most approaches to the problem of time in quantum gravity
\footnote{A similar problem arises when discussing thermodynamics and
statistical physics in a curved spacetime. A recent interesting
discussion is \citen{Rov91e} which makes specific connections between
this problem of time and the one that arises in quantum gravity.}
seek to address this central issue by identifying an {\em
internal\/} time which is defined in terms of the system itself,
using either the gravitational field or the matter variables that
describe the material content of the universe. The various schemes
differ in the way such an identification is made and the point in the
procedure at which it is invoked. Some of these techniques inevitably
require a significant reworking of the quantum formalism itself.

\subsection{Current Research Programmes in Quantum Gravity}
An important question that should be raised at this point is the
relation of the problem of time to the various research programmes in
quantum gravity that are currently active. The primary distinction is
between approaches to quantum gravity that start with the {\em
classical\/} theory of general relativity (or a simple extension of
it) to which some quantisation algorithm is applied, and schemes
whose starting point is a {\em quantum\/} theory from which classical
general relativity emerges in some low-energy limit, even though that
theory was not one of the initial ingredients. Most of the standard
approaches to quantum gravity belong to the former category;
superstring theory is the best-known example of the latter.

    Some of the more prominent current research programmes in quantum
gravity are as follows (for recent reviews see Isham (1985, 1987,
1992) \nocite{Ish85,Ish87,Ish92} and \citen{Alv89}).
\bi
    \item {\em Quantum Gravity and the Problem of Time\/}. This
subject---the focus of the present paper---dates back to the earliest
days of quantum gravity research. It has been studied extensively in
recent years---mainly within the framework of the canonical
quantisation of the classical theory of general relativity (plus
matter)---and is now a significant research programme in its own
right. A major recent review is \citen{Kuc92a}. Earlier reviews, from
somewhat different perspectives, are \citen{BS88} and \citen{UW89}.
There are also extensive discussions in \citen{AS91} and in
\citen{HP-MZ92}; other relevant literature will be cited at the
appropriate points in our discussion.

    \item {\em The Ashtekar Programme\/}. A major technical
development in the canonical formalism of general relativity was the
discovery by Ashtekar (1986, 1987)\nocite{Ash86,Ash87} of a new set
of canonical variables that makes general relativity resemble
Yang-Mills theory in several important respects, including the
existence of non-local physical observables that are an analogue of
the Wilson loop variables of non-abelian gauge theory
\cite{RS90,Rov91d,Smo92}; for a recent comprehensive review of the
whole programme see \citen{Ash91}. This is one of the most promising
of the non-superstring approaches to quantum gravity and holds out
the possibility of novel non-perturbative techniques. There have also
been suggestions that the Ashtekar variables may be helpful in
resolving the problem of time (for example, in \citen{Ash91}[pages
191-204]). The Ashtekar programme is not discussed in these notes,
but it is a significant development and has important
implications for quantum gravity research in general.

    \item {\em Quantum Cosmology\/}. This subject was much studied in
the early days of quantum gravity and has enjoyed a renaissance in
the last ten years, largely due to the work of \citen{HH83} and
\citen{Vil88} on the possibility of constructing a quantum theory of
the creation of the universe. The techniques employed have been
mainly those of canonical quantisation, with particular emphasis on
minisuperspace (\ie finite-dimensional) approximations. The problem
of time is central to the subject of quantum cosmology and is often
discussed within the context of these models, as are a number of
other conceptual problems expected to arise in quantum gravity
proper. A major recent review is \citen{Hal91a} which also contains a
comprehensive bibliography (see also \citen{Hal90} and
\citen{Hal92b}).

    \item{\em Low-Dimensional Quantum Gravity\/}. Studies of gravity
in $1+1$ and $2+1$ dimensions have thrown valuable light on many of
the difficult technical and conceptual issues in quantum gravity,
including the problem of time (for example, Carlip (1990,
1991)\nocite{Car90,Car91}). The reduction to lower dimensions
produces major technical simplifications whilst maintaining enough of
the flavour of the $3+1$-dimensional case to produce valuable
insights into the full theory. Lower-dimensional gravity also has
direct physical applications. For example, idealised cosmic strings
involve the application of gravity in $2+1$ dimensions, and the
theory in $1+1$ dimensions has applications in statistical mechanics.
The subject of gravity in $2+1$ dimensions is reviewed in
\citen{Jac92a} whilst recent developments in $1+1$ dimensional
gravity are reported in \citen{Jac92b} and \citen{Tei92} (in the
proceedings of this Summer School).

    \item {{\em Semi-Classical Quantum Gravity\/}. Early studies of
this subject (for a review see \citen{Kib81}) were centered on the
equations
\beq
    G_{\a\b}(X,\g]=\bra{\psi}T_{\a\b}(X;\g,\op\f\,]\ket{\psi}
                                                \label{Gab=E(T)}
\eeq
where the source for the classical spacetime metric $\g$ is an
expectation value of the energy-momentum tensor of quantised matter
$\op\f$. More recently, equations of this type have been developed
from a WKB-type approximation to the quantum equations of canonical
quantum gravity, especially the Wheeler-DeWitt equation
(\S\ref{SSSec:WDE}). Another goal of this programme is to provide a
coherent foundation for the construction of quantum field theory in a
fixed background spacetime: a subject that has been of enduring
interest since Hawking's discovery of quantum-induced radiation from
a black hole.

    The WKB approach to solving the Wheeler-DeWitt equation is
closely linked to a semi-classical theory of time and will be
discussed in \S\ref{SSec:SemClass}.}

    \item{\em Spacetime Structure at the Planck Length\/}. There is a
gnostic subculture of workers in quantum gravity who feel that the
structure of space and time may undergo radical changes at scales of
the Planck length. In particular, the idea surfaces repeatedly that
the continuum spacetime picture of classical general relativity may
break down in these regions. Theories of this type are highly
speculative but could have significant implications for the question
of `time' which, as a classical concept, is grounded firmly in
continuum mathematics.

    \item{\em Superstring Theory\/}. This is often claimed to be the
`correct' theory of quantum gravity, and offers many new perspectives
on  the intertwining of general relativity and quantum theory. Of
particular interest is the recurrent suggestion that there exists
a minimal length, with the implication that many normal spacetime
concepts could break down at this scale. It is therefore unfortunate
that the current approaches to superstring theory are mainly
perturbative in character (involving, for example, graviton
scattering amplitudes) and are difficult to apply directly to the
problem of time. But the question of the implications of string
theory is an intriguing one, not least because of the very different
status assigned by the theory to important spacetime concepts such
as the diffeomorphism group.
\ei

    It is noteable that almost all studies of the problem of time
have been performed in the framework of `conventional' canonical
quantum gravity in which attempts are made to apply quantisation
algorithms to the field equations of classical general relativity.
However, the resulting theory is well-known to be perturbatively
non-renormalisable and, for this reason, much of the work on time has
used finite-dimensional models that are free of ultraviolet
divergences. Therefore, it must be emphasised that many of the
specific problems of time are {\em not\/} connected with the
pathological short-distance behaviour of the theory, and appear in an
authentic way in these finite-dimensional model systems .

    Nevertheless, the non-renormalisability is worrying and raises
the general question of how seriously the results of the existing
studies should be taken. The Ashtekar programme may lead eventually
to a finite and well-defined `conventional' quantum theory of
gravity, but quite a lot is being taken on trust. For example, any
addition of Riemann-curvature squared counter-terms to the normal
Einstein Lagrangian would have a drastic effect on the canonical
decomposition of the theory and could render irrelevant much of the
discussion involving the Wheeler-DeWitt equation.

    In practice, most of those who work in the field seem to believe
that, whatever the final theory of quantum gravity may be (including
superstrings), enough of the conceptual and geometrical structure of
classical general relativity will survive to ensure the relevance of
most of the general questions that have been asked about the meaning and
significance of `time'. However, it should not be forgotten that the
question of what constitutes a conceptual problem---such as the
nature of time---often cannot be decided in isolation from the
technical framework within which it is posed. Therefore, it is
feasible that when, for example, superstring theory is better
understood, some of the conceptual issues that appear now to be
genuine problems will be seen to be the result of asking ill-posed
questions, rather than reflecting a fundamental problem in nature
itself.

\subsection{Outline of the Paper}
We begin in \S\ref{Sec:Problem} by discussing the status of time in
general relativity and conventional quantum theory, and some of the
{\em prima facie\/} problems that arise when attempts are made to
unite these two, somewhat disparate, conceptual structures. As we
shall see, approaches to the problem of time fall into three
distinct categories---those in which time is identified before
quantising, those in which time is identified after quantising, and
`timeless' schemes in which no fundamental notion of time is
introduced at all. A brief description is given in
\S\ref{Sec:Problem} of the principle schemes together with the major
difficulties that are encountered in their implementation.

    Most approaches to the problem of time involve the canonical
theory of general relativity, and \S\ref{Sec:CanGR} is devoted to
this topic. Section \S\ref{Sec:TBefore} deals with the approaches
to the problem of time that involve identifying time before
quantising, while \S\ref{Sec:TAfter} addresses those schemes in
which the gravitational field is quantised first. Then follows an
account in \S\ref{Sec:Timeless} of the schemes that avoid
invoking time as a fundamental category at all. As we shall see,
attempts of this type become rapidly involved in general questions
about the interpretation of quantum theory and, in particular, of
the domain of applicability of the traditional Copenhagen approach. The
paper concludes with a short summary of the current situation and
some speculations about the future.

\subsection{Conventions}
The conventions and notations to be used are as follows. A
Lorentzian metric $\g$ on a four-dimensional spacetime manifold $\M$
is assumed to have signature $(-1,1,1,1)$. Lower case Greek indices
refer to coordinates on $\M$ and take the values $0,1,2,3$. We adopt
the modern differential-geometric view of coordinates as local {\em
functions\/} on the manifold. Thus, if $X^\a$, $\a=0\ldots3$ is a
coordinate system on $\M$ the {\em values\/} of the coordinates of a
point $Y\in\M$ are the four numbers $X^\a(Y)$, $\a=0\ldots3$.
However, by a familiar abuse of notation, sometimes we shall simply
write these numbers as $Y^\a$. Lower case Latin indices refer to
coordinates on the three-manifold $\Si$ and range through the values
$1,2,3$.

    The symbol $f:X\map Y$ means that $f$ is a map from the space $X$
into the space $Y$. In the canonical theory of gravity there are a
number of objects $F$ that are functions simultaneously of (i) a
point $x$ in a finite-dimensional manifold (such as $\Si$ or $\M$);
and (ii) a function $G:X\map Y$ where $X,Y$ can be a variety of
spaces. It is useful to adopt a notation that reflects this property.
Thus we write $F(x,G\,]$ to remind us that $G$ is itself a function.
If $F$ depends on $n$ points $x_1,\ldots x_n$ in a finite-dimensional
manifold, and $m$ functions $G_1,\ldots G_m$, we shall write
$F(x_1,\ldots x_n;G_1,\ldots G_m\,]$.


\section{QUANTUM GRAVITY AND THE PROBLEM OF TIME}
\label{Sec:Problem}
\subsection{Time in Conventional Quantum Theory}
The problem of time in quantum gravity is deeply connected with the
special role assigned to temporal concepts in standard theories of
physics. In particular, in Newtonian physics, time---the parameter
with respect to which change is manifest---is {\em external\/} to the
system itself. This is reflected in the special status of time in
conventional quantum theory:
\be
    \item Time is not a physical observable in the normal sense
since it is not represented by an operator. Rather, it is treated as
a background parameter which, as in classical physics, is used to
mark the evolution of the system. In particular, it provides the
parameter $t$ in the time-dependent Schr\"odinger equation
\beq
        i\hbar{d\psi_t\over dt}= \op H\psi_t.   \label{SE}
\eeq
This special property of time applies both to non-relativistic
quantum theory and to relativistic particle dynamics and quantum
field theory. It is the reason why the meaning assigned to the
time-energy uncertainty relation $\de t\,\de E\ge\half\hbar$ is quite
different from that pertaining to, for example, the position and the
momentum of a particle.

    \item  This view of time is related to the difficulty of
describing a truly {\em closed\/} system in quantum-mechanical terms.
Indeed, it has been cogently argued that the only physical states in
such a system are {\em eigenstates\/} of the Hamiltonian operator,
whose time evolution is essentially trivial \cite{PW83}. Of course,
the ultimate closed system is the universe itself.

    \item The idea of events happening at a single time plays a
crucial role in the technical and conceptual foundations of quantum
theory:
    \bi
        \item The notion of a {\em measurement\/} made at a
particular time is a fundamental ingredient in the conventional
Copenhagen interpretation. In particular, an {\em observable\/} is
something whose value can be measured at a fixed time. On the other
hand, a `history' has no direct physical meaning except in so far as
it refers to the outcome of a sequence of time-ordered measurements.

        \item One of the central requirements of the {\em scalar
product\/} on the Hilbert space of states is that it be conserved under
the time evolution \eq{SE}. This is closely connected to the unitarity
requirement that probabilities always sum to one.

        \item More generally, a key ingredient in the construction of
the Hilbert space for a quantum system is the selection of a
complete set of observables that are required to {\em commute\/} at
a fixed value of time.
    \ei

    \item These ideas can be extended to systems that are compatible
with special relativity: one simply replaces the unique time system
of Newtonian physics with the set of relativistic inertial reference
frames.  The quantum theory can be made independent of a choice of
frame if it carries a unitary representation of the Poincar\'e
group. In the case of a relativistic quantum field theory, this is
closely related to the requirement of microcausality, \ie
\beq
        [\,\op\f(X),\op\f(X')\,]    = 0     \label{microcausal}
\eeq
for all spacetime points $X$ and $X'$ that are spacelike separated.
\ee

    The background Newtonian time appears explicitly in the
time-dependent Schr\"odinger equation \eq{SE}, but it is pertinent to
note that such a time is truly an abstraction in the sense that no {\em
physical\/} clock can provide a precise measure of it \cite{UW89}.
For suppose there is some quantum observable $T$ that can serve as a
`perfect' physical clock in the sense that, for some initial state,
its observed values increase monotonically with the abstract time
parameter $t$. Since $\op T$ may have a continuous spectrum, let us
decompose its eigenstates into a collection of normalisable vectors
$\ket{\tau_0},\ket{\tau_1},\ket{\tau_2}\ldots$ such that
$\ket{\tau_n}$ is an eigenstate of the projection operator onto the
interval of the spectrum of $\op T$ centered on $\tau_n$. Then to
say that $T$ is a perfect clock means:
\be
    \item For each $m$ there exists an $n$ with $n>m$ and $t>0$ such
that the probability amplitude for $\ket{\tau_m}$ to evolve to
$\ket{\tau_n}$ in Newtonian time $t$ is non-zero (\ie the clock has
a non-zero probability of running forwards with respect to the
abstract Newtonian time $t$). This means that
\beq
    f_{mn}(t):=\bra{\tau_n}U(t)\ket{\tau_m} \ne 0   \label{Def:fmn(t)}
\eeq
where $U(t):=e^{-it\op H/\hbar}$.

    \item For each $m$ and for all $t>0$, the amplitude to evolve
from $\ket{\tau_m}$ to $\ket{\tau_n}$ vanishes if $m>n$ (\ie the
clock never runs backwards).
\ee
\citen{UW89} show that these conditions are
incompatible with the physical requirement that the energy of the
system be positive. This follows by studying the function $f_{mn}(t)$,
$m>n$, in \eq{Def:fmn(t)} for complex $t$ and with $m>n$. Since $\op
H$ is bounded below, $f_{mn}$ is holomorphic in the lower-half plane
and hence cannot vanish on any open real interval unless it vanishes
identically for all $t$ whose imaginary part is less than, or equal
to, zero. However, the requirement that the clock never runs
backwards is precisely that $f_{mn}(t)=0$ for all $t>0$, and hence
$f_{mn}(t)=0$ for all real $t$. But then, if $m<n$,
\beq
    f_{mn}(t)=\bra{\tau_n}U(t)\ket{\tau_m}=
        \bra{\tau_m}U(t)^\dagger\ket{\tau_n}^* = f_{nm}^*(-t)
\eeq
which, as we have just shown, vanishes for all $t>0$. Hence the clock
can never run forwards in time, and so perfect clocks do
not exist.  This means that any physical clock always has a small
probability of sometimes running backwards in abstract Newtonian
time.

    An even stronger requirement on $T$ is that it should be a
`Hamiltonian time observable' in the sense that
\beq
        [\,\op T,\op H\,] = i\hbar.                     \label{CR:TH}
\eeq
This implies at once that $U(t)\ket{T}=\ket{T+t}$ where $\op
T\ket{T}=T\ket{T}$, which is precisely the type of behaviour that is
required for a perfect clock. However, it is well-known that
self-adjoint operators satisfying (exponentiable) representations of
\eq{CR:TH} necessarily have spectra equal to the whole of $\R$, and
hence \eq{CR:TH} is manifestly incompatible with the requirement
that $\op H$ be a positive operator.

    This inability to represent abstract Newtonian time with any
genuine physical observable is a fundamental property of quantum
theory. As we shall see in \S\ref{SSec:RefFluid}, it is reflected in
one of the major attempts to solve the problem of time in quantum
gravity.

\subsection{Time in a $\DM$-invariant Theory}
The special role of time in quantum theory suggests strongly that it
(or, more generally, the system of inertial reference frames) should
be regarded as part of the {\em a priori\/} classical background that
plays such a crucial role in the Copenhagen interpretation of the
theory. But this view becomes highly problematic once general
relativity is introduced into the picture; indeed, it is one of the
main sources of the problem of time in quantum gravity.

    One way of seeing this is to focus on the idea that the equations
of general relativity transform covariantly under changes of
spacetime coordinates, and physical results are meant to be
independent of choices of such coordinates. However, `time' is
frequently regarded as a coordinate on the spacetime manifold $\M$
and it might be expected therefore to play no fundamental
role in the theory. This raises several important questions:
\be
     \item If time is indeed merely a coordinate on $\M$---and hence
of no direct physical significance---how does the, all-too-real,
property of change emerge from the formalism?

    \item {Like all coordinates, `time' may be defined only in a
local region of the manifold. This could mean that:
    \bi
        \item the time variable $t$ is defined only for some finite
range of values for $t$; or
        \item the subset of points $t$=const may not be a complete
three-dimensional submanifold of the spacetime.
    \ei
        How are such global problems to be handled in the quantum
theory?}

     \item Is this view of the coordinate nature of time compatible
with normal quantum theory, based as it is on the existence of a
universal, Newtonian time?
\ee
The global issues are interesting, but they are not central to the
problem of time and therefore in what follows I shall assume the
topology of the spacetime manifold $\M$ to be such that global time
coordinates can exist. If $\M$ is equipped with a Lorentzian metric
$\g$, such a global time coordinate can be regarded as the parameter
in a foliation of $\M$ into a one-parameter family of spacelike
hypersurfaces: the hypersurfaces of `equal-time'. This useful picture
will be employed frequently in our discussions.

    A slightly different way of approaching these issues is to note
that the equations of general relativity are covariant with respect
to the action of the group $\DM$ of diffeomorphisms (\ie smooth and
invertible point transformations) of the spacetime manifold $\M$. We
shall restrict our attention to diffeomorphisms with compact support,
by which I mean those that are equal to the unit map outside some
closed and bounded region of $\M$. Thus, for example, a
Poincar\'e-group transformation of Minkowski spacetime is not deemed
to belong to $\DM$. This restriction is imposed because the role of
transformations with a non-trivial action in the asymptotic regions
of $\M$ is quite different from those that act trivially. It must
also be emphasised that $\DM$ means the group of active point
transformations of $\M$. This should not be confused with the
pseudo-group which describes the relations between overlapping pairs
of coordinate charts (although of course there is a connection
between the two).

    Viewed as an active group of transformations, the diffeomorphism
group $\DM$ is analogous in certain respects to the gauge group of
Yang-Mills theory. For example, in both cases the groups are
associated with field variables that are non-dynamical, and with a
canonical formalism that entails constraints on the canonical
variables. However, in another, and very important sense the two
groups are quite different. Yang-Mills transformations occur at a
fixed spacetime point whereas the diffeomorphism group moves points
around. Invariance under such an active group of transformations robs
the individual points in $\M$ of any fundamental ontological
significance. For example, if $\f$ is a scalar field on $\M$ the
value $\f(X)$ at a particular point $X\in\M$ has no invariant
meaning. This is one aspect of the Einstein `hole' argument that has
featured in several recent expositions \cite{EN87,Sta89}. It is
closely related to the question of what constitutes an {\em observable\/}
in general relativity---a surprisingly contentious issue that has
generated much debate over the years and which is of particular
relevance to the problem of time in quantum gravity. In the present
context, the natural objects that are manifestly $\DM$-invariant are
spacetime integrals like, for example,
\beq
    F[\g]:=\int_\M d^4X\,\big(-\det\g(X)\big)^{\half}\,
            R^{\a\b\gamma\de}(X,\g]\,R_{\a\b\gamma\de}(X,\g].
\eeq
Thus `observables' of this type are intrinsically non-local.

    These implications of $\DM$-invariance pose no real difficulty in
the classical theory since once the field equations have been
solved the Lorentzian metric on $\M$ can be used to give meaning to
concepts like `causality' and `spacelike separated', even if
these notions are not invariant under the action of $\DM$. However,
the situation in the quantum theory is very different. For example,
whether or not a hypersurface is spacelike depends on the spacetime
metric $\g$. But in any quantum theory of gravity there will
presumably be some sense in which $\g$ is subject to quantum
fluctuations. Thus causal relationships, and in particular the notion
of `spacelike', appear to depend on the quantum state. Does this mean
that `time' also is state dependent?

    This is closely related to the problem of interpreting a
microcausality condition like \eq{microcausal} in a quantum theory of
gravity. As emphasised by \citen{FH87}, for most pairs of points
$X,X'\in\M$ there exists at least one Lorentzian metric with respect
to which they are {\em not\/} spacelike separated, and hence, in so
far as all metrics are `virtually present' (for example, by being
summed over in a functional integral) the right hand side of
\eq{microcausal} is never zero. This removes at a stroke one of the
bedrocks of conventional quantum field theory!

    The same argument throws doubts on the use of canonical
commutation relations: with respect to what metric is the
hypersurface meant to be spacelike? This applies in particular to
the equation $[\op{g}_{ab}(x),\op{g}_{cd}(x')]=0$ which, in a
conventional reading, would imply that the canonical configuration
variable $g$ (a metric on a three-dimensional manifold $\Si$) can be
measured simultaneously at two points $x,x'$ in $\Si$. This
difficulty in forming a meaningful interpretation of commutation
relations also renders dubious any attempts to find a quantum gravity
analogue of the powerful $C^*$-algebra approach to conventional
quantum field theory.

    These rather general problems concerning time and the
diffeomorphism group might be resolved in several ways. One
possibility is to restrict attention to spacetimes $(\M,\g)$ that are
asymptotically flat. It is then possible to define {\em asymptotic\/}
quantities, including time and space coordinates, using the values of
fields in the asymptotic regions of $\M$. This works because
the diffeomorphism group transformations have compact support and
hence quantities of this type are manifestly invariant. In such a
theory, time evolution would be meaningful only in these regions and
could be associated with the appropriate generator of some sort of
asymptotic Poincar\'e group. DeWitt used the concept of an asymptotic
observable in his seminal investigations of covariant quantum gravity
(DeWitt 1965, 1967a, 1967b, 1967c)
\nocite{DeW65,DeW67a,DeW67b,DeW67c}.

    However, it is not clear to what extent this really solves the
problem of time. An asymptotic time might suffice for calculations of
graviton-graviton scattering amplitudes, but it is difficult to see
how it could be used in the quantum cosmological situations to which
so much attention has been devoted in recent years. One can also
question the extent to which such a notion of time can be related
operationally to what could be measured with the aid of real physical
clocks. This is particularly relevant in the light of the discussion
below of the significance of internal time coordinates.

    In any event, the present paper is concerned mainly with the problem
of time in the context of compact three-spaces, or of non-compact spaces
but with boundary data playing no fundamental role. In theories of this
type, the problem of time and the spacetime diffeomorphism group might be
approached in several different ways.
\bi
    \item {General relativity could be forced into a Newtonian
framework by assigning special status to some particular foliation of
$\M$. For example, a special background metric $\g$ could be
introduced to act as a source of preferred reference frames and
foliations. However, the action of a diffeomorphism on $\M$ generally
maps a hypersurface into a different one, and hence insisting on
preserving the special foliation generates a type of symmetry
breaking in which the $\DM$ invariance is reduced to the group of
transformations that leave the foliation invariant. If the foliation
is tied closely to the properties of the metric $\g$, this group is
likely to be just the group of isometries (if any) of $\g$.

    There is an interesting point at issue at here. Some people (for
example, certain process philosophers) might argue that, since we do
in fact live in one particular universe, we should be free to exploit
whatever special characteristics it might happen to have. Thus, for
example, we could use the background $3^0$K radiation to define some
quasi-Newtonian universal time associated with the Robertson-Walker
homogeneous cosmology with which it is naturally associated.
\footnote{An interesting recent suggestion by \citen{Val92} is that a
preferred foliation of spacetime could arise from the existence of
nonlocal hidden-variables.}
On the other hand, theoretical physicists tend to want to consider
all {\em possible\/} universes under the umbrella of a single
theoretical structure. Thus there is not much support for the idea of
focussing on a special foliation associated with a contingent feature
of the actual universe in which we happen to find ourselves.}

    \item  At the other extreme, one might seek a new interpretation
of quantum theory in which the concept of time does not appear at all
(or, at the very least, plays no fundamental role). Our familiar
notion of temporal evolution will then have to `emerge' from the
formalism in some phenomenological way. As we shall see, some of the
most interesting approaches to the problem of time are of this type.

    \item Events in $\M$ might be identified using the positions of
physical particles. For example, if $\f$ is a scalar field on $\M$
the value of the field where a particular particle {\em
is\/}, is a $\DM$-invariant number and is hence an observable in the
sense discussed above. This idea can be generalised in several ways.
In particular, the value of $\f$ could be specified at that event
where some collection of fields takes on a certain set of values
(assuming there is a unique such event). These fields, or `internal
coordinates', might be specified using distributions of matter, or
they could be part of the gravitional field complex itself. Almost
all of the existing approaches to the problem of time involve ideas
of this type.
\ei

    This last point emphasises the important fact that a physical
definition of time requires more than just saying that it is a local
coordinate on the spacetime manifold. For example, in the real world,
time is often measured with the aid of a spatially-localised physical
clock, and this usually means the proper time along its worldline.
This quantity is $\DM$-invariant if the diffeomorphism group is
viewed as acting simultaneously on the points in $\M$ (and hence on
the world line) and the spacetime metric $\g$. More precisely, if the
beginning and end-points of the world-line are labelled using
internal coordinates, the proper time along the geodesic connecting
them is an intrinsic property.

    The idea of labelling spacetime events with the aid of physical
clocks and spatial reference frames is of considerable importance in
both the classical and the quantum theory of relativity. However, in
the case of the quantum theory the example of proper time raises
another important issue. The calculation of the value of an interval
of proper time involves the spacetime metric $\g$, and therefore it
has a meaning only {\em after\/} the equations of motion have been
solved (and of course these equations include a contribution from
the energy-momentum tensor of the matter from which the clock is
made). This causes no problem in the classical theory, but
difficulties arise in any theory in which the geometry of spacetime
is subject to quantum fluctuations and therefore has no fixed value.
There is an implication that time may become a quantum
operator: a problematic concept that is not part of standard
quantum theory .

\subsection{Approaches to the Problem of Time}
Most approaches to resolving the problem of time in quantum gravity
agree that `time' should be identified in terms of the internal
structure of the system rather than being regarded as any sort of
external parameter. Their differences lie in:
\bi
\item the way in which such an identification is made;

\item whether it is done before or after quantisation;

\item the degree to which the resulting entity resembles the familiar
time of conventional classical and quantum physics;

\item the role it plays in the final interpretation of the theory.
\ei
In turn, these differences are closely related to the general
question of how to handle the constraints on the canonical variables
that are such an intrinsic feature of the canonical theory of general
relativity (\S\ref{Sec:CanGR}).

    As discussed in \citen{Kuc92a}, the various approaches to the
problem of time in quantum gravity can be organised into three broad
categories:
\bd
    \item[I\ \ \ \,\,] The first class of schemes are those in which
an internal time is identified as a functional of the canonical
variables, and the canonical constraints are solved, {\em before\/}
the system is quantised. The aim is to reproduce something like a
normal Schr\"odinger equation with respect to this choice of time.
This category is the most conservative of the three since its central
assumption is that the construction of {\em any\/} coherent
quantum-theoretical structure requires something resembling the
external, classical time of standard quantum theory.

    \item[II\ \ \,\,] In the second type of scheme the procedure
above is reversed and the constraints are imposed at a quantum level
as restrictions on allowed state vectors and with time being identified
only {\em after\/} this step. The states can be written as
functionals $\Psi[g]$ of a three-geometry $g_{ab}(x)$ (the basic
configuration variable in the theory) and the most important of the
operator constraints is a functional differential equation for
$\Psi[g]$ known as the Wheeler-DeWitt equation \eq{WDE}. The notion
of time has to be recovered in some way from the solutions to this
central equation. The key feature of approaches of this type is that
the final probabilistic interpretation of the theory is made only
{\em after\/} the identification of time. Thus the Hilbert space
structure of the final theory may be related only very indirectly (if
at all) to that of the quantum theory with which the construction
starts.

    \item [III\ \,] The third class of scheme embraces a variety of
methods that aspire to maintain the timeless nature of general
relativity by avoiding any specific conception of time in the quantum
theory. Some start, as in II, by imposing the constraints at a
quantum level, others proceed along somewhat different lines, but
they all agree in espousing the view that it is possible to construct
a technically coherent, and {\em conceptually complete\/}, quantum
theory (including the probabilistic interpretation) without needing
to make any direct reference to the concept of time which, at most,
has a purely phenomenological status. It is this latter feature that
separates these schemes from those of category II.
\ed
These three broad categories can be further sub-divided into the
following ten specific approaches to the problem of time.

\bd
\item[I]{\bf Tempus ante quantum}
    \be
    \item {\em The internal Schr\"odinger interpretation\/}. Time and
space coordinates are \mbox{identified} as specific functionals of the
gravitational canonical variables, and are then separated from the
dynamical degrees of freedom by a canonical transformation. The
constraints are solved classically for the momenta conjugate to these
variables, and the remaining physical (\ie `non-gauge') modes of the
gravitational field are then quantised in a conventional way, giving rise
to a Schr\"odinger evolution equation for the physical states.

    \item {\em Matter clocks and reference fluids\/}. This is an
extension of the internal Schr\"odinger interpretation in which
matter variables coupled to the geometry are used to label spacetime
events. They are introduced in a special way aimed at facilitating
the handling of the constraints that yield the Schr\"odinger
equation.

     \item {\em Unimodular gravity\/}. This is a modification of
general relativity in which the cosmological constant $\lambda$ is
considered as a dynamical variable. A `cosmological' time is
identified as the variable conjugate to $\lambda$, and the
constraints yield the Schr\"odinger equation with respect to this
time. This approach can be treated as a special case of a reference
fluid.
\ee

\item[II]{\bf Tempus post quantum}
\be
    \item {\em The Klein-Gordon interpretation\/}. The Wheeler-DeWitt
equation is considered as an infinite-dimensional analogue of the
Klein-Gordon equation for a relativistic particle moving in a fixed
background geometry. The probabilistic interpretation of the theory
is based on the Klein-Gordon norm with the hope that it will be
positive on some appropriate subspace of solutions to the
Wheeler-DeWitt equation.

    \item {\em Third quantisation\/}. The problems arising from the
indefinite nature of the scalar product of the Klein-Gordon
interpretation are addressed by suggesting that the solutions
$\Psi[g]$ of the Wheeler-DeWitt equation are to be turned into
operators. This is analogous to the second quantisation of a
relativistic particle whose states are described by the Klein-Gordon
equation.

    \item {\em The semi-classical interpretation\/}. Time is deemed
to be a meaningful concept only in some semi-classical limit of the
quantum gravity theory based on the Wheeler-DeWitt equation. Using a
form of WKB expansion, the Wheeler-DeWitt equation is approximated by
a conventional Schr\"odinger equation in which the time variable is
extracted from the state $\Psi[g]$. The probabilistic interpretation
arises only at this level.
\ee

\item[III] {\bf Tempus nihil est}
    \be
    \item {\em The \naive Schr\"odinger interpretation\/}. The square
$|\Psi[g]|^2$ of a solution $\Psi[g]$ of the Wheeler-DeWitt equation
is interpreted as the probability density for `finding' a spacelike
hypersurface of $\M$ with the geometry $g$. Time enters as
an internal coordinate function of the three-geometry, and is
represented by an operator that is part of the quantisation of the
complete three-geometry.

    \item {\em The conditional probability interpretation\/}. This
can be regarded as a sophisticated development of the \naive
Schr\"odinger interpretation whose primary ingredient is the use of
conditional probabilities for the results of a pair of observables
$A$ and $B$. This is deemed to be correct even in the absence of any
proper notion of time; as such, it is a modification of the
conventional quantum-theoretical formalism. In certain cases, one of
the observables is regarded as defining an instant of time (\ie it
represents a physical, and therefore imperfect, clock) at which the
other variable is measured
\footnote{Or, perhaps, `has a value'; the
language used reflects the extent to which one favours operational
or realist interpretations of quantum theory. In practice, quantum
schemes of type III are particularly prone to receive a
`many-worlds' interpretation.}.
Dynamical evolution is then equated to the dependence of these
conditional probabilities on the values of the internal clock
variables.

    \item {\em Consistent histories approach\/}. This is based on a
far-reaching extension of normal quantum theory to a form that does
not require the conventional Copenhagen interpretation. The main
ingredient is a precise prescription from within the formalism itself
that says when it is, or is not, meaningful to ascribe a probability
to a {\em history\/} of the system. The extension to quantum gravity
involves defining the notion of a `history' in a way that avoids
having to make any direct reference to the concept of time. In its
current form, the scheme culminates in the hope that functional
integrals over spacetime fields may be well-defined, even in the absence of
a conventional Hilbert space structure.

    \item {\em The frozen time formalism\/}. Observables in quantum
gravity are declared to be operators that commute with all the
constraints, and are therefore constants of the motion. Attempts are
made to show that, although `timeless', such observables can
nevertheless be used to give a picture of dynamical evolution.
    \ee
\ed

\subsection{Technical Problems With Time}
\label{SSec:TProb}
In the context of either of the first two types of
scheme---I constrain before quantising, or II quantise before
constraining---a number of potential technical problems can be
anticipated. Some of the most troublesome are as follows (see
\citen{Kuc92a}).
\bi
    \item {\em The ultra-violet divergence problem\/.} Both schemes
involve complicated classical functions of fields defined at the same
spatial point. The perturbative non-renormalisability of quantum
gravity suggests that the operator analogues of these expressions
are very ill-defined. This pathology has no direct connection
with most aspects of the problem of time but it throws a big
question mark over some of the techniques used to tackle that
problem.

    \item {\em The operator-ordering problem\/.} Horrendous
operator-ordering difficulties arise when attempts are made to
replace the classical constraints and Hamiltonians with operator
equivalents. These difficulties cannot easily be separated from the
ultra-violet divergence problem.

    \item {\em The global time problem\/.} Experience with
non-abelian gauge theories suggests the existence of global
obstructions to making the crucial canonical transformations that
untangle the physical modes of the gravitational field from internal
spacetime coordinates.

    \item {\em The multiple-choice problem\/.} The Schr\"odinger
equation based on one particular choice of internal time may give a
different quantum theory from that based on another. Which, if any,
is correct? Can these different quantum theories be seen to be part of
an overall scheme that {\em is\/} covariant? A similar problem can be
anticipated with the identification of time in the solutions of the
Wheeler-DeWitt equation.

    \item{\em The Hilbert space problem\/.} Schemes of type I have
the big advantage of giving a natural inner product that is conserved
with respect to the internal time variables. This leads to a
straightforward interpretative framework. In the alternative
constraint quantisation schemes the situation is quite different.
The Wheeler-DeWitt equation is a second-order functional differential
equation, and as such presents familiar problems if one tries to
construct a genuine, positive-definite inner product on the space of
its solutions. This is the `Hilbert space problem'.

    \item{\em The spatial metric reconstruction problem.} The
classical separation of the canonical variables into physical and
non-physical parts can be inverted and, in particular, the metric
$g_{ab}$ can be expressed as a functional of the dynamical and
non-dynamical modes. The `spatial metric reconstruction problem' is
whether something similar can be done at the quantum level. This is
part of the general question of the extent to which classical
geometrical properties can be, or should be, preserved in the quantum
theory.

    \item{\em The spacetime problem\/.} If an internal space or time
coordinate is to operate within a conventional spacetime context, it
is necessary that, viewed as a function on $\M$, it be a {\em
scalar\/} field; in particular, it must not depend on any background
foliation of $\M$. However, the objects used in the canonical
approach to general relativity are functionals of the {\em
canonical\/} variables, and there is no {\em prima facie\/} reason
for supposing they will satisfy this condition. The spacetime problem
consists in finding functionals that do have this desirable property
or, if this is not possible, understanding how to handle the
situation and what it means in spacetime terms.

    \item{\em The problem of functional evolution\/.} This affects
both approaches to the canonical theory of gravity. In the scheme in
which time is identified before quantisation, the problem is the
possible existence of anomalies in the algebra of the local
Hamiltonians that are associated with the generalised Schr\"odinger
equation. Any such anomaly would render this equation inconsistent.
In quantisation schemes of type II (and in the appropriate schemes of
type III), the worry is potential anomalies in the quantum version of
the canonical analogue of the Lie algebra of the spacetime
diffeomorphism group $\DM$. In both cases, the consistency of the
classical evolution (guaranteed by the closing properties of the
appropriate algebras) is lost.
\ei
In theories of category III, time plays only a secondary role, and
therefore, with the exception of the first two, most of the problems
above are not directly relevant. However, analogues of several of
them appear in type III schemes, which also have additional
difficulties of their own. Further discussion is deferred to the
appropriate sections of the notes.


\section{CANONICAL GENERAL RELATIVITY}\label{Sec:CanGR}
\subsection{Introductory Remarks}
Most of the discussion of the problem of time in quantum gravity has
been within the framework of the canonical theory whose starting
point is a three-dimensional manifold $\Si$ which serves as a model
for physical space. This is often contrasted with the so-called
`spacetime' (or `covariant') approaches in which the basic
entity is a four-dimensional spacetime manifold $\M$. The virtues
and vices of these two different approaches have been the subject of
intense debate over the years and the matter is still far from being
settled. Not surprisingly, the problem of time looks very different
in the two schemes. Some of the claimed advantages of the canonical
approach to quantum gravity are as follows.

\be
    \item The spacetime approaches often employ formal
functional-integral techniques in which a number of serious
difficulties are swept under the carpet as problems concerned with
the integration measure, the contour of integration \etc. On the
other hand, canonical quantisation is usually discussed in an
operator-based framework, with the advantage that problems appear in
a more explicit and, perhaps, tractable way.

    \item The development of quantisation techniques that do not
depend on a background metric seems to be easier in the canonical
framework. This is particularly relevant to the problem of time.

    \item For related reasons, canonical quantisation is better
suited for discussing quantum cosmology, spacetime singularities and
similar topics.

    \item Canonical methods tend to place more emphasis on the
geometrical structure of general relativity. In particular, it is
easier to address the issue of the extent to which such structure is,
or should be, maintained in the quantum theory.

    \item Many of the deep conceptual problems in quantum gravity are
more transparent in a canonical approach. This applies in particular
to the problem of time and the, not unrelated, general question of
the domain of applicability of the interpretative framework of
conventional quantum theory.
\ee
It should be emphasised that some of these advantages arise only in
comparison with weak-field perturbation theory, and this does not
mean that canonical methods are intrinsically superior to those in
which spacetime fields are used from the outset. For example, there
could exist {\em bona fide\/} approaches to quantum gravity that
involve the calculation of a functional integral like
\beq
        Z = \int{\cal D}[g]\,e^{iS(g)}
\eeq
using methods other than weak-field perturbation theory. A
particularly interesting example is the consistent histories approach
to the problem of time discussed in \S\ref{SSec:CPI}.

\subsection{Quantum Field Theory in a Curved Background}
\label{SSec:QFTM}
\subsubsection{The Canonical Formalism}
In approaching the canonical theory of general relativity it is
helpful to begin by considering briefly the canonical quantisation
of a scalar field $\f$ propagating in a background Lorentzian metric
$\g$ on a spacetime manifold $\M$. The action for the system is
\beq
    S[\f]=-\int_\M d^4X\,\big(-\det\g(X)\big)^\half
        \left(\half\g^{\a\b}(X)\,\pa_\a\f(X)\,\pa_\b\f(X) +
            V\big(\f(X)\big)\right)                     \label{S:f}
\eeq
where $V(\f)$ is an interaction potential for $\f$ (which could
include a mass term).

    This system has a well-posed classical Cauchy problem only if
$(\M,\g)$ is a globally-hyperbolic pair \cite{HE73}. In particular,
this means that $\M$ is topologically equivalent to the Cartesian
product $\SiR$ where $\Si$ is some spatial three-manifold and $\R$ is
a global time direction. To acquire the notion of dynamical evolution
we need to {\em foliate\/} $\M$ into a one-parameter family of
embeddings $\Fo_t:\Si\map\M$, $t\in\R$, of $\Si$ in $\M$ that are
spacelike with respect to the background Lorentzian metric
$\g_{\a\b}$; the real number $t$ can then serve as a time parameter.
To say that an embedding $\Em:\Si\map\M$ is {\em spacelike\/} means
that the pull-back $\Em^*(\g)$---a symmetric rank-two covariant
tensor field on $\Si$---has signature $(1,1,1)$ and is positive
definite. We recall that the components of $\Em^*(\g)$ are
\footnote{The symbol $\Em^\a(x)$ means $X^\a\big(\Em(x)\big)$ where
$X^\a$, $\a=0,1,2,3$ is a coordinate system on $\M$. The quantity
defined by the left hand side of \eq{Def:Em*g} is independent of the
choice of such a system.}
\beq
    \big(\Em^*(\g)\big)_{ab}(x):=\g_{\a\b}(\Em(x))
        \,\Em^\a,_a(x)\,\Em^\b,_b(x)                \label{Def:Em*g}
\eeq
on the three-manifold $\Si$.

    The canonical variables are defined on $\Si$ and consist of the
scalar field $\f$ and its conjugate variable $\pi$ (a scalar
density) which is essentially the time-derivative of $\f$.
Classically these constitute a well-defined set of Cauchy data and
satisfy the Poisson bracket relations
\beqa
    \{\f(x),\f(x')\}   &=&0                         \label{PB:ff}\\
    \{\pi(x),\pi(x')\} &=&0                         \label{PB:pipi}\\
    \{\f(x),\pi(x')\}  &=&\de(x,x').                \label{PB:fpi}
\eeqa
The Dirac $\de$-function is defined such that the smeared fields
satisfy
\beq
    \{\f(f),\pi(h)\}= \int_\Si d^3x\,f(x)\,h(x)
\eeq
where the test-functions $f$ and $h$ are respectively a scalar
density and a scalar
\footnote{We are using a definition of the Dirac delta function
$\de(x,x')$ that is a scalar in $x$ and a scalar density in $x'$. Thus, if
$f$ is a scalar function on $\Si$ we have, formally, $f(x)=\int_\Si
d^3x'\,\de(x,x')f(x')$.}
on the three-manifold $\Si$.

    The dynamical evolution of the system is obtained by constructing
the Hamiltonian $H(t)$ in the usual way from the action \eq{S:f} and
the given foliation of $\M$. The resulting equations of motion are
\beqa
    {\pa\f(x,t)\over\pa t} &=& \big\{\f(x),H(t)\big\}       \label{EM:f}\\
    {\pa\pi(x,t)\over\pa t}&=& \big\{\pi(x),H(t)\big\}.     \label{EM:pi}
\eeqa
Note that these give the evolution with respect to the time parameter
associated with the specified foliation. However, the physical fields
can be evaluated on any spacelike hypersurface, and this should not
depend on the way the hypersurface happens to be included in a
particular foliation. Thus it should be possible to write the
physical fields as functions $\f(x,\Em\,]$, $\pi(x,\Em\,]$ of
$x\in\Si$ and functionals of the embedding functions $\Em$. Indeed,
the Hamiltonian equations of motion \eqs{EM:f}{EM:pi}) are valid for
all foliations and imply the existence of four functions
$h_\a(x,\Em\,]$ of the canonical variables such that $\f(x,\Em\,]$
and $\pi(x,\Em\,]$ satisfy the functional differential equations
\beqa
    {\de\f(x,\Em\,]\over\de \Em^\a(x')}&=&
       \big\{\f(x,\Em\,],h_\a(x',\Em\,]\big\}       \label{FEM:f}\\
{\de\pi(x,\Em\,]\over\de \Em^\a(x')}&=&
      \big\{\pi(x,\Em\,],h_\a(x',\Em\,]\big\}       \label{FEM:pi}
\eeqa
that describe how these fields change under an infinitesimal
deformation of the embedding $\Em$.

\subsubsection{Quantisation of the System}
The formal canonical quantisation of this system follows the usual
rule of replacing Poisson brackets with operator commutators. Thus
\eqs{PB:ff}{PB:fpi} become
\beqa
    [\,\op\f(x),\op\f(x')\,]   &=& 0                      \label{CR:ff}\\
    {[}\,\op\pi(x),\op\pi(x')\,] &=& 0                    \label{CR:pipi}\\
    {[}\,\op\f(x),\op\pi(x')\,]  &=& i\hbar\,\de(x,x')    \label{CR:fpi}
\eeqa
which can be made rigorous by smearing and exponentiating in the
standard way.

    The next step is to choose an appropriate representation of this
operator algebra. One might try to emulate elementary wave mechanics
by taking the state space of the quantum field theory to be a set of
functionals $\Psi$ on the topological vector space $E$ of all
classical fields, with the canonical operators defined as
\beqa
    \big(\op\f(x)\Psi\big)[\f] &:=& \f(x)\Psi[\f]   \label{Def:op-f}\\
    \big(\op\pi(x)\Psi\big)[\f] &:=& -i\hbar{\de\Psi[\f]\over\de\f(x)}.
                                                    \label{Def:op-pi}
\eeqa
Thus the inner product would be
\beq
    \brac{\Psi}{\Phi}=\int_E d\m[\f]\,\Psi^*[\f]\,\Phi[\f]
                                                    \label{SP:E}
\eeq
and, for a normalised function $\Psi$,
\beq
    \Prob(\f\in B;\Psi)=\int_B d\m[\f]\,|\Psi[\f]|^2
                                                    \label{Pr:finB}
\eeq
is the probability that a measurement of the field on $\Si$ will find
it in the subset $B$ of $E$.

   This analysis can be made rigorous after smearing the fields with
functions from an appropriate test function space. It transpires that
the support of a state functional $\Psi$ is typically on {\em
distributions\/}, rather than smooth functions, so that the inner
product is really
\beq
    \brac{\Psi}{\Phi}=\int_{E'} d\m[\f]\,\Psi^*[\f]\,\Phi[\f]
                                                        \label{SP:E'}
\eeq
where $E'$ denotes the topological dual of $E$. Furthermore, there
is no infinite-dimensional version of Lebesgue measure, and hence if
\eq{Def:op-pi} is to give a self-adjoint operator it must be
modified to read
\beq
    \big(\op\pi(x)\Psi\big)[\f] := -i\hbar{\de\Psi[\f]\over\de\f(x)}
                                    +i\r(x)\Psi[\f]      \label{Def:op-pi-r}
\eeq
where $\r(x)$ is a function that compensates for the weight factor in
the measure $d\mu$ used in the construction of the Hilbert space of
states $L^2(E',d\mu)$.

    The dynamical evolution of the system can be expressed in the
Heisenberg picture as the commutator analogue of the Poisson bracket
relations \eqs{EM:f}{EM:pi}. Alternatively, one can adopt the
Schr\"odinger picture in which the time evolution of state vectors is
given by
\footnote{As usual, the notation $H(t;\op\f,\op\pi\,]$ must be taken
with a pinch of salt. The classical fields $\f(x)$ and $\pi(x)$ can
be replaced in the Hamiltonian $H(t;\f,\pi\,]$ with their operator
equivalents $\op\f(x)$ and $\op\pi(x)$ only after a careful
consideration of operator ordering and regularisation.}
\beq
    i\hbar{\pa\Psi(t,\f]\over\pa t}=H(t;\op\f,\op\pi\,]\,\Psi(t,\f]
                                                    \label{SE:f}
\eeq
or, if the classical theory is described by the functional equations
\eqs{FEM:f}{FEM:pi}, by the functional differential equations
\beq
    i\hbar{\de\Psi[\Em,\f]\over\de\Em^\a(x)}=
        h_\a(x;\Em,\op\f,\op\pi\,]\,\Psi[\Em,\f].     \label{FSE:f}
\eeq
However, note that the steps leading to \eq{SE:f} or \eq{FSE:f} are
valid only if the inner product on the Hilbert space of states is $t$
(resp.\ $\Em$) independent. If not, a compensating term must be added
to \eq{SE:f} (resp.\ \eq{FSE:f}) if the scalar product
$\brac{\Psi_t}{\Phi_t}_t$ (resp.\ $\brac{\Psi_\Em}{\Phi_\Em}_\Em$) is
to be independent of $t$ (resp.\ the embedding $\Em$). This is
analogous to the quantum theory of a particle moving on an
$n$-dimensional Riemannian manifold $Q$ with a time-dependent
background metric $g(t)$. The natural inner product
\beq
    \brac{\psi_t}{\phi_t}_t:=\int_Q d^nq\,\big(\det g(q,t)\big)^\half\,
          \psi^*_t(q)\,\phi_t(q)                    \label{SE:Q}
\eeq
is not preserved by the \naive Schr\"odinger equation
\beq
    i\hbar{d\psi_t\over dt}=\op H(t)\psi_t
\eeq
because the time derivative of the right hand side of \eq{SE:Q}
acquires an extra term coming from the time-dependence of the metric
$g(t)$. In the Heisenberg picture the weight function $\rho(q)$
becomes time-dependent.

    The major technical problem in quantum field theory on a curved
background $(\M,\g)$ is the existence of infinitely many unitarily
inequivalent representations of the canonical commutation relations
\eqs{CR:ff}{CR:fpi}. If the background metric $\g$ is static, the
obvious step is to use the timelike Killing vector to select the
representation. This gives rise to a consistent one-particle picture
of the quantum field theory. In other special, but non-static, cases
(for example, a Robertson-Walker metric) there may be a `natural'
choice for the representations that gives rise to a picture of
particle creation by the background metric, and for genuine astrophysical
applications this may be perfectly adequate.

    The real problems arise if one is presented with a generic metric
$\g$, in which case it is not at all clear how to proceed. A minimum
requirement is that the Hamiltonians $\op H(t)$, or the Hamiltonian
densities $\op h_\a(x,\Em\,]$, should be well-defined. However, there
is an unpleasant possibility that the representations could be $t$
(resp.\ $\Em$) dependent, and in such a way that those corresponding
to different values of $t$ (resp.\ $\Em$) are unitarily inequivalent,
in which case the dynamical equations \eq{SE:f} (resp.\ \eq{FSE:f})
are not meaningful. This particular difficulty can be overcome by
using a $C^*$-algebra approach, but the identification of
physically-meaningful representations remains a major problem.

\subsection{The Arnowitt-Deser-Misner Formalism}\label{SSec:ADM}
\subsubsection{Introduction of the Foliation}
Let us consider now how these ideas extend to the canonical formalism
of general relativity itself. The early history of this subject was
grounded in the seminal work of Dirac
(1958a,1958b)\nocite{Dir58a,Dir58b} and culminated in the
investigations by Arnowitt, Deser and Misner (1959a, 1959b, 1960a,
1960b, 1960c, 1960d, 1961a, 1961b, 1962).
\nocite{ADM59a,ADM59b,ADM60a,ADM60b,ADM60c,ADM60d,ADM61a,ADM61b,ADM62}
These original studies involved the selection of a specific
coordinate system on the spacetime manifold, and some of the global
issues were thereby obscured. A more geometrical, global approach
was developed by Kucha\v{r} in a series of papers Kucha\v{r} (1972,
1976a, 1976b, 1976c, 1977, 1981a)
\nocite{Kuc72,Kuc76a,Kuc76b,Kuc76c,Kuc77,Kuc81a} and this will
be adopted here. The treatment will be fairly cursory since the main
aim of this course is to develop the conceptual and structural
aspects of the problem of time. A more detailed account of the
technical issues can be found in the forthcoming work \citen{IK94}.

    The starting point is a four-dimensional manifold $\M$ and a
three-dimensional manifold $\Si$ that  play the roles of physical
spacetime and three-space respectively. The space $\Si$ is assumed
to be compact; if not, some of the expressions that follow must be
augmented by surface terms. Furthermore, the topology of $\M$ is
assumed to be such that it can be foliated by a one-parameter family
of embeddings $\Fo_t:\Si\map\M$, $t\in\R$, of $\Si$ in $\M$ (of
course, then there will be many such foliations). As in the case of
the scalar field theory, this requirement imposes a significant {\em
a priori\/} topological limitation on $\M$ since (by the definition
of a foliation) the map $\Fo:\SiR\map\M$, defined by
$(x,t)\mapsto\Fo(x,t):=\Fo_t(x)$, is a diffeomorphism of $\SiR$ with $\M$.

  Since $\Fo$ is a diffeomorphism from $\SiR$ to $\M$, its inverse
$\Fo^{-1}:\M\map\SiR$ is also a diffeomorphism and can be written in
the form
\beq
    \Fo^{-1}(X) = \big(\s(X),\tau(X)\big) \in\SiR
\eeq
where $\s:\M\map\Si$ and $\tau:\M\map\R$. The map $\tau$ is a global
{\em time function\/} and gives the natural time parameter
associated with the foliation in the sense that $\tau(\Fo_t(x))=t$
for all $x\in\Si$. However, from a physical point of view such a
definition of `time' is rather artificial (how would it be
measured?) and is a far cry from the notion of time employed in the
construction of real clocks. This point is not trivial and we shall
return to it later.

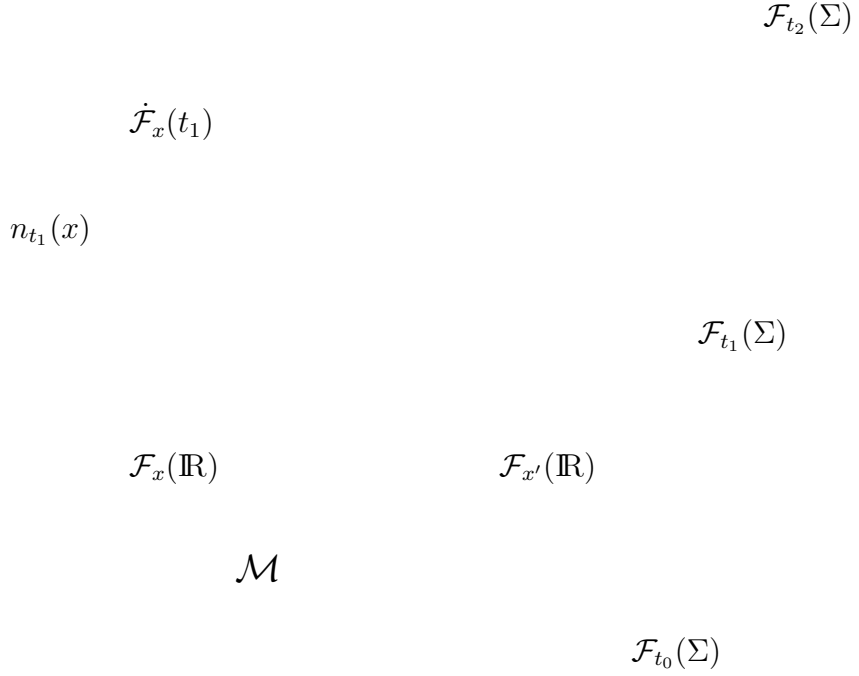
\begin{figure}[t]
\begin{picture}(400,300)(0,0)
    \put(400,300){$\Fo_{t_2}(\Si)$}
    \put(160,260){$\dot\Fo_x(t_1)$}
    \put(115,220){$n_{t_1}(x)$}
    \put(375,180){$\Fo_{t_1}(\Si)$}
    \put(160,130){$\Fo_x(\R)$}
    \put(200,90){\large $\M$}
    \put(300,130){$\Fo_{x'}(\R)$}
    \put(350,60){$\Fo_{t_0}(\Si)$}
\end{picture}
\caption{The flow lines of the foliation of $\M$}\label{Fig:flow-fol}
\end{figure}

    Note that for each $x\in\Si$ the map $\Fo_x:\R\map\M$ defined
by $t\mapsto\Fo(x,t)$ is a curve in $\M$ and therefore has a
one-parameter family of tangent vectors on $\M$, denoted
$\dot\Fo_x(t)$, whose components are
$\dot\Fo_x^\a(t)=\pa\Fo^\a(x,t)/\pa t$. The flow lines of the
ensuing vector field (known as the {\em deformation\/} vector
\footnote{The deformation vector is a hybrid object in the sense
that if $x$ is a point in the three-space $\Si$ the vector
$\dot\Fo_x(t)$ lies in the tangent space $T_{\Fo(x,t)}\M$ at the
point $\Fo(x,t)$ in the four-manifold $\M$. Such an object is best
regarded as an element of the space $T_{\Fo_t}\E$ of vectors tangent
to the infinite-dimensional manifold $\E$ of embeddings of $\Si$ in
$\M$ at the particular embedding $\Fo_t$. This way of viewing things
is quite useful technically but I shall not develop it further
here.}
of the foliation) are illustrated in Figure~\ref{Fig:flow-fol}
which also shows the normal vector $n_{t_1}$ on the hypersurface
$\Fo_{t_1}(\Si)\subset\M$. In general, if $\Em:\Si\map\M$ is a
spacelike embedding, the {\em normal\/} vector field $n$ to $\Em$ is
defined by the equations
\beqa
    n_\a(x,\Em\,]\,\Em^\a,_a(x) &=& 0                   \label{Def:n1}\\
\g^{\a\b}(\Em(x))\,n_\a(x,\Em\,]\,n_\b(x,\Em\,] &=& -1  \label{Def:n2}
\eeqa
for all $x\in\Si$. Equation \eq{Def:n1} defines what it means to say
that $n$ is {\em normal\/} to the hypersurface $\Em(\Si)$, while \eq{Def:n2}
is a normalisation condition on $n$ and emphasises that this vector
is timelike with respect to the Lorentzian metric on $\M$. The minus
sign on the right hand side of \eq{Def:n2} reflects our choice of
signature on $(\M,\g)$ as $(-1,1,1,1)$.

    In considering the dynamical evolution, a particularly
interesting quantity is the functional derivative of
$g_{ab}(x,\Em\,]$ with respect to $\Em$ projected along the normal
vector $n$. A direct calculation \cite{HKT74,Kuc74} shows that
\beq
    n^\a(x,\Em\,]\,{\de\over\de \Em^\a(x)}g_{ab}(x',\Em\,]=
    -2K_{ab}(x,\Em\,]\,\de(x,x')                      \label{Del-g-Em}
\eeq
where $K$ is the {\em extrinsic curvature\/} of the hypersurface
$\Em(\Si)$ defined by
\beq
K_{ab}(x,\Em\,]:=-{}^4\nabla_\a n_\b(x,\Em\,]\,\Em^\a,_a(x)\,\Em^\b,_b(x)
                                                    \label{Def:K}
\eeq
where ${}^4\nabla_\a n_\b(x,\Em\,]$ denotes the covariant derivative
obtained by parallel transporting the cotangent vector
$n(x,\Em\,]\in T^*_{\Em(x)}\M$ along the hypersurface $\Em(\Si)$
using the Lorentzian four-metric $\g$ on $\M$. It is straightforward
to show that $K$ is a symmetric tensor, \ie
$K_{ab}(x,\Em\,]$=$K_{ba}(x,\Em\,]$.

\subsubsection{The Lapse Function and Shift Vector}
The next step is to decompose the deformation vector into two
components, one of which lies along the hypersurface $\Fo_t(\Si)$
and the other of which is parallel to $n_t$. In particular we can
write
\beq
    \dot\Fo^\a(x,t)=\Np(x,t)\,\g^{\a\b}(\Fo(x,t))\,n_\b(x,t) +
          N^a(x,t)\,\Fo^\a,_a(x,t)                  \label{Def:LS}
\eeq
where we have used $n_\b(x,t)$ instead of the more clumsy notation
$n_\b(x,\Fo_t\,]$. The quantities $\Np(t)$ and $N^a(t)$ are known
respectively as the {\em lapse\/} function and the {\em shift\/}
vector associated with the embedding $\Fo_t$.

    Note that, like the normal vector, the lapse and shift depend on
both the spacetime metric $\g$ and the foliation. This relationship
can be partially inverted with, for a fixed foliation, the lapse and
shift functions being identified as parts of the metric tensor $\g$.
This can be seen most clearly by studying the `pull-back'
$\Fo^*(\g)$ of $\g$ by the foliation $\Fo:\SiR\map\M$ in coordinates
$X^\a$, $\a=0\ldots3$, on $\SiR$ that are adapted to the product
structure in the sense that $X^{\a=0}(x,t)=t$, and
$X^{\a=1,2,3}(x,t)=x^{a=1,2,3}(x)$ where $x^a$, $a=1,2,3$ is some
coordinate system on $\Si$. The components of $\Fo^*(\g)$ are
\beqa
    (\Fo^*\g)_{00}(x,t) &=& N^a(x,t)\,N^b(x,t)\,g_{ab}(x,t) -
            \big(\Np(x,t)\big)^2                        \label{PuB:g00}\\
    (\Fo^*\g)_{0a}(x,t) &=& N^b(x,t)\,g_{ab}(x,t)       \label{PuB:g0a}\\
    (\Fo^*\g)_{ab}(x,t) &=& g_{ab}(x,t)                 \label{PuB:gab}
\eeqa
where $g_{ab}(x,t)$ is shorthand for $g_{ab}(x,\Fo_t\,]$ and is given by
\beq
    g_{ab}(x,t):=(\Fo_t^*\,\g)_{ab}(x)
    =\g_{\a\b}(\Fo(x,t))\,\Fo^\a,_a(x,t)\,\Fo^\b,_b(x,t)\label{PuB:gab=}.
\eeq

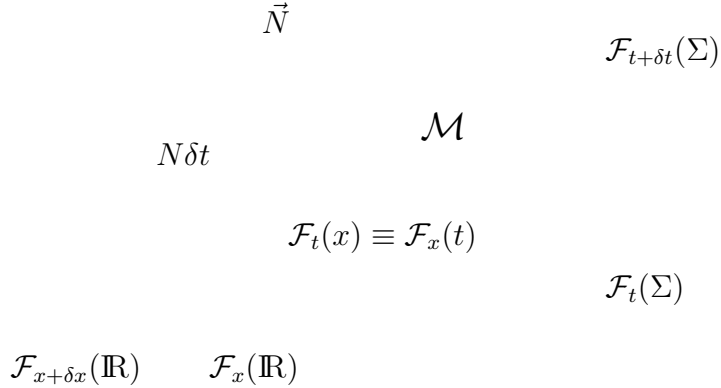
\begin{figure}[t]
\begin{picture}(400,250)(0,0)
    \put(125,100){$\Fo_{x+\de x}(\R)$}
    \put(200,100){$\Fo_{x}(\R)$}
    \put(350,130){$\Fo_t(\Si)$}
    \put(230,150){$\Fo_t(x)\equiv\Fo_x(t)$}
    \put(180,180){$\Np\de t$}
    \put(280,190){\large $\M$}
    \put(220,230){$\vec{N}$}
    \put(350,220){$\Fo_{t+\de t}(\Si)$}
\end{picture}
\caption{The lapse function and shift vector of a foliation}
\label{Fig:lapse-shift}
\end{figure}

    We see from \eq{Def:LS} that the lapse function and shift vector
provide information on how a hypersurface with constant time
parameter $t$ is related to the displaced hypersurface with constant
parameter $t+\de t$ as seen from the perspective of an enveloping
spacetime. More precisely, for a given Lorentzian metric $\g$ on $\M$
and foliation $\Fo:\SiR\map\M$, the lapse function specifies the
proper time separation $\de_\perp\tau$ between the hypersurfaces
$\Fo_t(\Si)$ and $\Fo_{t+\de t}(\Si)$ measured in the
direction normal to the first hypersurface:
\beq
    \de_\perp\tau(x) = \Np(x,t)\de t.                   \label{de-tau-N}
\eeq
The shift vector $\N(x)$ determines how, for each $x\in\Si$, the
point $\Fo_{t+\de t}(x)$ in $\M$ is displaced with respect to the
intersection of the hypersurface $\Fo_{t+\de t}(\Si)$ with the normal
geodesic drawn from the point $\Fo_{t}(x)\equiv\Fo_{x}(t)$. If this
intersection point can be obtained by evaluating $\Fo_{t+\de t}$ at a
point on $\Si$ with coordinates $x^a+\de x^a$ then
\beq
    \de x^a(x) = -N^a(x,t)\de t                          \label{de-xa-Na}
\eeq
as illustrated in Figure~\ref{Fig:lapse-shift}.

    It is clear from \eqs{PuB:g00}{PuB:g0a} that the shift vector and
lapse function are essentially the $\g_{0a}$ and $\g_{00}$ parts of
the spacetime metric. It was in terms of this coordinate language
that the original ADM approach was formulated. As we shall see
shortly, the Einstein field equations do not lead to any dynamical
development for these variables.

\subsubsection{The Canonical Form of General Relativity}
The canonical analysis of general relativity proceeds as follows. We
choose some reference foliation $\For:\SiR\approx\M$ and consider
first the situation in which $\M$ carries a {\em given\/} Lorentzian
metric $\g$ that satisfies the vacuum Einstein field equations
$G_{\a\b}(X,\g]=0$ and is such that each leaf $\For_t(\Si)$ is a
hypersurface in $\M$ that is spacelike with respect to $\g$. The
problem is to find a set of canonical variables for this system and
an associated set of first-order differential equations that
determine how the variables evolve from one leaf to another and whose
solution recovers the given metric $\g$. Of course, in practice this
procedure is used in the situation in which $\g$ is unknown and is to
be determined by solving the Cauchy problem using given Cauchy data.

    The Riemannian metric on $\Si$ defined by \eq{PuB:gab=} plays the
role of the basic configuration variable. The rate of change of
$g_{ab}$ with respect to the label time $t$ is related to the
extrinsic curvature \eq{Def:K} $K$ for the embeddings
$\For_t:\Si\map\M$ by
\beq
 K_{ab}(x,t)={1\over 2\Np(x,t)}\big(-\dot g_{ab}(x,t)+
    L_{\N}g_{ab}(x,t)\big)          \label{g-dot}
\eeq
where $L_{\N}g_{ab}$ denotes the Lie derivative
\footnote{This quantity is equal to $N_{a|b}+N_{b|a}$ where $|$ means
covariant differentiation using the induced metric $g_{ab}$ on
$\Si$.} of $g_{ab}$ along the shift vector field $\N$ on $\Si$.

    The key step in deriving the canonical form of the action
principle is to pull-back the Einstein Lagrangian density by the
foliation $\For:\SiR\map\M$ and express the result as a function of
the extrinsic curvature, the metric $g$, and the lapse vector and
shift function. This gives
\beqa
    (\For)^*\big(|\g|^\half R[\g]\big) = \Np\dg^\half
        \big(K_{ab}K^{ab}- (K^a_a)^2+R[g]\big)           \nonumber\\
    -2{d\over dt}\big(\dg^\half {K_a}^a\big) -
        \big(\dg^\half({K_a}^a N^b-g^{ab}\Np,_a)\big),_b
\eeqa
where $R[g]$ and $\dg$ denote respectively the curvature scalar and
the determinant of the metric on $\Si$.

    The spatial-divergence term vanishes because $\Si$ is assumed to
be compact. However, the term with the total time-derivative must be
removed by hand to produce a genuine action principle. The resulting
`ADM' action for matter-free gravity can be written as
\beq
 S[g,\Np,\N] = {1\over\k^2}\int dt\int_\Si d^3x\,
    \Np\dg^\half\big(K_{ab}K^{ab} -({K_a}^a)^2+R[g]\big)\label{S:ADM}
\eeq
where $\k^2:=8\pi G/c^2$. This action is to be varied with respect
to the Lorentzian metric $(\For)^*(\g)$ on $\SiR$, and hence with
respect to the paths of spatial geometries, lapse functions, and
shift vectors associated via \eqs{PuB:g00}{PuB:gab} with
$(\For)^*(\g)$. The extrinsic curvature $K$ is to be thought of as
the explicit functional of these variables given by \eq{g-dot}.

    The canonical analysis now proceeds in the standard way except
that, since \eq{S:ADM} does not depend on the time derivatives of
$\Np$ or $\N$, in performing the Legendre transformation to the
canonical action, the time derivative $\pa g_{ab}/\pa t$ is replaced
by the conjugate variable $p^{ab}$ but $\Np$ and $\N$ are left
untouched. The variable conjugate to $g_{ab}$ is
\beq
    p^{ab}:= {\de S\over\de \dot g_{ab}} =   -
        {\dg^\half\over\k^2}(K^{ab}-g^{ab}{K_c}^c)   \label{Def:p}
\eeq
which can be inverted in the form
\beq
    K^{ab}=-\k^2\dg^{-\half}(p^{ab}-\half g^{ab}{p_c}^c). \label{K-from-p}
\eeq
The action obtained by performing the Legendre transformation is
\beq
    S[g,p,\Np,\N]=\int dt\int_\Si d^3x\, \big(p^{ab}\dot g_{ab} -
          \Np\Hp - N^a\Ha\big)                     \label{S:can-g}
\eeq
where
\beqa
    \Ha(x;g,p] &:= &-2{p_a{}^b}_{|b}(x)             \label{Def:Ha}\\
    \Hp(x;g,p] &:=& \k^2{\cal G}_{ab\,cd}(x,g]\,p^{ab}(x)\,p^{cd}(x) -
        {\dg^\half(x)\over\k^2}\, R(x,g]           \label{Def:Hperp}
\eeqa
in which
\beq
    {\cal G}_{ab\,cd}(x,g]:=
        \half\dg^{-\half}(x)\big(g_{ac}(x)g_{bd}(x)+g_{bc}(x)g_{ad}(x)
           -g_{ab}(x)g_{cd}(x)\big)             \label{Def:DeW-metric}
\eeq
is the `DeWitt supermetric' on the space of three-metrics
\cite{DeW67a}. The functions $\Ha$ and $\Hp$ of the canonical
variables $(g,p)$ play a key role in the theory and are known as the
{\em supermomentum\/} and {\em super-Hamiltonian\/} respectively.

    It must be emphasised that the variables $g$, $p$, $\Np$ and
$\N$ are treated as independent in varying the action \eq{S:can-g}.
In particular, $g_{ab}(x,t)$ is no longer viewed as the restriction
of a spacetime metric $\g$ to a particular hypersurface $\Fo_t(\Si)$
of $\M$. Rather, $t\mapsto g_{ab}(x,t)$ is any path in the space
$\RS$ of Riemannian metrics on $\Si$, and similarly $t\mapsto
p^{ab}(x,t)$ (resp.\ $\Np(x,t)$, $N^a(x,t)$) is any path in the space
of all contravariant, rank-2 symmetric tensor densities (resp.\
functions, vector fields) on $\Si$. Consistency with the earlier
discussion is ensured by noting that the equation obtained by varying
the action \eq{S:can-g} with respect to $p^{ab}$
 \beq
    \dot g_{ab}(x,t) = {\de H[\Np,\N]\over\de p^{ab}(x,t)}
                                                \label{EE-can-gab},
\eeq
can be solved algebraically for $p^{ab}$ and, with the aid of
\eq{K-from-p}, reproduces the relation \eq{g-dot} between the
time-derivative of $g_{ab}$ and the extrinsic curvature. Here, the
functional
\beq
    H[\Np,\N](t):=\int_\Si d^3x\left(\Np\Hp +N^a\Ha\right) \label{Def:H[N]}
\eeq
of the canonical variables $(g,p)$ acts as the Hamiltonian of the
system.

    Varying the action \eq{S:can-g} with respect to $g_{ab}$ gives
the dynamical equations
\beq
\dot p^{ab}(x,t) = -{\de H[\Np,\N]\over\de g_{ab}(x,t)}\label{EE-can-pab}
\eeq
while varying $N^a$ and $\Np$ leads respectively to
\beq
    \Ha(x;g,p]=0                                    \label{EE-can-Ha}
\eeq
and
\beq
    \Hp(x;g,p]=0                                    \label{EE-can-Hperp}
\eeq
which are {\em constraints\/} on the canonical variables $(g,p)$.

    Note that the action \eq{S:can-g} contains no time derivatives of
the fields $\Np$ or $\N$ which appear manifestly as Lagrange
multipliers that enforce the constraints \eq{EE-can-Hperp} and
\eq{EE-can-Ha}. This confirms the status of the lapse function and
shift vector as non-dynamical variables that can be specified as
{\em arbitrary\/} functions on $\SiR$; indeed, this must be done in
some way before the dynamical equations can be solved (see
\S\ref{SSSec:ENDV}).

    The set of equations \eq{EE-can-gab}, \eq{EE-can-pab},
\eq{EE-can-Ha} and \eq{EE-can-Hperp} are completely equivalent to
Einstein's equations $G_{\a\b}=0$ in the following sense \cite{FM79}:

\bi
    \item Let $\g$ by any Lorentzian metric on $\M$ that satisfies
the  vacuum Einstein equations $G_{\a\b}(X,\g]=0$ and let $t\mapsto
\Fo_t:\Si\map\M$ be a one-parameter family of spacelike embeddings of
$\Si$ in $\M$ that foliates $\M$ with associated lapse function and
shift vector given by \eq{Def:LS}. Then the family of induced metrics
$t\mapsto g(t)$ and momenta $t\mapsto p(t)$ (where $p(t)$ is computed
from $\g$ using \eq{g-dot} and \eq{Def:p}) satisfies the equations
\eq{EE-can-gab}, \eq{EE-can-pab}, \eq{EE-can-Ha} and
\eq{EE-can-Hperp}.

    \item Conversely, if $t\mapsto \Fo_t$ is a spacelike foliation of
$(\M,\g)$ such that the evolution and constraint equations above
hold, then $\g$ satisfies the vacuum field equations.
\ei
In effect, equations \eq{EE-can-gab} and \eq{EE-can-pab} reproduce
the projections of the Einstein \mbox{equations} that are tangent to the
hypersurfaces $t=${\em const.}, while \eq{EE-can-Ha} and
\eq{EE-can-Hperp} reproduce the normal projections of these
equations,  $n^\a\,n^\b\,G_{\a\b}=0$ and $n^\a\,G_{\a a}=0$.

\subsubsection{The Constraint Algebra}
\label{SSSec:ConsAlg}
The classical canonical algebra of the system is expressed by the
basic Poisson brackets
\beqa
    \big\{g_{ab}(x),g_{cd}(x')\big\}&=&0            \label{PB:gg}\\
    \big\{p^{ab}(x),p^{cd}(x')\big\}&=&0            \label{PB:pp}\\
    \big\{g_{ab}(x),p^{cd}(x')\big\}&=&\de^c_{(a}\de^d_{b)}\,\de(x,x')
                                                    \label{PB:gp}
\eeqa
which can be used to cast the dynamical equations into a canonical
form in which the right hand sides of \eq{EE-can-gab} and
\eq{EE-can-pab} are $\big\{g_{ab}(x),H[\Np,\N]\big\}$ and
$\big\{p^{ab}(x),H[\Np,\N]\big\}$ respectively. However, the
constraints \eq{EE-can-Ha} and \eq{EE-can-Hperp} imply that not all
the variables $g_{ab}(x)$ and $p^{cd}(x)$ are
independent, and so, as in all systems with constraints, the Poisson
bracket relations \eqs{PB:gg}{PB:gp} need to be used with care.

    A crucial property of the canonical formalism is the closure of
the Poisson brackets of the super-Hamiltonian and supermomentum,
computed using \eqs{PB:gg}{PB:gp}; \ie the set of constraints is
{\em first class\/}. This is contained in the fundamental relations
\cite{Dir65}
\beqa
\big\{\Ha(x),\H_b(x')\big\} &=& -\H_b(x)\,\partial_a^{x'}\de(x,x') +
     \Ha(x')\,\partial_b^x\de(x,x')                     \label{PB:HaHb}\\
\big\{\Ha(x),\Hp(x')\big\} &=&
        \Hp(x)\,\partial_a^x\de(x,x')                   \label{PB:HaH}\\
\big\{\Hp(x),\Hp(x')\big\} &=&g^{ab}(x)\,\Ha(x)\,
    \partial^{x'}_b\de(x,x')-                           \nonumber\\
 &\ &\ \ \ \ g^{ab}(x')\,\Ha(x')\,\partial_b^x\de(x,x').\label{PB:HH}
\eeqa
Using the smeared variables
\footnote{The integrals are well-defined since both $\Ha$ and $\Hp$
are densities on $\Si$.}
\beq
   H[N] :=\int_\Si d^3x\,N(x)\,\Hp(x),\ \
        H[\N]:=\int_\Si d^3x\,N^a(x)\,\Ha(x),
\eeq
where $N$ and $\N$ are any scalar function and vector field on $\Si$,
these equations can be written as
\beqa
    \big\{H[\N_1],H[\N_2]\,\big\} &=& H\big([\,\N_1,\N_2\,]\big)
                                                \label{PB:Sm-HaHb}\\
    \big\{H[\N],H[N]\,\big\} &=& H[L_{\N}\,N]   \label{PB:Sm-HaH}\\
    \{H[N_1],H[N_2]\,\} &=& H[\N]               \label{PB:Sm-HH}
\eeqa
where, in \eq{PB:Sm-HH}, $N^a(x):=g^{ab}(x)\big(N_1(x)N_2(x),_b -
N_2(x)N_1(x),_b\big)$.

\noindent
The geometrical interpretation of these expressions is as follows.
\be

    \item The Lie algebra of the diffeomorphism group $\DS$ is
generated by the vector fields on $\Si$ with minus the commutator of
a pair of vector fields
\beq
    [\,N_1,N_2\,]^a := N_1^b\,N_2^a,_b-N_2^b\,N_1^a,_b
\eeq
playing the role of the Lie bracket. Thus \eq{PB:Sm-HaHb} shows that
the map $\N\map-H[\N]$ is a homomorphism of the Lie algebra of $\DS$
into the Poisson bracket algebra of the theory. A direct calculation
of the Poisson brackets of $H[\N]$ with $g_{ab}(x)$ and $p^{cd}(x)$
confirms the interpretation of $-H[\N]$ as a generator of spatial
diffeomorphisms.

    \item Similarly, $H[N]$ can be interpreted as generating
deformations of a hypersurface {\em normal\/} to itself as embedded
in $\M$. However, note that, unlike the analogous statement for
$H[\N]$, this interpretation applies only {\em after\/} the field
equations have been solved.
\ee
Two important things should be noted about the `gauge' algebra
\eqs{PB:Sm-HaHb}{PB:Sm-HH}:
\be
    \item It is {\em not\/} the Lie algebra of $\DM$ even though this
was the invariance group of the original theory.

    \item The presence of the $g^{ab}$ factor in the right hand side
of \eq{PB:Sm-HH} means it is {\em not\/} a genuine Lie algebra at
all.
\ee
These two features are closely related since the Dirac algebra
\eqs{PB:Sm-HaHb}{PB:Sm-HH} is essentially the Lie algebra of $\DM$
{\em projected\/} along, and normal to, a spacelike hypersurface.
The significance of this in the quantum theory will emerge later.

    Even though the Dirac algebra is not a genuine Lie algebra, it
still generates an action on the canonical variables $(g,p)$. This is
obtained by integrating the infinitesimal changes of the form
\beqa
    \de_{(N,\N)}\,g_{ab}(x):=\big\{g_{ab}(x),H[N]+H[\N]\big\}
                                            \label{Def:de-gab}\\
    \de_{(N,\N)}\,p^{ab}(x):=\big\{p^{ab}(x),H[N]+H[\N]\big\}
                                            \label{Def:de-pab}
\eeqa
for arbitrary infinitesimal smearing functions $N$ and $\N$. We shall
refer to the set of all such trajectories in the phase space $\cal S$
of pairs $(g,p)$ as the {\em orbits\/} of the Dirac algebra. Note
that, because of the first-class nature of the constraints
$\Ha(x)=0=\Hp(x)$, the subspace ${\cal S}_{\cal C}\subset{\cal S}$ on
which the constraints are satisfied is mapped into itself by the
transformations \eqs{Def:de-gab}{Def:de-pab}. On this constraint
surface, an orbit consists of the set of all pairs $(g,p)$ that can
be obtained from spacelike slices of some specific Lorentzian metric
$\g$ on $\M$ that satisfies the vacuum Einstein equations.

    A peculiar feature of general relativity is that an orbit on the
constraint surface includes the {\em dynamical evolution\/} of a pair
$(g,p)$ with respect to any choice of lapse function and shift
vector. Indeed, the dynamical equations \eq{EE-can-gab} and
\eq{EE-can-pab} are simply a special case of the transformations
above. This has an important implication for the notion of an
`observable'. In a system with first-class constraints, an observable
is normally defined to be any function on the phase space of the system whose
Poisson bracket with the constraints vanishes weakly, \ie it vanishes
on the constraint surface. In the present case, this means that $A$
is an observable if and only if
\beq
        \{\,A,H[N,\N]\,\} \approx 0                \label{Def:obs}
\eeq
for all $N$ and $\N$, where $H[N,\N]:=H[N]+H[\N]$. Thus any such
quantity is a {\em constant\/} of the motion with respect to
evolution along the foliation associated with $N$ and $\N$. We shall
deal later with the quantum analogue of this situation which is
deeply connected with the problem of time in quantum gravity.

\subsubsection{The Role of the Constraints}
\label{SSSec:RoleCon}
The constraints \eqs{EE-can-Ha}{EE-can-Hperp} are of major importance
in both the classical and the quantum theories of gravity, and it is
useful at this point to gather together various results concerning
them.
\be
    \item The constraints are consistent with the equations of motion
\eq{EE-can-gab} and \eq{EE-can-pab} in the sense of being
automatically maintained in time. For example,
\beq
    {d\Hp(x)\over dt} = \big\{\H_\perp(x),H[\Np,\N]\big\}
\eeq
and the right hand side vanishes on the constraint surface ${\cal
S}_C$ in phase space by virtue of the closing nature of the algebra
\eqs{PB:HaHb}{PB:HH}.

    \item The constraints lead to a well-posed Cauchy problem for the
dynamical equations \cite{HE73,FM79} once the undetermined
quantities $\Np$ and $\N$ have been fixed in some way (see below).

    \item If a Lorentzian metric $\g$ on the spacetime $\M$ satisfies
the vacuum Einstein equations $G_{\a\b}(X,\g]=0$ then the constraint
equations \eqs{EE-can-Ha}{EE-can-Hperp} are satisfied on all
spacelike hypersurfaces of $\M$. (It is understood that $p^{ab}$ is to be
computed from the given spacetime geometry using the definition
\eq{Def:K} of the extrinsic curvature $K_{ab}$ and the relation
\eq{K-from-p} between $p^{ab}$ and $K^{ab}$.)

    \item Conversely, let $(\M,\g)$ be a Lorentzian spacetime with the
property that the constraint equations \eqs{EE-can-Ha}{EE-can-Hperp}
are satisfied on every spacelike hypersurface. Then $\g$ necessarily
satisfies all ten Einstein field equations $G_{\a\b}(X,\g]=0$.
\ee
This last result is highly significant since it means the dynamical
aspects of the Einstein equations are already contained in the
constraints alone. This plays a crucial role in the Dirac
quantisation programme (see \S\ref{SSec:CQG-QBC}) and also in the
Hamilton-Jacobi approach to the classical theory.

     The proof is rather simple. For any given foliation of $\M$, the
Hamiltonian constraints $\H_\perp\big(x;g(t),p(t)\big]=0$ are equivalent
to the spacetime equation
\beq
     n^\a(x,t)\, n^\b(x,t)\,G_{\a\b}(\Fo(x,t),\g]=0
\eeq
where $n^\a$ is the vector normal to the hypersurface $\Fo_t(\Si)$ at the
point $\Fo(x,t)$ in $\M$. Now, at any given point $X\in\M$, and for any
(normalised) timelike vector $m$ in the tangent space $T_X\M$, the
foliation can be chosen so that $m$ is the normal vector at that point.
Thus, by ranging over all possible foliations, we find that for all
$X\in\M$ and for all timelike vectors $m$ at $X$
\beq
             m^\a(X)\,m^\b(X)\,G_{\a\b}(X,\g]=0.
\eeq

     Now suppose $m_1,m_2$ is any pair of timelike vectors. Then
$m_1+m_2$ is also timelike and hence
$(m_1+m_2)^\a(m_1+m_2)^\b\,G_{\a\b}=0$. However,
$m_1^\a\,m_1^\b\,G_{\a\b}$=$m_2^\a\,m_2^\b\,G_{\a\b}=0$ and hence, since
$G_{\a\b}=G_{\b\a}$, we see that
\beq
         m_1^\a\,m_2^\b\,G_{\a\b}=0.            \label{Gtt=0}
\eeq
Now let $\s$ be any spacelike vector. Then there exist timelike
vectors $m_1,m_2$ such that $\s=m_1-m_2$. It follows from (\ref{Gtt=0})
that for any timelike $m$,
\beq
         m^\a\,\s^\b\,G_{\a\b}=0.            \label{Gts=0}
\eeq
Finally, a similar result shows that for any pair of spacelike vectors
$\sigma_1,\sigma_2$,
\beq
         \s_1^\a\,\s_2^\b\,G_{\a\b}=0.           \label{Gss=0}
\eeq
However, any vector $u\in T_p\M$ can be written as the sum of a spacelike
and a timelike vector, and hence for all $X\in\M$ and $u,v\in T_X\M$ we
have
\beq
         u^\a\,v^\b\,G_{\a\b}[\g]=0,
\eeq
which means precisely that $G_{\a\b}[\g]=0$.

     Note that:
\be
    \item The super-Hamiltonian constraints
\footnote{I have written $\H_\perp(x,\Em]$ to emphasise that the
variables $g_{ab}$ and $p^{ab}$ that appear in $\H_\perp$ are
computed from the given Lorentzian four-metric $\g$ using the
spacelike embedding $\Em:\Si\map\M$.}
$\H_\perp(x,\Em]=0$ for all spacelike embeddings $\Em$ are sufficient
by themselves: it is not necessary to impose the supermomentum
constraints $\H_a(x,\Em]=0$ in addition.

    \item There is a similar result for electromagnetism: if the
initial-value equations ${\rm div}\vec{E}=0$ hold in every inertial
frame (\ie on  every spacelike hyperplane), the electromagnetic field
must necessarily evolve according to the dynamical Maxwell
equations.
\footnote{Karel Kucha\v{r}, private communication.}
\ee

\subsubsection{Eliminating the Non-Dynamical Variables}
\label{SSSec:ENDV}
I wish now to return to the canonical equations of motion
\eqs{EE-can-gab}{EE-can-Hperp} and the status of the non-dynamical
degrees of freedom. These variables must be removed in some way
before the equations of motion can be solved. As emphasised already,
these redundant variables include $\Np$ and $\N$ which must be fixed
before the abstract label time $t$ in the foliation can
be related to a physical quantity like proper time. However, it is
important to note that these are not the only variables in the theory
that are non-dynamical. This is clear from a count of degrees of
freedom. The physical modes of the gravitational field should
correspond to $2\times\infty^3$ configuration
\footnote{The notation $n\times\infty^m$ refers to a field complex
with $n$ components defined on an $m$-dimensional manifold.}
variables (the two circular polarisations of a gravitational wave;
the two helicity states of a graviton), or $4\times\infty^3$ phase
space variables. However, if a specific coordinate system
\footnote{In most of the discussion in this paper I shall neglect
possible global issues such as, for example, the fact that a
three-manifold $\Si$ that is compact cannot be covered by a single
coordinate system.}
is chosen on $\Si$ then, after eliminating $\Np$ and $\N$, we have
the $12\times\infty^3$ variables $\big(g_{ab}(x),p^{cd}(x)\big)$. In
principle, the four constraint equations \eq{EE-can-Ha} and
\eq{EE-can-Hperp} can be used to remove a further $4\times\infty^3$
variables, but this still leaves $8\times\infty^3$, which is
$4\times\infty^3$ too many.

    The physical origin of these spurious variables lies in the
$\DM$-invariance of the original Einstein action. The orbits of the
Dirac algebra on the phase space of pairs $(g,p)$ are similar in
many respects to the orbits of the Yang-Mills gauge group on the
space of vector-potentials and, as in that case, it is necessary to
impose some sort of `gauge'. This will have the desired effect of
eliminating the remaining $4\times\infty^3$ non-physical degrees of
freedom.

    The exact way in which these additional non-physical variables
are to be identified and removed depends on how the lapse function
and shift vector are fixed. The main possibilities are as follows.
\be
    \item
        {Set $\Np$ and $\N$ equal to specific functions
\footnote{To correspond to a proper foliation it is necessary that,
for all $(x,t)\in\SiR$, $\Np(x,t)\ne 0$ and, by convention, we
require the continuous function $\Np$ to satisfy $\Np(x,t)>0$. This
means that proper time increases in the same direction as the
foliation time label $t$; changing the sign of $\Np$ corresponds to
reversing the sign of $t$.}
on $\SiR$. If these functions are substituted into the dynamical
equations \eq{EE-can-gab} and \eq{EE-can-pab} they yield four
\mbox{equations} of the form $\dot F^A\big(x,t;g(t),p(t)\big]=0$,
$A=0,1,2,3$, whose solutions are $F^A(x,t;g(t),p(t)]$=$f^A(x)$ for
any arbitrary set of functions $f^A$ on $\Si$. Solving these
equations enables an extra  $4\times\infty^3$ variables to be
removed, as was required.

    This technique can be generalised in various ways. For example,
$\Np$ and $\N$ can be set equal to specific functionals $\Np(x,t;g,p]$
and $N^a(x,t;g,p]$ of the canonical variables. Another possibility
is to impose conditions on the time derivatives of $\Np$ and $\N$
rather than on $\Np$ and $\N$ themselves. This is necessary in the
context of the so-called `covariant gauges'.
    }

    \item
    {Another option is to start by imposing four conditions
\beq
    F^A(x,t;g(t),p(t)]=0                                \label{FA=0}
\eeq
which restrict the paths $t\mapsto\big(g(t),p(t)\big)$ in the phase
space. This method is popular in path-integral approaches to
canonical quantisation where it is implemented using a (formal)
functional $\de$-function (see \S\ref{SSec:CQG-CBQ}). The main
requirement is that each orbit of the Dirac group contains one, and
only one, of these restricted paths on the subspace of the phase
space defined by the initial value constraints
\eqs{EE-can-Ha}{EE-can-Hperp}. As we shall discuss later, there can
be global obstructions (in the phase space) to the existence of such
functions $F^A$. Care must also be taken to avoid eliminating genuine
physical degrees of freedom in addition to the non-dynamical modes.
This potential problem has been discussed at length by Teitelboim
(1982, 1983a, 1983b, 1983c, 1983d)
\nocite{Tei82,Tei83a,Tei83b,Tei83c,Tei83d}.

    Since the equations $F^A\big(x,t;g(t),p(t)\big]=0$ are valid for
all times, the time-derivative must also vanish. If the ensuing
equations $\dot F^A(x,t;g(t),p(t)]=0$ are substituted into the
dynamical equations \eq{EE-can-gab} and \eq{EE-can-pab}, there
results a set of four elliptic partial-differential equations for
$\Np$ and $\N$ that can be solved (in principle) to eliminate these
variables as specific functionals of $g$ and $p$.
    }

    \item The third approach to eliminating the non-dynamical
variables will play a major role in our discussions of time. This
involves the parametrisation of spacetime points by `internal'
space and time coordinates, defined as the values of various
functionals $\X^A(x;g,p]$, $A=0,1,2,3$ of the canonical variables,
which can be set equal to some fixed functions $\chi^A_t(x)$. In
effect, this leads to an equation of the type \eq{FA=0} and has the
same technical consequences although the physical interpretation is
more transparent.
\ee

\subsection{Internal Time}\label{SSec:IntTime}
\subsubsection{The Main Ideas}
The idea of specifying spacetime points by values of the fields was
emphasised in the canonical theory by \citen{BSW62}, and is of
sufficient importance to warrant a subsection in its own right. As
we shall see, this is related to the problem of rewriting
\eq{S:can-g} as a true canonical action for just the physical modes
of the gravitational field.

    The key idea is to introduce quantities $\T(x;g,p]$ and
$\Z^a(x;g,p]$, $a=1,2,3$, that can serve as the time and spatial
coordinates for a spacetime event that is associated (in a way yet to
be specified) with the point $x\in\Si$ and the pair of canonical
variables $(g,p)$; the collection of four functions $(\T,\Z^a)$ will
be written as $\X^A$, $A=0,1,2,3$. Strictly speaking, we should
worry about the fact that the three-manifold $\Si$ is compact and
hence cannot be covered by a single system of coordinates. The same
will be true for the spacetime manifold $\SiR$, and therefore we
should allow for more than one set of the internal spatial functions
$\Z^a$. This could be handled in various ways. One option is to work
directly with collections of spatial coordinate functions that can
cover $\Si$ globally. But this is complicated since the number of
coordinate charts used on a manifold is arbitrary provided only that
it is greater than, or equal to, the mininum number determined by the
manifold topology.

    A slightly more elegant (although ultimately equivalent) approach
is to realise that, since $\M$ is diffeomorphic to $\SiR$, a
spacetime event can be labelled by specifying (i) a value for a time
variable, and (ii) a {\em point\/} in the three-manifold $\Si$. Thus
we could try to extend the scheme above to include functions
$\Z(x;g,p]$ whose values lie in $\Si$, rather than in some
region of $\R^3$. When used in this way, I shall refer to $\Si$ as
the spatial {\em label\/} space and denote it by $\Si_l$. A
particular set of functions $\Z^a(x;g,p]$ can then be associated with
each set of local coordinate functions $x^a$ on $\Si_l$ according to
the identification $\Z^a(x;g,p]:=x^a\big(\Z(x;g,p]\big)$. I shall not
become too involved in this subtlety here since we are mainly
concerned with broad principles rather than technical details.
However, if desired, the discussion that follows can be generalised
to include this concept of a label space.

    To see how the idea of internal coordinates is used let us
consider first the case of a given Einstein spacetime $(\M,\g)$ where
we wish to use the functionals to locate an event $X$ in $\M$. The
key steps are as follows.
\be
    \item Fix a reference foliation $\For:\SiR\map\M$ with spacelike
leaves and write its inverse $(\For)^{-1}:\M\map\SiR$ as
\beq
  (\For)^{-1}(X):= \big(\s^\rref(X),\tau^\rref(X)\big)\in\SiR
\eeq
where $\s^\rref:\M\map\Si$, and where $\tau^\rref:\M\map\R$ is a
global reference-time function on $\M$. This illustrates the general
sense in which space and time are not intrinsic spacetime structures
but are put into $\M$ by hand. Thus an instant of time, labelled by
the number $t$, is the hypersurface
\beq
  \For_t(\Si):=\{\For_t(x)|x\in\Si\} = (\tau^\rref)^{-1}\{t\},
\eeq
and a point in space $x$ is the worldline
\beq
  \For_x(\R):=\{\For_x(t)|t\in\R\}=(\s^\rref)^{-1}\{x\}
\eeq
where we recall that $\For_t:\Si\map\M$ and $\For_x:\R\map\M$ are
defined by $\For_t(x):=\For(x,t)$ and $\For_x(t):=\For(x,t)$
respectively. This is illustrated in Figure~\ref{Fig:Tint}.
\begin{figure}[t]
\begin{picture}(400,400)(0,0)
    \put(200,380){$\{Y\in\M|\Z^a(Y)=y^a\}$}
    \put(340,370){$\For_{x=\s^\rref(X)}(\R)$}
    \put(400,300){$\For_{t_2}$}
    \put(250,200){$X=\For(x,t)$}
    \put(375,180){$\For_{t=\tau^\rref(X)}(\Si)$}
    \put(350,120){$\{Y\in\M|T(Y)=T\}$}
    \put(210,80){\large $\M$}
    \put(330,60){$\For_{t_0}(\Si)$}
\end{picture}
\caption{The foliation of $\M$ and the hypersurface
      $T(Y)=T$ of constant internal time.}
\label{Fig:Tint}
\end{figure}

    \item Construct the particular leaf of the reference foliation
that passes through the given event $X$. According to the notation
above, this embedding of $\Si$ in $\M$ is $\For_{t=\tau^\rref(X)}$.
Let $x\in\Si$ be such that
$\For_{t=\tau^\rref(X)}(x)\equiv\For_x\big(\tau^\rref(X)\big)=X$, \ie
$x=\s^\rref(X)$. Compute:
    \be
    \item the induced metric $g_{ab}(x,\For_{t=\tau^\rref(X)}\,]$ on
$\Si$ using \eq{Def:Em*g};
    \item the induced momentum $p^{cd}(x,\For_{t=\tau^\rref(X)}\,]$,
using the definition \eq{Def:K} of the extrinsic curvature
$K_{ab}(x,\For_{t=\tau^\rref(X)}\,]$ and the relation \eq{Def:p}
between $p$ and $K$.
    \ee

    \item Use these values of $g$ and $p$ in the definitions of $\T$
and $\Z^a$ to ascribe time and space coordinates to the event $X$.
\ee
In summary, the `internal time' coordinate of the point $X\in\M$ is
defined to be
\beq
    T(X) := \T\big(\s^\rref(X);g[\Fo^\rref_{t=\tau^\rref(X)}\,],
        p[\For_{t=\tau^\rref(X)}\,]\,\big],         \label{Def:T(X)}
\eeq
while the spatial labels are
\beq
    Z^a(X) :=\Z^a\big(\s^\rref(X);g[\For_{t=\tau^\rref(X)}\,],
        p[\For_{t=\tau^\rref(X)}\,]\,\big].         \label{Def:Za(X)}
\eeq
Alternatively, one can write
\beqa
    T\big(\For(x,t)\big)  &=& \T\big(x;g(t),p(t)\,\big]     \\
    Z^a\big(\For(x,t)\big) &=&\Z^a\big(x;g(t),p(t)\,\big]
\eeqa
where $g(t)$ and $p(t)$ denote the metric and momentum induced from
$\g$ on the hypersurface $\For_t(\Si)$ of $\M$.

    Note that to be consistent such an identification of internal
coordinates requires the functions $\T$ and $\Z^a$ to be chosen such
that the following two conditions are satisfied with respect to the
given Lorentzian metric $\g$ on $\M$:
\be
    \item For all values of the constant number $T\in\R$, the `instant
of time' $\{Y\in\M|T(Y)=T\}$ must be a {\em spacelike\/} subspace of
$(\M,\g)$.

    \item For all $\vec y\in\R^3$, the set $\{Y\in\M|Z^a(Y)=y^a\}$
must be a {\em timelike\/} subspace
\footnote{Strictly speaking, this will be true only locally if the
$Z^a$ are local coordinate functionals. This problem can be overcome
by using the labelling space $\Si_l$ and requiring that, for all
points $y\in\Si_l$, the set $\{Y\in\M|\Z(Y)=y\}$ is timelike.}
of $(\M,\g)$.
\ee

\subsubsection{Reduction to True Canonical Form}
\label{SSec:RTCF}
In the context of the general canonical theory---where there is no
fixed spacetime metric---the internal time and space functionals are
used to specify a set of non-dynamical degrees of freedom. This is
part of the programme to reduce the theory to true canonical form.
The key steps are as follows.

    1.\ Perform a canonical transformation
\beq
   \big(g_{ab}(x),p^{cd}(x)\big)\mapsto
        \big(\X^A(x),\P_B(x);\f^r(x),\pi_s(x)\big)          \label{CT}
\eeq
in which the $12\times\infty^3$ variables
$\big(g_{ab}(x),p^{cd}(x)\big)$ are mapped into
\bi
    \item the four functions $\X^A(x)$ specifying a
particular choice of internal space and time coordinates;
    \item their four conjugate momenta $\P_B(x)$;
    \item the two modes $\f^r(x)$, $r=1,2$, which represent the
physical degrees of freedom of the gravitational field;
    \item their conjugate momenta $\pi_s(x)$, $s=1,2$. \ei  The
statement that $\P_B$ are the momenta conjugate to the four internal
coordinate variables $\X^A$ means they satisfy the Poisson bracket
relations (computed using the basic Poisson brackets
\eqs{PB:gg}{PB:gp}) \beqa
    \{\X^A(x),\X^B(x')\}&=&0                        \label{PB:XAXB} \\
    \{\P_A(x),\P_B(x')\}&=&0                        \label{PB:PAPB} \\
    \{\X^A(x),\P_B(x')\}&=&\de^A_B\,\de(x,x')       \label{PB:XAPB}
\eeqa
while the corresponding relations for the `physical'
\footnote{To avoid confusion it should be emphasised that $\f^r$ and
$\pi_s$ may {\em not\/} be observables in the sense of satisfying
\eq{Def:obs}.}
variables $(\f^r,\pi_s)$ are
\beqa
    \{\f^r(x),\f^s(x')\}&=&0                        \label{PB:frfs} \\
    \{\pi_r(x),\pi_s(x')\}&=&0                      \label{PB:prps} \\
    \{\f^r(x),\pi_s(x')\}&=&\de^r_s\,\de(x,x').     \label{PB:frps}
\eeqa
It is also assumed that all cross brackets of $\X^A$ and $\P_B$ with
$\f^r$ and $\pi_s$ vanish:
\beqa
\lefteqn{0=\{\f^r(x),\X^A(x')\}= \{\f^r(x),\P_B(x')\}=} \nonumber\\
& &\ \ \{\pi_s(x),\X^A(x')\}= \{\pi_s(x),\P_B(x')\}.    \label{PB:fpi-XP}
\eeqa
Note that some or all of these relations may need to be generalised
to take account of the global topological properties of the function
spaces concerned. There may also be global obstructions to some of the
steps.

    2.\ Express the super-Hamiltonian and supermomentum as
functionals of these new canonical variables and write the canonical
action \eq{S:can-g} as
\beq
    S[\f,\pi,\Np,\N,\X,\P\,]=\int dt\int_\Si d^3x\big(\P_A\dot\X^A
        +\pi_r\dot\f^r-\Np\Hp-N^a\Ha\big)     \label{S:can-trans}
\eeq
where all fields are functions of $x$ and $t$, and where $\Hp$ and
$\Ha$ are rewritten as functionals of $\X_A$, $\P_B$, $\f^r$ and
$\pi_s$. Note that $\f^r$, $\pi_s$, $\X^A$, $\P_B$, $\Np$ and $\N$
are all to be varied in \eq{S:can-trans} as independent functions.
The fields $\X^A$ are interpreted geometrically as defining an
embedding of $\Si$ in $\M$ via the parametric equations
\beqa
    T&=&\T(x)                                           \\
    Z^a&=&\Z^a(x)
\eeqa
where $\T(x):=\X^0(x)$ and $\Z^a(x):=\X^a(x)$, $a=1,2,3$. The
conjugate variables $P_B(x)$ can be viewed as the energy and
momentum densities of the gravitational field measured on this
hypersurface. Note however that the spacetime picture that lies
behind these interpretations is defined only {\em after\/} solving
the equations of motion and reconstructing a Lorentzian metric on
$\M$.

    3.\ Remove $4\times\infty^3$ of the $8\times\infty^3$ non-dynamical
variables by solving
\footnote{There may be global obstructions to this step.}
the constraints $\Hp(x)=0$ and $\Ha(x)=0$ for the variables $\P_A(x)$
in the form
\beq
    \P_A(x) + h_A(x;\X,\f,\pi]=0.                    \label{Pm+h=0}
\eeq

    4.\ Remove the remaining $4\times\infty^3$ non-dynamical
variables by `deparametrising' the canonical action functional
\eq{S:can-trans} by substituting into it the solution \eq{Pm+h=0} of
the initial value equations. This gives
\beq
    S[\f,\pi] = \int dt\int_\Si d^3x\,\big\{\pi_r(x,t)\dot\f^r(x,t)
        -h_A\big(x;\chi_t,\f(t),\pi(t)\big]\dot\chi^A_t(x)\big\}
                                                    \label{S:can-red}
\eeq
which can be shown to give the correct field equations for the
physical fields $\f^r$ and $\pi_s$. Note that, in \eq{S:can-red}, the
four quantities $\X^A$ are no longer to be varied but have instead
been set equal to some {\em prescribed\/} functions $\chi_t^A$ of
$x$. This is valid since, after solving the constraints, the
remaining dynamical equations of motion give no information about how
the variables $\X^A$ evolve in parameter time $t$. In effect, in
terms of the original internal coordinate functionals, we have
chosen a set of conditions
\beq
    \X^A\big(x;g(t),p(t)\big]=\chi_t^A(x)             \label{gauge-con}
\eeq
that restrict the phase-space paths $t\mapsto\big(g(t),p(t)\big)$
over which the action is to be varied. In this way of looking at
things, \eq{gauge-con} can be regarded as additional constraints
which, when added to the original constraints, make the entire set
second-class.

    The lapse function and shift vector play no part in
this reduced variational principle. However, it will be necessary to
reintroduce them if one wishes to return to a genuine spacetime
picture. As discussed in \S\ref{SSSec:ENDV}, this can be done by
solving the additional set of Einstein equations that are missing
from the set generated by the reduced action. These are elliptic
partial-differential equations for  $\Np$ and $\N$.

    The equations of motion derived from the reduced action
\eq{S:can-red} are those corresponding to the Hamiltonian
\beq
    H_{\rm true}(t) := \int_\Si d^3x\,\dot\chi^A_t(x)\,
        h_A\big(x;\chi_t,\f(t),\pi(t)\big]          \label{Htrue}
\eeq
and can be written in the form
\beqa
    {\partial\f^r(x,t)\over\partial t}&=& \big\{\f^r(x,t),
        H_{\rm true}(t)\big\}_{\rm red}             \label{EM:grav-red-f}\\
    {\partial\pi_s(x,t)\over\partial t}&=& \big\{\pi_s(x,t),
        H_{\rm true}(t)\big\}_{\rm red}             \label{EM:grav-red-pi}
\eeqa
where $\{\,,\,\}_{\rm red}$ denotes the Poisson bracket evaluated
using only the physical modes $\f^r$ and $\pi_s$. Thus, at least
formally, the system has been reduced to one that looks like the
conventional field theory discussed earlier in section
\S\ref{SSec:QFTM} and epitomised by the dynamical equations
\eqs{EM:f}{EM:pi}. A particular choice for the four functions
$\chi^A_t$ (using some coordinate system $x^a$ on $\Si$) is
\beqa
      \chi^0_t(x) &=& t                                 \label{chi0}\\
      \chi^a_t(x) &=&x^a                                \label{chia}
\eeqa
for which the true Hamiltonian \eq{Htrue} is just the integral over
$\Si$ of $h_0$. Thus we arrive at the fully-reduced form of
canonical general relativity as derived in the work of Arnowitt,
Deser and Misner.

\subsubsection{The Multi-time Formalism}
Although a particular time label $t$ (corresponding to some reference
foliation of $\M$ into a one-parameter family of hypersurfaces) has
been used in the above, this label is in fact quite arbitrary. This
is reflected in the arbitrary choice of the functions $\chi^A_t$ in
the `gauge conditions' \eq{gauge-con}. Indeed, the physical fields
$(\f^r,\pi_s)$ ought to depend only on the hypersurface in $\M$ on
which they are evaluated, not on the way in which that hypersurface
happens to be included in a particular foliation. Thus it should be
possible to write the physical fields $\f^r(x,\X\,]$, $\pi_s(x,\X\,]$
as functions of $x\in\Si$ and functionals of the internal coordinates
$\X^A$ which are now regarded as {\em arbitrary\/} functions of
$x\in\Si$ rather than being set equal to some fixed set.

    This is indeed the case. More precisely, the Hamiltonian equations
of motion \eqs{EM:grav-red-f}{EM:grav-red-pi}, which hold for all
choices of the functions $\chi^A$, imply that there exist Hamiltonian
densities $h_A(x,\X\,]$ such that $\f^r(x,\X\,]$ and
$\pi_s(x,\X\,]$ satisfy the functional differential equations
\beqa
    {\de\f^r(x,\X\,]\over\de \X^A(x')}&=& \big\{\f^r(x,\X\,],
        h_A(x',\X\,]\big\}_{\rm red}                \label{FEM:grav-f}\\
    {\de\pi_s(x,\X\,]\over\de\X^A(x')}&=& \big\{\pi_s(x,\X\,],
        h_A(x',\X\,]\big\}_{\rm red}                \label{FEM:grav-pi}
\eeqa
which is a gravitational analogue
\footnote{However, note that the quantities  $\Em^\a$ in
\eqs{FEM:f}{FEM:pi} are the components of an embedding map
$\Em:\Si\map\M$ with respect  some (unspecified) coordinate system on
$\M$, whereas the quantities $\X^A$ in \eqs{FEM:grav-f}{FEM:grav-pi}
are coordinate functions on $\M$.}
of the scalar field equations \eqs{FEM:f}{FEM:pi}.

    This so-called `bubble-time' or `multi-time' canonical
formalism is of considerable significance and will play an important
role in our discussions of the problem of time in quantum gravity
\cite{Kuc72}.


\section{IDENTIFY TIME BEFORE QUANTISATION}\label{Sec:TBefore}
\subsection{Canonical Quantum Gravity: Constrain Before Quantising}
\label{SSec:CQG-CBQ}
\subsubsection{Basic Ideas}
We shall now discuss the first of those interpretations of type I in
which time is identified as a functional of the geometric canonical
variables {\em before\/} the application of any quantisation
algorithm. These approaches to the problem of time are conservative
in the sense that the quantum theory is constructed in an essentially
standard way. In particular, `time' is regarded as part of the {\em a
priori\/} background structure used in the formulation of the quantum
theory. There are two variants of this approach. The first starts
with the fully-reduced formalism, and culminates in a quantum version
of the Poisson-bracket dynamical equations
\eqs{EM:grav-red-f}{EM:grav-red-pi}; the second starts with the
multi-time dynamical equations \eqs{FEM:grav-f}{FEM:grav-pi} and
ends with a multi-time Schr\"odinger equation.

    In the first approach, the problem is to quantise the first-order
action principle \eq{S:can-g} in which the canonical variables
$g_{ab}(x)$ and $p^{cd}(x)$ are subject to the constraints
$\H_a(x;g,p]=0=\H_\perp(x;g,p]$ where $\H_a$ and $\H_\perp$ are given
by \eq{Def:Ha} and \eq{Def:Hperp} respectively, and where all
non-dynamical variables are removed before quantisation. The account
that follows is very brief and does not address the very difficult
technical question of whether it is really feasible to construct a
consistent quantum theory of gravity in this way by applying some
quantisation algorithm to the classical theory of general relativity.

    The main steps are as follows.

    1.\ Impose a suitable `gauge'. As explained in \S\ref{SSSec:ENDV},
this can be done in several ways:
\bi
\item Set the lapse function $\Np$ and shift vector $\N$ equal to
specific functions (or functionals) of the remaining canonical variables.

\item {\em Or}, impose a set of conditions $F^A(x,t;g,p]=0$, $A=0,1,2,3$.

\item {\em Or}, identify a set $\X^A(x;g,p]$, $A=0,1,2,3$ of internal
spacetime coordinates and set these equal to some set of {\em
fixed\/} functions $\chi^A_t(x)$.
\ei
In all three cases, the outcome is that $\Np$ and $\N$ drop out of
the formalism, as do $8\times\infty^3$ of the $12\times\infty^3$
canonical variables.

    2.\ Construct a canonical action (\cf \eq{S:can-red}) to
reproduce the dynamical equations (\cf
\eqs{EM:grav-red-f}{EM:grav-red-f}) of the remaining
$4\times\infty^3$ `true' canonical variables
$\big(\f^r(x),\pi_s(x)\big)$.

    3.\ Impose canonical commutation relations on these variables and
then proceed as in any standard
\footnote{This last step is rather problemetic since the reduced
phase space obtained by the construction above is topologically
non-trivial, and hence the fields $\f^r(x)$ and $\pi_s(x)$ are only
{\em local\/} coordinates on this space. This means the naive
canonical commutation relations should be replaced by an operator
algebra that respects the global structure of the system
\cite{Ish84}.}
quantum theory. In particular, we obtain a Schr\"odinger equation of
the form
\beq
    i\hbar{\pa\Psi_t\over\pa t}=\op H_{\rm true}\Psi_t  \label{SE:grav}
\eeq
where $\op H_{\rm true}$ is the quantised version of the Hamiltonian
\eq{Htrue} for the physical modes introduced in the classical Poisson
bracket evolution equations \eqs{EM:grav-red-f}{EM:grav-red-pi}.

    A variant of this operator approach uses a formal, canonical
path-integral to compute expressions like
\footnote{This is only intended as an example. As defined,
$Z$ is just a number, and to produce physical predictions it is
necessary to include external sources, or background fields, in the
standard way.}
\beq
Z=\int\Pi_{x,t}{\cal D}g_{ab}(x,t)\,{\cal D}p^{cd}(x,t)\,{\cal D}N(x,t)\,
{\cal D}N^a(x,t)\,\de[F^A]\,\big|\det\{F^A,\H_B\}\big|\,
                    e^{{i\over\hbar}S[g,p,N,\N]}     \label{Def:Z}
\eeq
where $S[g,p,N,\N\,]$ is given by \eq{S:can-g}, $\H_{A=0}$ denotes
$\Hp$, and where the gauge conditions $F^A(x,t;g,p\,]=0$ are enforced
by the functional Dirac delta-function. Note that because $S$ is a
linear functional of $N$ and $\N$, the functional integrals over
these variables in \eq{Def:Z} formally generate functional
delta-functions
\footnote{There is a subtlety here, since the classical lapse
function $\Np$ is always positive. If the same condition is imposed
on the variable $N$ in the functional integral it will not lead to a
delta function.}
which impose the constraints $\Hp(x)=0=\Ha(x)$.

\subsubsection{Problems With the Formalism}
Neither the operator nor the functional-integral programmes are easy
to implement, and both are unattractive for a number of reasons. For
example:
\bi
    \item  The constraints $\Ha(x;g,p]=0=\Hp(x;g,p]$ cannot be
solved in a closed form
\footnote{The Ashtekar variables might change this situation as they
lead to powerful new techniques for tackling the constraints
\cite{Ash86,Ash87}. For two recent developments in this direction see
\citen{NR92} and \citen{MM92}.}.
Weak-field perturbative methods could be used but these throw little
light on the problem of time and inevitably founder on the problem
of non-renormalisability.

    \item The programme violates the geometrical structure of general
relativity by removing parts of the metric tensor. This makes it very
difficult to explore the significance of spacetime geometry in the
quantum theory.

    \item Choosing a gauge breaks the gauge invariance of the theory
and, as usual, the final quantum results must be shown to be
independent of the choice. In the functional-integral approach, this
is the reason for the functional Jacobian $|{\rm det}\{F^A,\H_B\}|$
in \eq{Def:Z}. However, this integral is so ill-defined that a
proper demonstration of gauge invariance is as hard as it is in the
operator approaches. The issue of gauge invariance is discussed at
length by \citen{Bar91}.
\ei

\subsection{The Internal Schr\"odinger Interpretation}\label{SSec:ISI}
\subsubsection{The Main Ideas}
The classical dynamical evolution described above employs the
single-time Poisson bracket equations of the form
\eqs{EM:grav-red-f}{EM:grav-red-pi} and, in the operator form, leads
to the gravitational analogue \eq{SE:grav} of the Schr\"odinger
equation \eq{SE:f}. However, the resulting description of the
evolution is in terms of a fixed foliation of spacetime, whereas a
picture that is more in keeping with the spirit of general relativity
would be one that involves an analogue of the functional-differential
equations \eqs{FEM:f}{FEM:pi} and the associated multi-time
Schr\"odinger equation \eq{FSE:f} that describe evolution along the
deformation of arbitrary hypersurfaces of $\M$.

    The internal Schr\"odinger interpretation refers to such a
generalisation in which one quantises the multi-time version of the
canonical formalism of classical general relativity. As always in
canonical approaches to general relativity, the first step is to
introduce a reference foliation $\For:\SiR\map\M$ which is used to
define the canonical variables. However, it is important to
appreciate that this should serve only as an intermediate tool in the
development of the theory and, like a basis set in a vector space
theory, it should not appear in the final results.

    The key steps in constructing the internal Schr\"odinger
interpretation are as follows:
\be
    \item Pick a set of classical functions $\T(x;g,p],\Z^a(x;g,p]$
that can serve as internal time and space coordinates.

    \item Perform the canonical transformation \eq{CT}
\beq
     \big(g_{ab}(x),p^{cd}(x)\big)\mapsto
        \big(\X^A(x),\P_B(x); \f^r(x),\pi_s(x)\big)    \label{2CT}
\eeq
in which $\big(\f^r(x),\pi_s(x)\big)$ are identified with the
$4\times\infty^3$ physical canonical modes of the gravitational
field.

    \item Solve the super-Hamiltonian constraint $\Hp(x)=0$ for the
variables $\P_A(x)$ in the form \eq{Pm+h=0}
\beq
          \P_A(x) + h_A(x;\X,\f,\pi]= 0.               \label{2Pm+h=0}
\eeq

    \item Identify $h_A(x;\X,\f,\pi]$ as the set of Hamiltonian
densities appropriate to a multi-time version of the classical
dynamical equations as in \eqs{FEM:grav-f}{FEM:grav-pi}.

    \item Quantise the system in the following steps:
    \be

    \item Replace the Poisson bracket relations
\eqs{PB:frfs}{PB:frps} with  the canonical commutation relations
\beqa
    [\,\op\f^r(x),\op\f^s(x')\,]     &=&0               \label{CR:frfs} \\
    {[}\,\op\pi_r(x),\op\pi_s(x')\,] &=&0               \label{CR:prps} \\
    {[}\,\op\f^r(x),\op\pi_s(x')\,]  &=&i\hbar\de^r_s\,\de(x,x').
                                                        \label{CR:frps}
\eeqa
    \item Try to construct quantum Hamiltonian densities $\op h_A(x)$
by the usual substitution rule in which the fields $\f^r$ and $\pi_s$
in the classical expression $h_A(x;\X,\f,\pi]$ are replaced by their
operator analogues $\op\f^r$ and $\op\pi_s$.

    \item The classical multi-time evolution equations
\eqs{FEM:grav-f}{FEM:grav-pi} can be quantised in one of two ways.
The first is to replace the classical Poisson brackets expressions
with Heisenberg-picture operator relations. The second is to use a
Schr\"odinger picture in which the constraint equation \eq{2Pm+h=0}
becomes the multi-time Schr\"odinger equation
\beq
    i\hbar{\de\Psi_\X\over\de \X^A(x)}=h_A(x;\X,\op\f,\op\pi\,]\Psi_\X
                                                        \label{FSE:grav}
\eeq
for the state vector $\Psi_\X$.
    \ee
\ee
This fundamental equation can be justified from two slightly
different points of view. The first is to claim that, since the
classical Hamiltonian associated with the internal coordinates $\X^A$
is $h_A$, the Schr\"odinger equation \eq{FSE:grav} follows from the
basic time-evolution axiom of quantum theory, suitably generalised to
a multi-time situation. However, note that this step is valid only if
the inner product on the Hilbert space of states is $\X$-independent.
If not, a compensating term must be added to \eq{FSE:grav} if the
scalar product $\la\Psi_\X,\Phi_\X\ra_\X$ is to be independent
of the internal coordinate functions. This is analogous to the
situation discussed earlier for the functional Schr\"odinger equation
\eq{FSE:f} for a scalar field theory in a background spacetime.

    Another approach to deriving \eq{FSE:grav} is to start with the
classical constraint equation \eq{2Pm+h=0} and then impose it at the
quantum level as a constraint on the allowed state vectors,
\beq
    \big(\op\P_A(x)+h_A(x;\op X,\op\f,\op\pi\,]\big)\Psi=0. \label{op-Pa=h}
\eeq
The Schr\"odinger equation \eq{FSE:grav} is reproduced if one makes
the formal substitution
\beq
    \op\P_A(x) = -i\hbar{\de\over\de\X^A(x)}.       \label{Def:op-Pa}
\eeq
This approach has the advantage of making the derivation of the
Schr\"odinger equation look somewhat similar to that of the
Wheeler-DeWitt equation. However, once again this is valid only if
the scalar product is $\X$-independent. If not, it is necessary to
replace \eq{Def:op-Pa} with
\beq
    \op\P_A(x)=-i\hbar{\de\over\de\X^A(x)}+F_A(x,\X]\label{Def:op-Pa-comp}
\eeq
where the extra term $F_A(x,\X]$ is chosen to compensate the
$\X$-dependence of the scalar product.

    Considerable care must be taken in interpreting \eq{Def:op-Pa} or
\eq{Def:op-Pa-comp}. These equations might appear to come from
requiring the existence of self-adjoint operators $\op\X^A(x)$ and
$\op P_B(x')$ that satisfy canonical commutation relations
\beqa
    [\,\op\X^A(x),\op\X^B(x')\,]  &=&0              \label{CR:XAXB}\\
    {[}\,\op\P_A(x),\op\P_B(x')\,]&=&0              \label{CR:PAPB}\\
    {[}\,\op\X^A(x),\op\P_B(x')\,]&=&i\hbar\de^A_B\,\de(x,x')
                                                    \label{CR:XAPB}
\eeqa
which are the quantised versions of the classical Poisson-bracket
relations \eqs{PB:XAXB}{PB:XAPB}. However, the example of elementary
wave mechanics shows that such an interpretation is misleading. In
particular, the $i\hbar\pa/\pa t$ that appears in the time-dependent
Schr\"odinger equation cannot be viewed as an operator defined on the
physical Hilbert space of the theory.

\subsubsection{The Main Advantages of the Scheme}
Many problems arise in the actual implementation of this programme,
but let us first list the main advantages.
\be
    \item We have argued in the introductory sections that one of the
main sources of the problem of time in quantum gravity is the special
role played by time in normal physics; in particular, its position
as something that is external to the system and part of the {\em a
priori\/} classical background of the Copenhagen interpretation of
quantum theory. The canonical transformation \eq{2CT} applied {\em
before\/} quantisation could be said to cast quantum gravity into the
same mould.

    \item The relative freedom from interpretational problems is
epitomised by the ease with which the theory can be equipped (at
least formally) with a Hilbert space structure. In particular, since
$\big(\f^r(x),\pi_s(x)\big)$ are genuine physical modes of the
gravitational field, it is fully consistent with conventional quantum
theory to insist that (once suitable smeared) the corresponding
operators are {\em self-adjoint\/}. This is certainly not the case
for the analogous statement in the Dirac, constraint-quantisation,
scheme in which all $12\times\infty^3$ variables $\big(g_{ab}(x),
p^{cd}(x)\big)$ are assigned operator status. It is then consistent
to follow the earlier discussion in \S\ref{SSec:QFTM} of quantum
field theory in a background spacetime and define the quantum states
as functionals $\Psi[\f]$ of the configuration variables $\f^r(x)$
with the canonical operators $\op\f^r(x)$ and $\op\pi_s(x)$ defined
as the analogue of \eq{Def:op-f} and \eq{Def:op-pi} respectively. Of
course, the cautionary remarks made earlier about measures on
infinite-dimensional spaces apply here too.

    \item This Hilbert space structure provides an unambiguous
probabilistic interpretation of the theory. Namely, if the normalised
state is $\Psi$, and if a measurement is made of the true
gravitational degrees of freedom $\f^r$ on the hypersurface
represented by the internal coordinates $\X^A$, the probability of
the result lying in the subset $B$ of field configurations is
(\cf \eq{Pr:finB})
\beq
     \Prob(\Psi;\f\in B) = \int_Bd\m(\f)\,\big|\Psi[\X,\f]\big|^2.
\eeq

    \item The associated inner product also allows the construction
of a meaningful notion of a quantum {\em observable\/}. This is
simply any well-defined operator functional $\op
F=F[\X,\op\f,\op\pi]$ of the true gravitational variables
$\op\f^r(x)$, $\op\pi_s(x)$ and of the internal spacetime coordinates
$\X^A(x)$ that is self-adjoint in the inner product. The usual rules
of quantum theory allow one to ask, and in principle to answer,
questions about the spectrum of $\op F$ and the probabilistic
distributions of the values of the observable $F$ on the hypersurface
defined by $\X$.
\ee

\subsubsection{A Minisuperspace Model}
\label{SSSec:RW}
One of the basic questions in this programme is how to select the
internal embedding variables and, in particular, the internal time
variable $\T(x)$. Three general options have been explored:
    \bi
    \item {\em Intrinsic time\/} $\T(x,g]$. The internal-time value
of a spacetime event is constructed entirely from the internal metric
$g_{ab}(x)$ carried by the hypersurface of the reference foliation
that passes through the event.

    \item {\em Extrinsic time\/} $\T(x;g,p]$. To identify the time
coordinate of an event, one needs to know not only the intrinsic
metric on the hypersurface, but also its extrinsic curvature, and
hence how the hypersurface would appear from the perspective of a
classical spacetime constructed from the pair $(g,p)$.

    \item {\em Matter time\/} $\T(x;g,p,\f,p_\f]$. Time is not
constructed from geometric data alone but from matter fields $\f$
coupled to gravity, or from a combination of matter and gravity
fields.
\ei
Very little is known about how to choose an internal time in the full
theory of quantum gravity although it is noteable that the use of an
extrinsic time is natural in the linearised theory and was  discussed
in some of the original series of ADM papers \cite{ADM60a,ADM62}, and
in \citen{Kuc70}. This type of time appears naturally also in the
theory of quantised cylindrical gravitational waves
\cite{Kuc71,Kuc73,Kuc92a}, and in the Ashtekar formalism
\cite{Ash91}.

    The full theory is very complicated and most studies have been
in the context of so-called {\em minisuperspace\/} models in
which only a finite number of the gravitational degrees of freedom
are invoked in the quantum theory, the remainder being eliminated by
the imposition of symmetries on the spatial metric. These models
represent homogeneous (but, in general, anisotropic) cosmologies, of
which the various Bianchi-type vacuum spacetimes (in which the
spatial three-manifolds are assorted three-dimensional Lie groups)
have been the most widely studied. The requirement of homogeneity
limits the allowed hypersurfaces to a priviledged foliation in which
each leaf can be labelled by a single time variable.

    Restricting things in this way has the big advantage that the
worst quantum field theory problems---in particular,
non-renormalisability---do not arise, which means that some of the
conceptual issues can be discussed in a technical framework that is
mathematically well-defined. The main limitation of these models is
the absence of any proper perturbative scheme into which they can be
made to fit. At the classical level, models of this type correspond
to exact solutions of Einstein's field equations, but this not true
in the corresponding quantum theory since, for example, the infinite
set of neglected modes presumably possess zero-point fluctuations.
Some of these issues have been discussed in detail by \citen{KR89} to
which the reader is referred for further information. However, we
shall not get too involved with these particular issues here since
our main use of minisuperspace models is to illustrate various
specific features of the problem of time, and this they do quite
well.

    A simple example is when the spacetime geometries are restricted
to be one of the familiar homogeneous and isotropic Robertson-Walker
metrics
\beq
 g_{\a\b}(X)\,dX^\a\otimes dX^\b := -\Np(t)^2\,dt\otimes dt +a(t)^2
                \w_{ab}(x)\,dx^a\otimes dx^b
\eeq
where $\w_{ab}(x)$ is the metric for a three-space of constant
curvature $k$. The case $k=1$ is associated with $\Si$ being a
three-sphere, while $k=0$ and $k=-1$ correspond to the flat and
hyperbolic cases respectively. Note that the lapse vector $\N$ does
not appear in this expression.

     Spacetimes of this type produce a non-vanishing Einstein tensor
$G_{\a\b}$ and therefore require a source of matter. Much work in recent
times has employed a scalar field $\f$ for this purpose, and we shall use
this here. It can be shown \cite{KV74,BI75} that the coupled equations
\footnote{There is no specific information in these notes about how
the canonical formalism for general relativity is extended to include
matter but, generally speaking, nothing very dramatic happens. In
particular, although the supermomentum and super-Hamiltonian
constraints acquire contributions from the matter variables, the
crucial Dirac algebra is still satisfied. However, the details can be
quite complicated, especially for systems with internal degrees of
freedom; the subject as a whole has been analysed comprehensively in
a series of papers by Kucha\v{r} (1976a, 1976b, 1976c, 1977).
\nocite{Kuc76a,Kuc76b,Kuc76c,Kuc77}}
for the variables $a$, $\f$ and $\Np$ can all be derived
from the first-order action
\footnote{For simplicity, I have chosen units in which $8\pi G/c^4=1$.}
\beq
    S[a,p_a,\f,p_\f,\Np]=\int dt\,\big(p_a\dot a +
                                p_\f\dot\f-\Np H_{\rm RW}\big)
\eeq
where $p_a,a,p_\f,\f$ and $\Np$ are functions of $t$ and are to
be varied independently of each other. The super-Hamiltonian in this
model is
\beq
    H_{\rm RW} := {-p_a^2\over24a}-6ka +{p_\f^2\over2a^3}+a^3V(\f).
                                                        \label{Def:HRW}
\eeq

    A natural choice for an intrinsic time in the $k=1$ model is the
radius $a$ of the universe. Thus, according to the discussion above,
the reduced Hamiltonian for this system can be found by solving the
constraint
\beq
    {-p_a^2\over24a} -6a +{p_\f^2\over 2a^3} +a^3V(\f)=0 \label{HRW=0}
\eeq
for the variable $p_a$. Hence
\beq
    h_a = \pm\sqrt24\Big(-6t^2+{p_\f^2\over2t^2}+ t^4V(\f)\Big)^\half.
                                                        \label{RW-H(a)}
\eeq
A simple extrinsic time is $t:=p_a$ which (in the simple case
$V(\f)=0$) leads to the Hamiltonian
\beq
    h_{p_a}=\pm\bigg({-{1\over12}t^2\pm({1\over144}t^4+48p_\f^2)^\half
                       \over 24}\bigg)^\half,           \label{RW-H(pa)}
\eeq
while a natural definition using the matter field is $t:=\f$, which
yields
\beq
h_\f = \pm\big({a^2p_a^2\over12}+12a^4-2a^6V(t)\big)^\half.
                                                        \label{RW-H(f)}
\eeq

\subsubsection{Mean Extrinsic Curvature Time}
One of the few definitions of internal time that has been studied in
depth is one in which spacetime is foliated by hypersurfaces of
constant mean intrinsic curvature. The $\T$ functional is
\beq
    \T(x;g,p]:=\textstyle{{2\over3}}\dg^{-\half}(x)\,p^a{}_a(x)
                                                    \label{T-York}
\eeq
which is conjugate to
\beq
    \P_\T(x;g,p]:= -\dg^\half(x)                        \label{PT-York}
\eeq
in the sense that
\beq
   \big\{\T(x),\P_\T(x')\big\}=\de(x,x').
\eeq
The use of \eq{T-York} as an internal time variable in the spatially
compact case was developed in depth by York and his collaborators
\cite{Yor72a,Yor72b,SY78,Yor79,IM84}. An interesting account of its
use in $2+1$ dimensions is Carlip (1990, 1991)\nocite{Car90,Car91}.

     The first step in implementing the quantisation programme is to
extend the definitions \eqs{T-York}{PT-York} to a full canonical
transformation of the type \eq{2CT}. One possible example is
\beq
 \big(g_{ab}(x),p^{cd}(x)\big)\mapsto\big(\T(x),\s_{ab}(x),
            \P_\T(x),\pi^{cd}(x)\big)               \label{CT-York}
\eeq
where the `conformal metric' $\s_{ab}(x)$ and its 'conjugate'
momentum $\pi^{cd}(x)$ are defined by
\beqa
    \s_{ab}  &:=& \dg^{-{1\over3}}g_{ab},              \\
    \pi^{ab} &:=& \dg^{{1\over3}}\big(p^{ab}-
                                \textstyle{{1\over3}}p\,g^{ab}\big).
\eeqa
Note that $\det\s_{ab}=1$ and $\pi^{ab}$ is traceless, and so each
field corresponds to $5\times\infty^3$ independent variables.
\footnote{This is reflected in the fact that the Poisson bracket of
$\s_{ab}(x)$ with $\pi^{cd}(x')$ is proportional to the {\em
traceless\/} version of the Kronecker-delta. Thus these variables
are not a conjugate pair in the strict sense: to get such variables
it would be necessary first to solve the second-class constraints
$\det\s_{ab}=1$ and $\pi^c_c=0$. Alternatively, these variables
could be employed as part of an over-complete set in a
group-theoretical approach to the quantum theory.}
The next step therefore should be to remove $3\times\infty^3$
variables from each by identifying the spatial parts $\Z^a(x;g,p]$
of the embedding variables plus their conjugate momenta. However,
since we are concerned primarily with the problem of time, we shall
not perform this (rather complex) step but concentrate instead on
what is already entailed by the use of the preliminary canonical
transformation \eq{CT-York}. The complete analysis is contained in
the papers by York {\em et al\/} cited above.

    In terms of these new variables $\T$, $\P_\T$, $\s_{ab}$ and
$\pi^{ab}$, the super-Hamiltonian constraint $\Hp=0$ becomes the
equation for $\Phi:=(-\P_\T)^{1\over6}$
\beq
{1\over\k^2}\left(\triangle_\s\Phi -\textstyle{{1\over8}}\,R[g]\right) +
        \k^2\left(\textstyle{{1\over8}}\pi^{ab}\pi_{ab}\Phi^{-7} -
        \textstyle{{3\over 64}}\T^2\Phi^5\right)=0   \label{Hperp-York}
\eeq
where $\triangle_\s$ is the Laplacian operator constructed from the
metric $\s_{ab}$. This is a non-linear, elliptic,
partial-differential equation which, in principle, can be solved for
$\Phi$ (and hence for $\P_\T=-\dg^\half$) as a functional of the
remaining canonical variables
\beq
     \P_\T(x) + h(x;\T,\s,\pi]=0.
\eeq
In the quantum theory this leads at once to the desired functional
Schr\"odinger equation
\beq
     i\hbar{\de\Psi[\T,\s]\over\de\T(x)} =
        \big(h(x;\T,\op\s,\op\pi\,]\Psi\big)[\T,\s].
\eeq

\subsubsection{The Major Problems}
As remarked earlier, considerable conceptual advantages accrue from
using the internal Schr\"odinger interpretation of time, and it is
therefore unfortunate that the scheme is riddled with severe
technical difficulties. These fall into two categories. The first are
those quantum-field theoretic problems that are
expected to arise in any theory that we know from weak-field
perturbative analysis to be non-renormalisable. In particular, the
Hamiltonian densities $h_A(x;\X,\f,\pi]$ are likely to be highly
non-linear functions of the fields $\f^r(x),\pi_s(x)$ and we shall
encounter the usual problems when trying to define products of
quantum fields evaluated at the same point.

    The second class of problems are those that more closely involve
the question of time itself and are not directly linked to the
non-renormalisability of the theory. Indeed, many such problems arise
already in minisuperspace models which, having only a finite number
of degrees of freedom, are free from the worst difficulties of
quantum field theory. I shall concentrate here mainly on this second
type of problem, using the classification mentioned briefly at the
end of \S\ref{SSec:TProb}; see also \citen{Ish92} and \citen{Kuc92a}.

\medskip\noindent
{\em The Global Time Problem\/.}
It is far from obvious that it is possible to perform a canonical
transformation \eq{2CT} with the desired characteristics. This raises
the following questions:
\be
    \item What properties must the functionals $\X^A(x;g,p]$ possess
in order to serve as internal spacetime coordinates?

    \item If such functions do exist, is it possible to perform a
canonical transformation of the type \eq{2CT} which is such that:
    \be
    \item the constraints $\H_\perp(x;g,p]=0$ and $\H_a(x;g,p]=0$ can
be  solved globally (on the phase space) for $\P_A$ in the form
\eq{2Pm+h=0}; and
    \item there is a unique such solution?
    \ee
\ee
One example of the first point is the requirement hat the internal space and
tim
coordinates produce a genuine foliation of the spacetime. In
particular, if $t\mapsto\big(g(t),p(t)\big)$ is any curve in the
Cauchy data of a solution to the vacuum Einstein equations, we
require $d\T(x;g(t),p(t)]/dt>0$. Of course, if $\g$ is static,
this condition can never be satisfied, which means that spacetimes of
this type do not admit internal time functions. This is hardly
surprising, but it does show that we are unlikely to find
conditions on $\T$ that are to be satisfied on {\em all\/}
spacetimes.

    The existence of functionals satisfying these conditions in
various minisuperspace models has been studied by H\'aj\'\i\v{c}ek (1986,
1988, 1989, 1990a, 1990b).\nocite{Haj86,Haj88,Haj89,Haj90a,Haj90b} He
showed that these finite-dimensional models generically exhibit
global obstructions to the construction of such functionals, and the
evidence suggests that the problem may get worse as the number of
degrees of freedom increases. See also the comments in \citen{Tor92a}.

    This is not really surprising since, if $\chi^A$ are to serve as
proper `gauge functions', the subspace of the space $\cal S$ of all
$(g,p)$ satisfying the equations $\X^A(x,g,p]=\chi^A(x;g,p]$ (and
also the constraints $\Hp(x)=0=\Ha(x)$) must intersect each orbit of
the Dirac algebra just once. This is analogous to gauge-fixing in
Yang-Mills theory and therefore one anticipates the presence of a
Gribov phenomenon in the form of obstructions to the
construction of such a global gauge \cite{Sin78}. This possibility
arises because the topological structure of the space of physical
configurations is non-trivial. However, this structure seems not to
have been studied systematically in the full theory of general
relativity, and so the subject warrants further investigation.

\medskip\noindent
{\em The Spatial Metric Reconstruction Problem\/.}
At a classical level the canonical transformation \eq{CT} can be
inverted and, in particular, the metric $g_{ab}(x)$ can be expressed
as a functional $g_{ab}(x;\X,\P,\f,\pi]$ of the embedding variables
$(\X^A,\P_B)$ and the physical degrees of freedom $(\f^r,\pi_s)$. The
question is whether something similar can be done at the quantum
level. In particular:
\bi
    \item is it possible to make sense of an expression like
$g_{ab}(x;\X,\op\P,\op\f,\op\pi\,]$ with $\op\P_A$ being replaced by
$-\op h_A$?;

    \item if so, is there any sense in which this operator looks like
an operator version of a Riemannian {\em metric\/}?
\ei
The first question is exceptionally difficult to answer. Even
classically, $g_{ab}(x)$ is likely to be a highly non-linear and
non-local function of the physical modes $(\f^r,\pi_s)$, and
therefore  intractable problems of operator ordering and infinite
operator-products can be expected in the quantum theory. One might
attempt to answer the second question by showing that, if they can be
defined at all, the operators $g_{ab}(x;\X,\op\P,\op\f,\op\pi\,]$
and ${p_a}^b(x;\X,\op\P,\op\f,\op\pi\,]$ satisfy the {\em affine\/}
commutation relations \eqs{ACR:gg}{ACR:gp} that arise in the
quantisation schemes of type II where all $6\times\infty^3$ modes
$g_{ab}(x)$ are afforded operator status (see \S\ref{SSec:CQG-QBC}).

\medskip\noindent
{\em The Definition of the Operators $\op h_A(x)$\/.}
Even apart from the question of ultra-violet divergences, many
problems appear when trying to construct operator equivalents of the
Hamiltonian/momentum densities $h_A(x;\X,\f,\pi]$ that arise as the
solutions \eq{2Pm+h=0} of the initial-value constraints for the
conjugate variables $\P_A(x)$. For example:

    1.\ As mentioned earlier, the solution for $\P_A(x)$ may exist
only locally in phase space, and there may be more than one such
solution. In the latter case it might be possible to select a
particular solution on `physical grounds' (such as the requirement
that the physical Hamiltonian is a positive functional of the
canonical variables) but the status of such a step is not clear since
it means that certain classical solutions to the field equations are
deliberately excluded.

     Simple examples of these phenomena can be seen in the
Robertson-Walker model \eqs{RW-H(a)}{RW-H(f)}. The constraint
\eq{HRW=0} is quadratic in the conjugate variable $p_a$ and therefore
has the {\em two\/} solutions in \eq{RW-H(a)}. Note also that for a
number of typical potential functions $V(\f)$ the expression under
the square root in \eq{RW-H(a)} is negative for sufficiently large
$t$, and the range of such values depends on the values of the
canonical variables. Thus even classically the constraint can be
solved by a {\em real\/} $p_a$ only in a restricted region of phase
space and values of $t$.

    2.\ Even if it does exist, the classical solution for
$h_A(x;\X,\f,\pi]$ is likely be a very complicated expression of the
canonical variables. Indeed, it may exist only in some implicit
sense: a good example is the solution $\Phi$ of the elliptic
partial-differential equation \eq{Hperp-York} that determines $\P_\T$
in the case of the mean extrinsic curvature time. The solution may
also be a very non-local function of the canonical variables.

    These properties of the classical solution pose various problems
at the quantum level. For example:
    \bi
       \item Operator ordering is likely to be a major difficulty.
This is particularly relevant to the problem of functional evolution
discussed below.

        \item The operator that represents a physical quantum
Hamiltonian is required to be self-adjoint and positive. In a simple
model, the positivity requirement may involve just selecting a
particular solution to the constraints, but self-adjointness and
positivity are very difficult to check in a situation in which even
the classical expression is only an implicit function of the
canonical variables.

        \item As remarked already, the constraint equations may well
be algebraic equations for $\P_A$ whose solution therefore involves
taking roots of some operator $\op K$. This can be done with the aid
of the spectral theorem provided $\op K$ is a positive, self-adjoint
operator---which takes us back to the preceding problem. If $\op K$
is {\em not\/} positive then even if the quantum Hamiltonian exists
it is not self-adjoint, and so the time evolution becomes
non-unitary.
    \ei

    3.\ The solutions to the constraints are likely to be explicit
functionals of the internal spacetime coordinate functions, which
gives rise to a time-dependent Hamiltonian. This effect, which can be
seen clearly in the minisuperspace example, has several implications:
\bi
    \item A time-dependent Hamiltonian means that energy can be fed
into, or taken out of, the quantum system. In normal physics, this
happens whenever the system is not closed, with the time dependence
of the Hamiltonian being determined by the environment to which the
system couples. However, a compact three-manifold $\Si$ (the
`universe') has no external environment, and so the time-dependence
seems a little odd.

    \item If $h(t)$ is time-dependent, the Schr\"odinger equation
\beq
         i\hbar{d\psi_t\over dt}=\op h(t)\psi_t
\eeq
does not lead to the simple second-order equation
\beq
       -\hbar^2{d^2\psi_t\over dt^2} = \big(\op h(t)\big)^2\psi_t
\eeq
because of the extra term involving the time derivative of $\op
h(t)$. In the case of quantum gravity this means that the functional
Schr\"odinger equation does {\em not\/} imply the Wheeler-DeWitt
equation. This is not necessarily a bad thing but it does illustrates
the potential inequivalence of different approaches to the canonical
quantisation of  gravity.

    \item As remarked already, roots of operators can be handled if
the object whose root is being taken is self-adjoint and positive.
However, in general this will be true only for certain ranges of $t$
(which can depend on the values of the true canonical
variables). Thus a proper Hamiltonian operator may exist only for a
limited range of time values. A clear example of this is the
Robertson-Walker model \eq{RW-H(a)}.
\ei
These difficulties in defining the operators $\op h_A(x)$ may seem
collectively to constitute a major objection to the internal
Schr\"odinger interpretation. However, the apparent failure to
identify a completely satisfactory set of internal spacetime
coordinates is not necessarily a disaster: it might reflect something
of genuine physical significance. For example, much work has been
done in recent years on quantum theories of the creation of the
universe and, in any such theory, something peculiar must necessarily
happen to time near the origination point. A good example is the
schemes of \citen{HH83} and \citen{Vil88}, in both of which there is
a sense in which time becomes imaginary. Perhaps the tendency to
produce a non-unitary evolution in the internal time variables is a
reflection of this effect.

\medskip\noindent
{\em The Multiple Choice Problem\/.}
Generically, there is no geometrically natural choice for the
internal spacetime coordinates and, classically, all have an equal
standing. However, this classical cornucopia becomes a real problem
at the quantum level since there is no reason to suppose that the
theories corresponding to different choices of time will agree.

    The crucial point is that two different choices of internal
coordinates are related by a canonical transformation and, in this
sense, are classically of equal validity. However, one of the central
properties/problems of the quantisation of any non-linear system is
that, because of the well-known Van-Hove phenomenon
\cite{Gro46,VanH51}, most classical canonical transformations {\em
cannot\/} be represented by unitary operators while, at the same
time, maintaining the irreducibility of the canonical commutation
relations. This means that in quantising a system it is always
necessary to select some preferred sub-algebra of classical
observables which is to be quantised \cite{Ish84}. One analogue of
this situation in quantum gravity is precisely the dependence of the
theory on the choice of internal time. It must be emphasised that
this phenomenon is not related to ultra-violet divergences, or other
pathologies peculiar to quantum field theories, but arises already in
finite-dimensional systems. A good example is the Robertson-Walker
model used above to illustrate various types of internal time. The
three Hamiltonians \eq{RW-H(a)}, \eq{RW-H(pa)} and \eq{RW-H(f)} are
associated with different quantum theories of the same classical
system.

\medskip\noindent {\em The Problem of Functional Evolution\/.}
The problem of functional evolution is concerned with the key
question of the consistency of the dynamical evolution generated by
the constraints; in particular, the preservation of the constraints
as the system evolves in time. Much more than most of the other
problems of time, the issue of functional evolution depends
critically on our ability to construct quantum gravity as a
consistent quantum {\em field\/} theory by giving sense to the
infinite collection of Hamiltonians and their commutation relations.

    At the classical level, there is no problem. The consistency of
the original constraints \eq{EE-can-Ha}, \eq{EE-can-Hperp} with the
dynamical equations \eq{EE-can-gab}, \eq{EE-can-pab} follows from the
first-class nature of the constraints, \ie their Poisson brackets
vanish on the constraint subspace by virtue of the Dirac algebra
\eqs{PB:HaHb}{PB:HH}. Similarly, the consistency of the internal time
dynamical equations \eqs{FEM:grav-f}{FEM:grav-pi} is ensured by the
Poisson bracket relations between $h_A(x)$ with $h_B(x')$. These
results mean that the classical evolution of the system from one
initial hypersurface to another is independent of the family of
hypersurfaces chosen to interpolate between them.

    In the internal Schr\"odinger interpretation, the analogous
requirement on the quantum operators is
\beq
    {\de\op h_A(x,\X\,]\over\de\X^B(x')}-
        {\de\op h_B(x,\X\,]\over\de\X^A(x')} +
            {1\over i\hbar}[\,\op h_A(x,\X\,],\op h_B(x',\X\,]\,]=0.
                                                    \label{CR:h1h2}
\eeq
If this fails, the functional Schr\"odinger equation \eq{FSE:grav}
breaks down.

    It is clear that these conditions can be checked only if the
operators concerned are defined properly, which raises the entire
gamut of problems in quantum gravity, including those of
operator-ordering and ultra-violet divergences. Not surprisingly,
very little can be said about this problem in the full theory. The
best that can be done is to elucidate some of the issues on
sufficiently simple (and hence almost trivial) systems. It is
essential however that these systems have an {\em infinite\/} number
of degrees of freedom, \ie we have to deal with a genuine {\em
field\/} theory for the phenomenon to arise at all. In particular,
minisuperspace models tell us nothing about this particular problem
of time.

     One useful example is the functional evolution of a
parametrised, massless scalar field propagating on a
$1+1$-dimensional flat cylindrical Minkowskian background
\footnote{It should be possible to generalise these results to any
parametrised free field theory on a flat four-dimensional background,
but this does not seem to have been done.}
spacetime: a system that has been studied
\footnote{Similarly, there exists a canonical transformation which
casts the super-Hamiltonian and supermomentum constraints of a
bosonic string moving in a $d$-dimensional target space into those of
a parametrised theory of $d-2$ independent scalar fields propagating
on a two-dimensional Minkowskian background. This enables the
functional evolution problem to be posed, and solved, for the bosonic
string \cite{KT89,KT91b}.}
in some detail by Kucha\v{r} (1988, 1989a, 1989b). There is a
canonical formulation of this theory which casts the
super-Hamiltonian and supermomentum constraints into the same form as
those of the midisuperspace model of cylindrical gravitational waves.
This links the functional evolution problem in the parametrised field
theory with that in quantum gravity proper \cite{Tor91}.

    In this particular case it has been shown that with the aid of
careful regularisation and renormalisation the apparent anomaly in
the commutator of the $\op h_A(x)$ operators can be removed, and
hence a consistent quantum evolution attained. However, in the full
theory of quantum gravity the problem of functional evolution is
particularly difficult to disentangle from the ambiguities generated
by ultra-violet divergences, and very little is known about its
solution.

\medskip\noindent
{\em The Spacetime Problem\/.}
We have seen how in the canonical version of general relativity
`time' is to be viewed as a function variable rather than the single
parameter of Newtonian physics. In classical geometrodynamics we are
required to locate an event $X$ in an Einstein spacetime $(\M,\g)$
using the canonical data on an embedding that passes through $X$.
Given a particular choice $(\T,\Z^a)$ of internal time and space
functions, the coordinates associated with the point $X$ were given
in \eq{Def:T(X)} and \eq{Def:Za(X)} as
\beq
    T(X) := \T\big(\s^\rref(X);g[\Fo^\rref_{t=\tau^\rref(X)}\,],
                p[\For_{t=\tau^\rref(X)}\,]\,\big],   \label{2Def:T(X)}
\eeq
and
\beq
    Z^a(X) :=\Z^a\big(\s^\rref(X);g[\For_{t=\tau^\rref(X)}\,],
                p[\For_{t=\tau^\rref(X)}\,]\,\big].   \label{2Def:Za(X)}
\eeq

    These internal embedding variables have some peculiar properties.
In particular, if a different reference foliation is used, so that
the hypersurface $\For_{t=\tau^\rref(X)}:\Si\map\M$ passing through
the same point $X$ in $\M$ is different, then the `time' value $T(X)$
alloted to the event will generally not be the same. Thus the time of
an event depends not just on the event itself, but also on a choice
of spatial hypersurface passing through the event.

    Viewed from a spacetime perspective, this feature is pathological
since coordinates on a manifold are local scalar functions. It is
physically highly undesirable since it means that the results and the
interpretation of the theory depend on the choice of the reference
foliation $\For$. Having to choose a specific background foliation
is equivalent to introducing a Newtonian-type universal time
parameter, which is something we wish to avoid since there is no
natural place in general relativity for such a field-independent
reference system. Indeed, our goal is to construct a formalism in
which $\For$ drops out entirely from the final result.

    This undesirable behaviour of the internal time $\T(x;g,p]$ can
be avoided only if it has a vanishing Poisson bracket with the
generator of `tilts' or `bends' of the hypersurface. In particular,
we require \cite{Kuc76b,Kuc82,Kuc91a}
\beq
         \big\{\T(x),H[N]\,\big\}=0                 \label{PB:THN=0}
\eeq
for all test functions $N$ that vanish at a point $x\in\Si$. Of
course, the same limitation should also be imposed on the internal
spatial coordinates $\Z^a(x;g,p]$. The search for internal spacetime
coordinates that satisfy \eq{PB:THN=0} constitutes the {\em spacetime
problem\/}.

    The requirement \eq{PB:THN=0} is rather strong. It clearly
excludes any intrinsic time like $R(x,g]$ that is a local functional
of the three-geometry alone. It also excludes many obvious choices of
an extrinsic time. For example, the time function \eq{T-York} is
certainly not a spacetime scalar: when a hypersurface is bent around
a given event in a vacuum Einstein spacetime, the mean extrinsic
curvature ${2\over3}\dg(x)^{-\half}\,g^{ab}(x)\,p_{ab}(x)$ changes,
even if the event remains the same. Thus the canonical coordinate
\eq{T-York} cannot be turned into a coordinate on spacetime.

    There do exist functionals $\T(x;g,p]$ of the canonical data that
{\em are\/} local spacetime scalars. For example, take the square
${}^{(4)}R_{\a\b\gamma\de}(X,\g]\,{}^{(4)}R^{\a\b\gamma\de}(X,\g]$ of
the Riemann curvature tensor of a vacuum Einstein spacetime, and
reexpress it in terms of the canonical data on a spacelike
hypersurface.  But note that not all functionals $\T(x;g,p]$ of this
type can serve as time functions. Two necessary conditions are:
\be
    \item $\big\{\T(x),\T(x')\big\}=0$;

    \item for any given Lorentzian metric $\g$ on $\M$ that satisfies
the vacuum Einstein field equations, the hypersurfaces of equal $\T$
time must be {\em spacelike\/}. (These hypersurfaces can be evaluated
using \eqs{2Def:T(X)}{2Def:Za(X)} with any convenient reference
foliation $\For$; the answer will be independent of the choice of
$\For$ for internal spacetime functionals that are compatible with
the spacetime problem.)
\ee
However, even if these conditions are satisfied, it is still
necessary to split off $\T$ from the rest of the canonical variables
by a canonical transformation \eq{CT}, and this is by no means a
trivial task. In fact, Kucha\v{r} and I are not aware of a single
concrete example of a decomposition of the canonical variables based
on a local scalar time function $\T(x;g,p]$. Of course, another
possibility is to construct scalars from {\em matter\/} fields in the
theory, and this is the topic of the next subsection.

\subsection{Matter Clocks and Reference Fluids}
\label{SSec:RefFluid}
\subsubsection{The Basic Ideas}
We have seen that it is difficult to produce a satisfactory definition of
time using only the canonical variables of the gravitational field. In
particular, there is nothing in the canonical formalism itself to provide
insight into how the internal spacetime functionals $\T(x;g,p]$ and
$\Z^a(x;g,p]$ are to be selected. However, in practice, location in time
and in space is not performed in this way. Real physical clocks are made of
matter with definite properties---an observation that has generated a
recent flurry of interest in the idea of `quantum clocks' with the hope
that they may lead to a more tractable approach to the problem of time.

    In a sense, the idea of matter clocks has already been implicit
in what has been said so far. For example, one of the dynamical
variables in the simple minisuperspace model \S\ref{SSSec:RW} is the
spatially-homogeneous scalar field $\f$, and this can be used to
define time. Indeed, \eq{RW-H(f)} is the Hamiltonian obtained by
selecting $t=\f$ as a classical time variable. However, a `quantum
clock' does not mean an arbitrary collection of particle or
matter-field variables, but rather a device whose self-interaction
and coupling to the gravitational field are deliberately optimised to
serve as a measure of time. The important question is the extent to
which the ability to measure (or, more precisely, to define) time using
one of these systems is compatible with its realisation as a real
physical entity. In particular, it must have a positive energy: a
property that, as emphasised in \S\ref{Sec:Problem}, is far from
being trivial

    Two different types of system are feasible. The first is a
point-particle clock that can measure time only at points along its
worldline. The second is a cloud of such clocks that fills the space
$\Si$ and which can therefore provide a global measure of time and
spatial position. The origin of this idea lies in the old classical
notion of a reference fluid, and was first applied to quantum gravity
in a major way by \citen{DeW62} (see also \citen{DeW67a}) who
discussed a gravitational analogue of the famous Bohr-Rosenfeld
analysis of the measurability of the quantised electromagnetic field
\cite{BR33,BR78}.

    In schemes of this type, the cloud of clocks is regarded as a
realistic material medium with a Lagrangian that accurately describes
its  physical properties. An important question is the precise sense
in which the ensuing structure corresponds to a genuine coordinate
system on the spacetime manifold. A different approach is to {\em
start\/} with a fixed set of coordinate conditions imposed on the
spacetime metric $\g$, and then implement them by appending the
conditions to the action with a family of Lagrange multipliers. These
extra terms in the action are then parametrised (\ie made invariant
under the action of $\DM$) and interpreted as the source terms for a
special type of matter. Schemes of this type have their origin in the
general problem of understanding if, and how, the full spacetime
diffeomorphism group $\DM$ (rather than its projections in the form
of the Dirac algebra) should be represented in the canonical theory
of gravity  \cite{BK72,SS83,IK85a,IK85b,Kuc86,LW90}.

\subsubsection{The Gaussian Reference Fluid}
The work of Kucha\v{r} and his collaborators has been especially
significant and I shall illustrate it here with the example discussed in
\citen{KT91a} of a Gaussian reference fluid. Other examples are
harmonic coordinate conditions \cite{KT91c,SK92} and the $K=${\em
const} slicing condition \cite{Kuc92b}. \citen{KT89} handle the
conformal, harmonic and light-cone gauges in the bosonic string in
the same way, and the canonical treatment of two-dimensional induced
quantum gravity is discussed in \citen{Tor89}.

    The main steps in the development of the Gaussian reference fluid
are as follows.

    1.\  The Gaussian coordinate conditions are $\g^{00}(X)=-1$ and
$\g^{0a}(X)=0$, $a=1,2,3$, but a more covariant-looking expression
can be obtained by introducing a set of spacetime functions $\T$,
$\Z^a$, $a=1,2,3$ and imposing the Gaussian conditions in the form
\beqa
    \g^{\a\b}(X)\,\T,_\a(X)\,\T,_\b(X)   &=& -1         \label{Gauss-1}\\
    \g^{\a\b}(X)\,\T,_\a(X)\,\Z^a,_\b(X) &=& 0.         \label{Gauss-2}
\eeqa
These should be viewed as a (spacetime-coordinate independent) set
of partial differential equations for the functions $(\T,\Z^a)$, a
set of whose solutions (there is some arbitrariness) are the Gaussian
coordinate functions on $\M$.

    2.\ The conditions \eqs{Gauss-1}{Gauss-2} are now added to the
dynamical system with the aid of Lagrange multipliers $M$ and $M_a$,
$a=1,2,3$. The extra term in the action is
\beq
    S_F[\g,M,\vec{M},\T,\vec\Z]:=\int_\M d^4X\,(-\det\g)^\half
        \big(-\textstyle{\half}M\left(\g^{\a\b}\T,_\a\,\T,_\b+1\right)+
            M_a\,\g^{\a\b}T,_\a\,\Z^a,_\b\big).          \label{Def:SF}
\eeq
Note that this action is invariant under coordinate transformations
on the $a$ superscript of $\Z^a$ provided that $M_a$ transforms
accordingly. This is consistent with the idea advanced in
\S\ref{SSec:IntTime} that objects like $\Z$ are best thought of as
taking their values in a {\em labelling\/} three-manifold $\Si_l$, in
which case the superscript $a$ refers to a coordinate system on
$\Si_l$. The Lagrange multiplier $M_a$ is then a hybrid object whose
domain is the spacetime manifold $\M$ but whose values lie in the
cotangent bundle to $\Si_l$.
\footnote{Thus $\vec{M}:\M\map T^*\Si_l$ with $\vec{M}(X)\in
T^*_{\Z(X)}\Si_l$. In general, if $v$ belongs to the cotangent space
$T^*_{\Z(X)}\Si_l$ of $\Si_l$, we interpret the symbol
$\Z^a,_\a(X)\,v_a$, $\a=0\ldots 3$ as the coordinate representation
on $\M$ of the pull-back $(\Z^*v)$ of $v$ to $\M$ by the map
$\Z:\M\map\Si_l$.}

    3.\ The action \eq{Def:SF} is defined covariantly on the
spacetime manifold $\M$, and is in `parametrised' form in the sense
that $(\T, \Z^a)$ are regarded as functions that can be varied
freely. These four extra variables can be interpreted as describing a
material system---the `Gaussian reference fluid'---that interacts
with the gravitational field and has its own energy-momentum tensor.
Variation of the action with respect to $\T$ and $\Z^a$ gives the
Euler hydrodynamic equations of the reference fluid, which turns out
to be a heat-conducting fluid \cite{KT91a}.

    4.\  A complete canonical analysis shows that the constraints of
the full theory have the form
\beqa
    \Pi_a(x)+\tilde{\H}_a(x;\T,\Z,g,p\,] &=&0       \label{Matt-Ha=0}\\
    \Pi_\T(x)+\tilde{\H}_\perp(x;\T,\Z,g,g\,]&=&0   \label{Matt-Hperp=0}
\eeqa
where $\tilde{\H}_a$ and $\tilde{\H}_\perp$ are linear combinations of
the usual gravitational supermomentum and super-Hamiltonian (but with
coefficients that depend on $\T$, $\Z$ and $g$), and where $\Pi_a(x)$
and $\Pi_\T(x)$ are the momenta conjugate to the variables $\Z^a(x)$
and $\T(x)$ respectively.

\subsubsection{Advantages and Problems}
The use of the Gaussian reference fluid has several big advantages
over a purely geometrical set of internal coordinate functions. In
particular:
\bi
    \item The most obvious property of the constraints
\eqs{Matt-Ha=0}{Matt-Hperp=0} is that they are {\em linear\/} in the
momenta $\Pi_a$ and $\Pi_\T$ of the matter-field coordinate
functions. In the quantum theory, this leads at once to a functional
Schr\"odinger equation of the type \eq{FSE:grav}, and hence to an
uncontentious intepretation of a state $\Psi[\,\T,\Z,g\,]$ as the
probability amplitude for measuring the three-metric $g_{ab}(x)$ on
the hypersurface in $\M$ associated with $(\T,\Z^a)$. Since the
constraints are manifestly linear in the reference-fluid momenta, the
Hamiltonian densities that appear in the Schr\"odinger equation are
local, explicit functions of the canonical variables. In addition,
the Hamiltonian is only quadratic in $p^{ab}(x)$.  These are
significant advantages over the situation that arises when purely
geometrical time and space functions are used.

    \item The spacetime problem is non-existent because the variables
$(\T,\Z^a)$ in the action \eq{Def:SF} are defined from the very
start as genuine spacetime scalar fields.
\ei

    These gains are attractive, and it is therefore unfortunate that
they are offset to some extent by several basic problems:
\bi
    \item The energy-momentum tensor of the matter fluid does not
satisfy the famous energy-conditions of general relativity, and
therefore the system cannot be regarded as physically realisable. On
the other hand, if one starts with a physically correct matter system
(such as, for example, the scalar field $\f$ in the minisuperspace
model in \S\ref{SSSec:RW}) the simple linear dependence on $\Pi_\T$
is lost. This suspension between Charybdis and Scylla is arguably an
inevitable consequence of the general problem of the non-existence of
physical Hamiltonian clocks (\S\ref{Sec:Problem}).

    \item In classical general relativity, it is well-known that
Gaussian coordinate conditions almost invariably breakdown somewhere,
and therefore Gaussian coordinates are defined only {\em locally\/}
on the spacetime manifold $\M$. This should be reflected in the
quantum theory at some point, but it is not clear where, or how.

    \item Even if the technical problems above can be overcome,
there still remains the issue of how fundamental are the Gaussian-type
coordinate conditions, and with what entities in the real world
the associated matter variables are to be identified. This is
particularly appropriate in discussions of early universe quantum
cosmology where there seems to be no room for a simple `phenomenological'
type of analysis.
\ei

\subsection{Unimodular Gravity}
One rather special example of a reference fluid is associated with
the unimodular coordinate condition
\beq
                \det\g_{\a\b}(X)=1
\eeq
that has often been used in discussions of classical general
relativity. If the procedure outlined above is applied to this
particular condition, the ensuing parametrised theory corresponds to
the usual theory of general relativity but with a cosmological
`constant' that is a dynamical variable, rather than a fixed
constant. The use of this theory as a possible solution to the
problem of time has been discussed in several recent papers, and
especially by Unruh and Wald
\cite{HT89,Unr88,Unr89,UW89,BY89,Kuc91b}.

    The super-Hamiltonian constraint of this modified theory is
\beq
            \lambda +|g(x)|^{-\half}\,\H_\perp(x) =0
\eeq
in which what would normally be the cosmological constant $\lambda$
appears as the momentum conjugate to a variable $\tau$ that is
identified as a `cosmological time'. The implication in the
quantum theory is that dynamical evolution with respect to $\tau$ is
described by the family of ordinary Schr\"odinger equations
\beq
    i\hbar{\pa\Psi(\tau,g\,]\over\pa\tau}=
|g(x)|^{-\half}\,\big(\op\H_\perp(x)\Psi\big)(\tau,g\,]   \label{SE:uni}
\eeq
parametrised by the point $x$ in $\Si$.

    The problem is how to interpret such a family of equations.
Kucha\v{r} studied this question with great care
\cite{Kuc91b,Kuc92a} and showed that the correct dynamical
Schr\"odinger equation is
\beq
    i\hbar{\pa\Psi(\tau,g\,]\over\pa\tau}=
        \Big(\int_\Si d^3x\,\dg^\half\Big)^{-\half}
            \int_\Si d^3x\,\big(\op\H_\perp(x)\Psi\big)(\tau,g\,].
\eeq
If this was the only equation satisfied by $\Psi(\tau,g\,]$ it would
be viable to interpret this function as the probability density for
the three-metric $g$ at a given value of the time $\tau$. However,
the effect of the existence of the family of equations \eq{SE:uni} is
that the state vector $\Psi$ must also satisfy the collection of
constraints
\beq
    |g(x)|^{-\half}\,\op\H_\perp(x)\big),_a\,\Psi(\tau,g\,]=0
                                    \label{Hperp-aPsi=0}
\eeq
where $x\in\Si$. But the three-geometry operator does not commute
with these constraints, and therefore the interpretation of
$\Psi(\tau,g]$ as a probability distribution for $g$ is not tenable.

    The geometrical origin of this problem is that the cosmological
time measures the four-volume enclosed between two embeddings of the
associated internal time functional $\T(x)$ but given one of the
embeddings the second is not determined uniquely by the value of
$\tau$: two embeddings that differ by a zero four-volume (something
that can happen easily in a spacetime with a {\em Lorentzian\/}
signature) cannot be separated in this way. The extra constraints
\eq{Hperp-aPsi=0} are to be interpreted as saying that the theory is
independent of this arbitrariness. For further details see the cited
papers by Kucha\v{r}.




%


\section{IDENTIFY TIME AFTER QUANTISATION}
\label{Sec:TAfter}
\subsection{Canonical Quantum Gravity: Quantise Before Constraining}
\label{SSec:CQG-QBC}
\subsubsection{The Canonical Commutation Relations for Gravity}
In approaches to the problem of time of category II, a quantum theory
is constructed without solving the constraints, which are then
imposed at the quantum level. The identification of `time' is made
{\em after\/} this process, and is used to give the final
physical interpretation of the theory, particularly the probabilistic
aspects. This final structure may be related only
loosely to the quantum structure with which the construction started.
As we shall see, this leads to a picture of quantum gravity that is
radically different from that afforded by the internal Schr\"odinger
interpretation.

    The starting point is the operator version of the Poisson-bracket
algebra \eqs{PB:gg}{PB:gp} in the form of the canonical commutation
relations
\beqa
    [\,\op g_{ab}(x),\op g_{cd}(x')\,]   &=& 0              \label{CR:gg}\\
    {[}\,\op p^{ab}(x),\op p^{cd}(x')\,] &=& 0              \label{CR:pp}\\
    {[}\,\op g_{ab}(x),\op p^{cd}(x')\,] &=&
               i\hbar\,\de^c_{(a}\de^d_{b)}\,\de(x,x')      \label{CR:gp}
\eeqa
of operators defined on the three-manifold $\Si$. Several things should be said
about this algebra:
\bi

    \item The classical object $g_{ab}(x)$ is not merely a symmetric
covariant tensor: it is also a {\em metric\/} tensor, \ie at each point
$x\in\Si$ the matrix $g_{ab}(x)$ is invertible with signature $(1,1,1)$. In
particular, for any non-vanishing vector-density field $v^a$ (of an
appropriate weight) we have
\beq
    g(v\otimes v):= \int_\Si d^3x\,v^a(x)\,v^b(x)\,g_{ab}(x)>0,
                                                    \label{gvv>0}
\eeq
and it is reasonable to require the corresponding quantum operators
$\op g(v\otimes v)$ to satisfy the analogous equations
\beq
    \op g(v\otimes v) > 0.                          \label{op-gvv>0}
\eeq
It is noteworthy that the canonical commutation relations
\eqs{CR:gg}{CR:gp} are {\em incompatible\/} with these geometrical
properties of $g_{ab}(x)$ provided that the smeared versions of the
operators $\op p^{cd}(x)$ are self-adjoint. In that case they
can be exponentiated to give unitary operators which when acting on $\op
g(v\otimes v)$ show that the spectrum of $\op g(v\otimes v)$ can take on
negative values.

    This problem can be partly remedied by replacing \eqs{CR:gg}{CR:gp} with
a set of {\em affine\/} relations
\cite{Kla70,Pil82,Pil83,Ish84,IK84a,IK84b,Ish92},
\beqa
    [\,\op g_{ab}(x),\op g_{cd}(x')\,]          &=&0    \label{ACR:gg}\\
    {[}\,{\op p_a}{}^b(x), {\op p_c}{}^d(x')\,] &=&
        i\hbar\big(\de_a^d\,{\op p_c}{}^b(x)-\de_c^b\,
            {\op p_a}{}^d(x)\big)\,\de(x,x')              \label{ACR:pp}\\
    {[}\,\op g_{ab}(x),{\op p_c}{}^d(x')\,]     &=&
        i\hbar\,\de_{(a}^d\op g_{b)c}(x)\,\de(x,x').    \label{ACR:gp}
\eeqa
At a classical level, the corresponding Poisson brackets are equivalent to
the \mbox{standard} canonical relations \eqs{PB:gg}{PB:gp} with
${p_c}^d(x):=g_{cb}(x)\,p^{bd}(x)$. However, the situation at the quantum
level is very different and, for example, there exist many representations of
th
affine relations \eqs{ACR:gg}{ACR:gp} in which the spectrum of the smeared
metric operator `almost' satisfies the operator
inequality \eq{op-gvv>0}; `almost' in the sense that the right hand side is
$\ge0$ rather than a strict inequality. This result is helpful, but it
provides only a partial resolution of the general question of the extent to
which the classical geometrical properties of $g_{ab}(x)$ can be, or should be,
captured in the quantum theory. This is how the {\em spatial metric
reconstruction\/} problem appears in this approach to quantum gravity.

    \item We are working in the `Schr\"odinger representation' in
which no time dependence is carried by the canonical variables. As
we shall see, there is more to this than meets the eye.

    \item Equation \eq{CR:gg} (or the affine analogue \eq{ACR:gg}) is
a form of microcausality. However, the functional form of the
constraints is independent of any foliation of spacetime, and
therefore it is not clear what this `microcausal' property means in
terms of the usual ideas of an `equal-time' hypersurface, or indeed
how the notion of spacetime structure (as opposed to spatial
structure) appears at all. At this stage, the most that can be said
is that, whatever the final spacetime interpretation may be,
\eq{CR:gg} implies that the points of $\Si$ are to be regarded as
spacelike separated.
\ei

\subsubsection{The Imposition of the Constraints}
The key question is how the constraint equations
\eqs{EE-can-Ha}{EE-can-Hperp} are to be handled. The essence of the Dirac
approach is to impose them as constraints on the physically allowed states
in the form
\beqa
    \Ha(x;\op g,\op p)\Psi &=& 0                    \label{HaPsi=0}\\
    \Hp(x;\op g,\op p)\Psi &=& 0.                   \label{HperpPsi=0}
\eeqa
We recall that, in the classical theory, the constraints are {\em
equivalent\/} to the dynamical equations in the sense that if they are
satisfied on all spatial hypersurfaces of a Lorentzian metric $\g$, then
$\g$ necessarily satisfies the Einstein vacuum field equations. This is
reflected in the quantum theory by the assumption that the operator
constraints \eqs{HaPsi=0}{HperpPsi=0} are the {\em sole\/} technical content
of the theory, \ie the dynamical evolution equations are {\em not\/}
imposed as well.

     This is closely related to the following fundamental observation.
According to the first-order action \eq{S:can-g}, the canonical Hamiltonian
\eq{Def:H[N]} associated with general relativity is
\beq
H[N,\N](t)=\int_\Si d^3x \big(N(x)\H_\perp(x) + N^a(x)\H_a(x)\big)
                                                \label{Def:H[N]'}
\eeq
where $N$ and $\N$ are regarded as external, $c$-number functions.
However, \eq{Def:H[N]'} has a rather remarkable implication for the putative
Schr\"odinger equation
\beq
     i\hbar{d\over dt}\Psi_t = \op H[N,\vec N](t)\Psi_t
\eeq
since if the state $\Psi_t$ satisfies the constraint equations
\eqs{HaPsi=0}{HperpPsi=0}, we see that it has no time dependence at all!
Similarly, it is not meaningful to speak of a `Schr\"odinger' or a
`Heisenberg' picture since the matrix elements between physical states of
a Heisenberg-picture field  will be the same as for the field in the
Schr\"odinger picture.

      This so-called `frozen formalism' caused much confusion when it
was first discovered since it seems to imply that nothing happens in a
quantum theory of gravity. Clearly this is some sort of quantum analogue of
the fact that classical observables (\ie those satisfying \eq{Def:obs}) are
constants of the motion. These days, this situation is understood to reflect
the
absence of any external time parameter in general relativity, and
therefore, in particular, the need to discuss the measurement of time with the
aid of functionals of the internal variables in the theory. This perspective
dominates almost all work on the problem of time in quantum gravity.

\subsubsection{Problems with the Dirac Approach}
Many problems arise when attempting to implement the Dirac scheme. For
example:
\be
    \item To what extent can, or should, the classical Poisson-bracket
algebra \eqs{PB:HaHb}{PB:HH} be maintained in the quantum theory? The
constraints \eq{Def:Ha} and \eq{Def:Hperp} are highly non-linear functions
of the canonical variables and involve non-polynomial products of field
operators evaluated at the same point. Thus we are lead inevitably to the
problems of regularisation, renormalisation, operator ordering, and
potential anomalies. This is the form taken by the {\em functional
evolution problem\/} in this approach to quantisation.

    \item  It is not clear what properties are expected of the constraint
operators $\op\H_\perp(x)$ and $\op\H_a(x)$; in particular, should they be
self-adjoint? Since one presumably starts with self-adjoint representations
of the canonical commutation representations \eqs{CR:gg}{CR:gp}, it is
perhaps natural to require self-adjointness for the super-Hamiltonian and
supermomentum operators. However, this has been challenged several times
and the issue is clearly of significance in discussing the
operator-ordering problem
\cite{Kom79a,Kom79b,Kuc86a,Kuc86b,Kuc87,HK90a,HK90b}. The possibility of
using a non-hermitian operator can be partly justified by noting that the
Hilbert space structure on the space that carries the representation of the
canonical algebra may be only distantly related to the Hilbert space
structure that ought to be imposed on the {\em physical\/} states (\ie those
that satisfy the constraints).

    \item More generally, what is the relation between these two Hilbert
spaces? This question has important implications for the problem of time.

    \item What is meant by an `observable' in a quantum theory of this type?
By analogy with the classical result \eq{Def:obs}, one might be tempted to
postulate that an operator $\op A$ defined on the starting Hilbert space
corresponds to an observable if
\beq
        [\,\op A, \op H[N,\N]\,]=0                      \label{Def:obs-qu}
\eeq
for all test functions $N$ and $\N$. However, it can be argued that it is
sufficient for \eq{Def:obs-qu} to be satisfied on the subspace of {\em
solutions\/} to the Dirac constraints \eqs{HaPsi=0}{HperpPsi=0}. This is the
quantum analogue of the fact that the classical condition \eq{Def:obs} for an
observable is a {\em weak\/} equality, \ie it holds only on the
subspace of the classical phase space given by the solutions $(g,p)$ to the
classical constraints $\Ha(x;g,p]=0=\Hp(x;g,p]$.
\ee

\subsubsection{Representations on Functionals $\Psi[g]$}
In attempting to find concrete representations of the canonical algebra
\eqs{CR:gg}{CR:gp} it is natural to try an analogue of the quantum scalar
field representations \eq{Def:op-f} and \eq{Def:op-pi}. Thus the state
vectors are taken to be functionals $\Psi[g]$ of Riemannian metrics $g$ on
$\Si$, and the canonical operators are defined as
\beqa
\big(\op g_{ab}(x)\Psi\big)[g] &:=&g_{ab}(x)\Psi[g] \label{Def:op-g}\\
\big(\op p^{cd}(x)\Psi\big)[g] &:=&-i\hbar{\de\Psi[g]\over\de g_{cd}(x)}.
                                                    \label{Def:op-p}
\eeqa
These equations have been used widely in the canonical approach to
quantum gravity although, even when suitably smeared, they do not define
proper self-adjoint operators because of the absence of any Lebesgue measure
on $\RS$ (\eg \citen{Ish92}) and, as remarked earlier, they are also
incompatible with the positivity requirement \eq{op-gvv>0}.

     Let us consider the Dirac constraints \eqs{HaPsi=0}{HperpPsi=0} in
this representation. From a physical perspective it is easy to see the
need for them. Formally, the domain space of the
state functionals is $\RS$, and to specify a metric $g_{ab}(x)$ at a
point $x\in\Si$ requires six numbers (the components of the metric in
some coordinate system). However, the true gravitational system should
have only {\em two\/} degrees of freedom per spatial point, and therefore
four of the six degrees of freedom need to be lost. This is precisely
what is achieved by the imposition of the constraints
\eqs{HaPsi=0}{HperpPsi=0}. As we saw in \S\ref{SSec:CQG-CBQ}, the same
counting argument applies if the system is reduced to true canonical form
before quantising.

      The easiest constraints to handle are those in the first set
\beq
        \big(\op H[\N]\Psi\big)[g] = 0.           \label{HaPsi=0'}
\eeq
We saw earlier that the classical functions $H[\N]$ are the infinitesimal
generators of the diffeomorphism group of $\Si$, and the same might be
expected to apply here. The key question is whether the operator-ordering
problems can be solved so that the algebra \eq{PB:Sm-HaHb} is preserved at
the quantum level. In practice, this is fairly straightforward;
indeed, one powerful way of {\em solving\/} the operator-ordering problem
for the $\op H[\N]$ generators is to insist that they form a self-adjoint
representation of the Lie algebra of $\DS$.

     The implications of \eq{HaPsi=0'} are then a straightforward
analogue  of those in conventional Yang-Mills gauge theories. The
group $\DS$ acts as a group of transformations on the space $\RS$ of
Riemannian metrics on $\Si$, with $f\in\DS$ sending $g\in\RS$ to
$f^*g$. Apart from certain technical  niceties, this leads to a
picture in which $\RS$ is fibered by the orbits of the $\DS$ action.
Then \eq{HaPsi=0'} implies that the state functional $\Psi$ is
constant (modulo possible $\theta$-vacuua effects) on the orbits of
$\DS$, and therefore passes to a function on the {\em superspace\/}
$\RS/\DS$ of $\DS$ orbits \cite{Mis57,Hig58}.

\subsubsection{The Wheeler-DeWitt Equation}
\label{SSSec:WDE}
We must consider now the final constraint $\op\H_\perp(x)\Psi=0$. Unlike the
constraint $H[\N]\Psi=0$, this has no simple group-theoretic interpretation
sinc
as remarked earlier, the presence of the explicit $g^{ab}(x)$ factor
on the right hand side of \eq{PB:HH} means that \eqs{PB:HaHb}{PB:HH} is
not a genuine Lie algebra. Thus the operator-ordering problem becomes
much harder. If we choose as a simple example the ordering in
which all the $p^{cd}$ variables are placed to the right of the $g_{ab}$
variables, the constraint \eq{HperpPsi=0} becomes
\beq
  -\hbar^2\k^2{\cal G}_{ab\,cd}(x,g]{\de^2\Psi[g]\over\de g_{ab}(x)\,\de
g_{cd}(x)}-{\dg^\half(x)\over\k^2}\,R(x,g]\Psi[g]=0     \label{WDE}
\eeq
where ${\cal G}_{ab\,cd}$ is the DeWitt metric defined in
\eq{Def:DeW-metric}.

     Equation \eq{WDE} is the famous Wheeler-DeWitt equation
\cite{Whe62,Whe64,DeW67a,Whe68}. It is the heart of the Dirac constraint
quantisation approach to the canonical theory of quantum gravity, and
everything
must be extracted from it. Needless to say, there are a number of
problems and questions that need to be considered. For example:

    1.\ The ordering chosen in \eq{WDE} is a simple one but there is no
particular reason why it should be correct. One popular alternative is to
write the `kinetic-energy' term as a covariant functional-Laplacian
using the DeWitt metric \eq{Def:DeW-metric}. This part of the operator is
then invariant under redefinitions of coordinates on $\RS$. Of course, a
key issue in discussing the ordering of $\op\H_\perp(x)$, $x\in\Si$, is
whether or not these operators are expected to be self-adjoint (the
ordering chosen in \eq{WDE} is certainly {\em not\/} self-adjoint in the
scalar product formally associated with the choice \eqs{Def:op-g}{Def:op-p}
for the canonical operators).

    2.\ The Wheeler-DeWitt equation contains products of functional
differential operators evaluated at the same spatial point and is therefore
likely to produce $\de(0)$ singularities when acting on a wide variety of
possible state functionals. Thus regularisation will almost certainly be
needed.

    3.\ A major question is how to approach the problem of solving
the Wheeler-DeWitt equation. One obvious tactic is to deal with it
as a functional differential equation {\em per se}. However, whether
or not this is valid depends on the general interpretation of the
constraint equations $\op\H_\perp(x)\Psi=0$, $x\in\Si$. If these
equations mean that $\Psi$ is a simultaneous eigenvector of
self-adjoint operators $\op\H_\perp(x)$, $x\in\Si$, with eigenvalue
$0$ then, as with eigenfunction problems for ordinary differential
operators, some sort of boundary value conditions need to be imposed
on $\Psi$, and the theory itself is not too informative about what
these might be. In practice, there has been a tendency to solve the
constraint  equation as a functional differential equation without
checking that $0$ is a genuine eigenvalue (\ie without worrying
about boundary conditions). This can lead to highly misleading
results---a fact that is often overlooked, especially in discussions
of minisuperspace approximations to the theory.

    4.\ One of the hardest problems is to decide what the Wheeler-DeWitt
equation means in physical terms.  In particular, the notions of `time' and
`time-evolution' must be introduced in some way. The central idea is one we
have mentioned several times already: time must be defined as an {\em
internal\/} property of the gravitational system (plus matter) rather than
being identified with some external parameter in the universe. We shall
return frequently to this important topic in our analysis of the various
approaches to the problem of time in quantum gravity.

\subsubsection{A Minisuperspace Example}
\label{SSSec:WDE-RW}
Further discussion of the Wheeler-DeWitt equation is assisted by having
access to the minisuperspace model discussed in \S\ref{SSSec:RW}. To
derive the Wheeler-DeWitt equation for this system we need first to
confront some of the problems mentioned above. In particular:
\bi
    \item The classical variable $a$ satisfies the inequality $a\ge0$.
How is this inequality to be implemented in the quantum theory? This is a
very simple example of the spatial metric reconstruction problem.

    \item The classical expression $p_a^2/24a$ in the super-Hamiltonian
\eq{Def:HRW} will lead to operator-ordering problems in the quantum
theory.
\ei
The first problem can be tackled in several different ways:
\be
    \item Ignore the problem and impose standard commutation relations
$[\,\op a,\op p_a\,]=i\hbar$ even though we know this leads to a spectrum
for $\op a$ which is the entire real line. This is the minisuperspace
analogue of taking the `\naive' commutation relations \eqs{CR:gg}{CR:gp}.
The Hilbert space
\footnote{This is only that part of the Hilbert space which refers to the
$a$-variable. The full Hilbert space is $L^2(\R^2,da\,d\f)$.}
will be $L^2(\R,da)$ with the operators defined in the usual way as
\beqa
    (\op a\psi)(a)   &:=& a\psi(a)                      \label{Def:op-a}\\
    (\op p_a\psi)(a) &:=& -i\hbar{d\psi(a)\over da}.    \label{Def:op-pa}
\eeqa
The problem in this approach is to give some physical meaning to the
negative values of $a$.

 \item Insist on using the Hilbert space $L^2(\R_+,da)$ of functions that
are concentrated on $\R_+$ but keep the definitions
\eqs{Def:op-a}{Def:op-pa}. The conjugate momentum $\op p_a$ is no longer
self-adjoint but, nevertheless, it is possible to arrange for the
super-Hamiltonian to be a self-adjoint function of $\op a$, $\op p_a$,
$\op p_a^\dagger$, $\op\f$ and $\op p_\f$.

 \item Perform a canonical transformation at the classical level to a new
variable $\Om$ defined by $a=e^\Om$. This new variable ranges freely over
the entire real line and can therefore be quantised as part of a
conventional set of commutation relations using the Hilbert space
$L^2(\R,d\Om)$. The conjugate variable is $p_\Om:=e^{-\Om}p_a$ and the
super-Hamiltonian is
\beq
H_{\rm RW} :=  e^{-3\Om}\left({-p_\Om^2\over24}+{p_\f^2\over2}\right)
                    -6ke^{\Om} +e^{3\Om}V(\f).
\eeq

\item Use the affine relation $[\,\op a,\op \pi_a\,]=i\hbar\op a$ which
is the minisuperspace analogue of the full set of affine relations
\eqs{ACR:gg}{ACR:gp}. The affine momentum $\pi_a$ is related classically
to the canonical momentum $p_a$ by $\pi_a:=ap_a$, and the
super-Hamiltonian becomes
\beq
    H_{\rm RW} :={1\over a^3}\left({-\pi_a^2\over24} +
                    {p_\f^2\over2}\right) - 6ka+a^3V(\f).
\eeq
A self-adjoint representation of the affine commutation relations can be
defined on the Hilbert space $L^2(\R_+,da/a)$ by
\beqa
    (\op a\psi)(a)    &:=& a\psi(a)                       \\
    (\op\pi_a\psi)(a) &:=& -i\hbar\,a{d\psi(a)\over da}.
\eeqa
\noindent
Note that the transformation $a:=e^\Om$ sets up an equivalence between
this approach and the previous one.
\ee
To get a sample Wheeler-DeWitt equation, let us choose method three with the
$\Om$ variable satisfying standard commutation relations. Then, ignoring
the operator-ordering problem, we get
\beq
    \left(\hbar^2e^{-3\Om}\Big({1\over24}{\pa^2\over\pa\Om^2}-
    {1\over2}{\pa^2\over\pa\f^2}\Big) -6ke^{\Om}
        +e^{3\Om}V(\f)\right)\psi(\Om,\f)=0.                \label{WDE-RW}
\eeq
This simple model illustrates many of the features of the full canonical
theory of general relativity and will be very useful in what follows. For a
recent treatment using path-integral techniques see \citen{LP91}.

\subsection{The Klein-Gordon Interpretation for Quantum Gravity}
\label{SSec:KGI}
\subsubsection{The Analogue of a Point Particle Moving in a Curved
Spacetime}
The key issue now is how the solutions to the Wheeler-DeWitt equation
\eq{WDE} are to be interpreted. This involves two related questions:
\be
    \item What inner product should be placed on the solutions?
    \item How is the notion of time evolution to be extracted from the
Wheeler-DeWitt equation?
\ee
One natural inner product might seem to be
\footnote{Here ${\cal D}g$ denotes the formal analogue of the Lebesgue
measure. No such measure really exists on $\RS$ and a more careful
discussion would need to take this into account.}
\beq
    \brac{\Psi}{\Phi}:=\int_{\RS}{\cal D}g\,\Psi^*[g]\,\Phi[g]
                                                \label{SP:naive}
\eeq
since this is the scalar product with respect to which, for example,
the canonical operators \eq{Def:op-g} and \eq{Def:op-p} are
self-adjoint (at least formally). Indeed, this is precisely what is
used in the so-called `\naive Schr\"odinger interpretation'
discussed in \S\ref{SSec:NSI}. However, the definition \eq{SP:naive}
can be applied to {\em any\/} functional of the three-metric $g$,
which as we shall see in \S\ref{SSec:NSI} is the cause of
considerable difficulties. In the present section we are concerned
rather with finding a scalar product that applies only to {\em
solutions\/} to the Wheeler-DeWitt equation.

    The central idea is to explore the analogy between the Wheeler-DeWitt
equation \eq{WDE} and the Klein-Gordon equation of a particle moving in a
curved space with an arbitrary, time-dependent potential \cite{DeW67a}.
The validity of this analogy can be seen especially clearly in the simple
minisuperspace model \eq{WDE-RW}.

    The point-particle model has been discussed in great depth in recent
years by Kucha\v{r}; here I shall sketch only the main ideas
\cite{Kuc91a,Kuc92a}. Consider a relativistic particle of mass $M$ moving in
a four-dimensional spacetime $(\M,\g)$ where $\g$ is a fixed Lorentzian metric.
The classical trajectories of the particle in $\M$ are parametrised by an
arbitrary real number $\tau$, and the theory is invariant under the
reparametrisation $\tau\mapsto\tau'(\tau)$. This invariance leads to the
constraint $H(X,P)=0$ with the super-Hamiltonian
\beq
    H(X,P):={1\over 2M}\g^{\a\b}(X)\,P_\a\,P_\b + V(X)  \label{Def:HPP}
\eeq
where $V(X)$ is the (positive) potential-energy term.

    In the quantum theory, this constraint becomes the Klein-Gordon
equation
\beq
    \big(\g^{\a\b}(X)\nabla_\a\nabla_\b+V(X)\big)\Psi(X)=0
                                                        \label{KGE}
\eeq
where a convenient choice has been made for the operator ordering of the
kinetic-energy term. The standard interpretation of this equation is based
on the pairing between any pair of solutions $\Psi$, $\Phi$ defined by
\beq
    \la\Psi,\Phi\ra_{\rm KG}:=\int_{\Em(\Si)} d\Si_\a(X)\,
        {1\over2i}\g^{\a\b}(X)\big(\Psi(X)^*\overrightarrow\pa_\b\Phi-
            \Psi(X)^*\overleftarrow\pa_\b\,\Phi(X)\big)   \label{SP:KG-M}
\eeq
where the integral is taken over the hypersurface $\Em(\Si)$ of $\M$
defined by an embedding $\Em:\Si\map\M$ that is spacelike with respect to
the background metric $\g$ on $\M$. Note that $d\Si_\a(X)$ is the directed
hypersurface volume-element in $\M$ defined by
\beq
        d\Si_\a(X):=\epsilon_{\a\b\gamma\de}\,
            dX^\b\wedge dX^\gamma\wedge dX^\de.  \label{Def:DSE-M}
\eeq

    It follows from the Klein-Gordon equation that
$\la\Psi,\Phi\ra_{\rm  KG}$ is independent of $\Em$, which suggests
that $\la\Psi,\Phi\ra_{\rm KG}$ might be a suitable choice for a
scalar product. However, as things stand, this is not viable since
the pairing is {\em not\/} positive definite. Indeed
\bi
 \item $\la\Psi,\Psi\ra_{\rm KG} = 0$ for all {\em real\/} functions
       $\Psi$;
 \item complex solutions to the Klein-Gordon equation exist for which
       $\la\Psi,\Psi\ra_{\rm KG} < 0$.
\ei

    The standard way of resolving this problem is to look for a timelike
vector field $U$ on $\M$ that is a Killing vector for the spacetime metric
$\g$ and is also such that the potential $V$ is constant along its flow lines.
A natural choice of time function $\tau(X)$ is then the parameter along
these flow lines, defined as a solution to the partial differential
equation
\beq
            U^\a(X)\pa_\a\tau(X)=1.
\eeq
If such a Killing vector exists, it follows at once that the energy
$E(X,P)=-P_\tau:=-U^\a(X)\,P_\a$ of the particle is a constant of the motion
with $\{E,H\}=0$. On quantisation this becomes
\beq
                [\,\op E,\op H\,]=0
\eeq
which, if all the operators are self-adjoint with respect to the original
inner product \eq{SP:naive}, means it is possible to find simultaneous
eigenstates of
\beq
        \op E:=i\hbar U^\a\pa_\a.            \label{Def:op-E}
\eeq
and the super-Hamiltonian \eq{Def:HPP}. It is therefore meaningful to
select those solutions of the Klein-Gordon equation that have
positive energy, and it is straightforward to see that the inner product
\eq{SP:KG-M} is {\em positive\/} on such solutions. Furthermore,
restricted to such solutions, the Klein-Gordon equation can be shown to
be equivalent to a conventional Schr\"odinger equation using the chosen
time parameter. This equation is obtained by factorising the
super-Hamiltonian in the form
\beq
            H= (P_\tau+h)(P_\tau-h).              \label{Fact-HM}
\eeq
which is equivalent to the constraint
\beq
                P_\tau+h=0                    \label{Fact-HM=0}
\eeq
on the subspace of the phase space of positive-energy solutions.
In effect, the Schr\"odinger equation is obtained by imposing this second
constraint as a constraint on allowed state vectors.

    The construction above forms the basis for a physically meaningful
interpretation of the quantum theory. However, if no suitable
Killing vector $U$ exists there is no consistent one-particle
quantisation of this theory.

\subsubsection{Applying the Idea to Quantum Gravity}
In trying to apply these ideas to the Wheeler-DeWitt equation, the key
observation is that the DeWitt metric \eq{Def:DeW-metric} on $\RS$ has a
hyperbolic character in which the conformal modes of
the metric play the role of time-like directions, \ie the transformation
$\g_{ab}(x)\mapsto F(x)\g_{ab}(x)$, $F(x)>0$, is a `time-like'
displacement in $\RS$. This suggests that it may be possible to choose
some internal time functional $\T(x,g]$ so that the
Wheeler-DeWitt equation can be written in the form
\beqa
\lefteqn{-\hbar^2\k^2\Big(\,{\de^2\over\de\T^2(x)}-
            {\cal F}^{R_1\,R_2}(x;\T,\s]
{\de^2\over\de\s^{R_1}(x)\,\de\s^{R_2}(x)}\Big)\Psi[\T,\s]} \nonumber\ \ \ \
\\
 & &   - {\dg^\half(x;\T,\s]\over\k^2}\,R(x;\T,\s]\Psi[\T,\s]=0
                                                \label{WDE:int-time}
\eeqa
where $\s^R(x,g]$, $R=1,\ldots,5$ denotes the $5\times\infty^3$
modes of the metric variables $g_{ab}(x)$ that remain after
identifying the $1\times\infty^3$ internal time modes $\T(x)$.

    The starting point is the formal pairing (the analogue of the
point-particle expression \eq{SP:KG-M})
\beq
\la\Psi,\Phi\ra :=i\prod_x\int_\Si d\Si^{ab}(x)\,\Psi^*[\g]
 \Big({\cal G}_{ab\,cd}(x,g]{\overrightarrow\de\over\de g_{cd}(x)}-
{\overleftarrow\de\over\de g_{cd}(x)}{\cal G}_{ab\,cd}(x,g]\Big)\Phi[\g]
                                                    \label{SP:WDE}
\eeq
between solutions $\Psi$ and $\Phi$ of the Wheeler-DeWitt equation. The
functional integral is over some surface in $\RS$ that is spacelike with
respect to the DeWitt metric \eq{Def:DeW-metric}, and $d\Si^{ab}(x)$ is
the directed surface-element in $\RS$ at the point $x\in\Si$. Of course,
considerable care would be needed to make this expression rigorous. For
example, it is necessary to take account of the $\DS$-invariance and then
project the inner product down to $\RS/\DS$, \ie we must also include the
action of the supermomentum constraints $\op\H_a(x)\Psi=0$. Note also that
the precise form of the scalar product depends on how the
operator-ordering problem in the Wheeler-DeWitt equation is solved.
However, the essential idea is clear. In particular, the expression
\eq{SP:WDE} has the important property of being invariant under
deformations of the `spatial' hypersurface in $\RS$. This is the
quantum-gravity analogue of the requirement in the normal
Klein-Gordon equation that the scalar product \eq{SP:KG-M} be time
independent.

    In the minisuperspace example, the Wheeler-DeWitt equation
\eq{WDE-RW} can be simplified by multiplying
\footnote{This step is contentious. It is true that the constraint
equation $\op\H_\perp\Psi=0$ is not formally affected by multiplying on the
left by any invertible operator, but this `renormalisation' of
$\op\H_\perp$ affects the total constraint algebra, and the implications of
this need to be considered at some point. Of course, it also affects the
hermiticity properties of the original constraint.}
both sides by $e^{3\Om}$ to give
\beq
\left(\hbar^2\Big({1\over24}{\pa^2\over\pa\Om^2}-
    {1\over2}{\pa^2\over\pa\f^2}\Big) - 6ke^{4\Om}+e^{6\Om}V(\f)\right)
        \psi(\Om,\f)=0.                     \label{WDE-RW'}
\eeq
The associated scalar product is simply
\beq
\la\psi,\phi\ra := i\int_{\Om={\rm const}} d\f\,
\left(\psi^*{\pa\phi\over\pa\Om} -\phi{\pa\psi^*\over\pa\Om}\right)
                                            \label{SP:WDE-RW}
\eeq
which is conserved in $\Om$-time by virtue of \eq{WDE-RW'}. Note that
\eq{WDE-RW'} has the anticipated form \eq{WDE:int-time} for the
Wheeler-DeWitt equation expressed using an appropriate internal time
$\T(x,g]$.

    Unfortunately, the right hand side of \eq{SP:WDE} cannot serve as a
genuine Hilbert space inner product because, as in the analogous case of
the point particle, it is not positive definite. Guided by the
point-particle example, one natural way of trying to resolve this problem is to
look for a vector field on $\RS$ that is a Killing vector for the DeWitt
metric \eq{Def:DeW-metric} and that scales the potential term
$\dg^\half(x)\,R(x,g]$ in an appropriate way. Sadly, Kucha\v{r} has
shown that $\RS$ admits no such vector \cite{Kuc81a,Kuc91a}, and hence
there is no possibility of defining physical states as an
analogue of the positive-frequency solutions of the normal Klein-Gordon
equation (but see \citen{FH89} for the asymptotically-flat case). However,
even if such a Killing vector did exist, there are other problems. For
example \cite{Kuc92a}:
\bi
 \item   The potential term $\dg^\half(x)\,R(x,g]$ can take on both
negative and positive values, whereas the potential $V(X)$ in the
point-particle super-Hamiltonian \eq{Def:HPP} was required to be
positive. Without this condition it is not possible to prove the
positivity of the Klein-Gordon scalar product restricted to
positive-energy solutions.

 \item   There are good physical reasons for selecting just the
positive-energy solutions for the point-particle, but the
justification for the analogous step in the gravitational case is not
clear. In particular, it is quite legitimate for the geometries along
a path in superspace to both expand and contract in volume, and this
means a classical solution to Einstein's equations can have either
sign of $E$. Therefore, there is no justification for picking just
the positive-frequency modes. Note that this objection applies
already to the simple minisuperspace model discussed above with the
scalar product \eq{SP:WDE-RW}.

\item The attempt to construct a Klein-Gordon interpretation of the
Wheeler-DeWitt equation entails the selection of some intrinsic time
functional $\T(x,g]$. However, as discussed earlier, any such choice
will necessarily fall foul of the {\em spacetime problem\/} whose
resolution requires an internal time to be a functional of
the conjugate momenta $p^{cd}(x)$ as well as the metric variables
$g_{ab}(x)$. Hence $\T(x,g]$ cannot be interpreted as a genuine
spacetime coordinate.

\item The feasability of a Klein-Gordon interpretation is dependent
on the fact that the classical super-Hamiltonian $\Hp(x;g,p]$ is {\em
quadratic\/} in the momentum variables $p^{ab}(x)$. This property is
lost if any powers of the Riemann curvature $R_{\a\b\gamma\de}(X,\g]$
are added to the classical spacetime action of the theory.
Expressions of this type are likely to arise as counter-terms in
almost any attempt to construct a proper quantisation of the
gravitational field (including superstring theory, but excepting the
Ashtekar programme in its current form) and it is important to have
some feel for how they change the situation. Several of the
approaches to the problem of time are sensitive to the precise form
of $\Hp$, but this is particularly so of the Klein-Gordon
interpretation.
\ei

    The problems above, plus the non-existence of a suitable Killing
vector on $\RS$, seem to form an immovable block to resolving the
Hilbert space problem of how to turn the solutions to the
Wheeler-DeWitt equation into a genuine Hilbert space.

\subsection{Third Quantisation}
\label{SSec:ThirdQ}
There have been several different reactions to the failure of the
Klein-Gordon approach to the Wheeler-DeWitt equation. In the case of
a relativistic particle with an external spacetime-dependent metric
or potential, the impossibility of isolating positive-frequency
solutions is connected with a breakdown of the one-particle
interpretation of the theory, and the standard resolution is to
second-quantise the system by turning the Klein-Gordon wave function
into a quantum field.

    It has been suggested several times that a similar process might be
needed in quantum gravity with $\Psi[g]$ becoming an operator $\op\Psi[g]$
in some new Hilbert space. This procedure is usually called `third
quantisation' since the original Wheeler-DeWitt equation is already the
result of a quantum field theory \cite{Kuc81b,Col88,GS88,McG88,McG89}.
However, it is unclear what this means, or if the problem of time can
really be solved in this way. Some of the many difficulties that
arise when trying to implement this programme are as follows.

    1.\ The approach to third quantisation that is closest to conventional
quantum field theory involves constructing a Fock space whose
`one-particle' sector is associated with the functionals $\Psi[g]$. But
this raises several difficulties:
    \bi
    \item What is the analogue of the one-particle Hilbert space
which is to form the basis for the Fock-space construction? The
difficulty is that the problem of time is closely related to the
Hilbert space problem, so where do we start? In the case of a
particle in the presence of a spacetime-dependent potential, one
common way of resolving this issue is to begin with a well-defined
free theory with a proper Hilbert space, and then to regard the
interactions with the background as a perturbation that can
annihilate and create the quanta of this theory. However, this
procedure is dubious if the spacetime dependence comes from the
spacetime metric $\g$ itself unless there is some sense in which one
can usefully write $\g_{\a\b}(X)$ as the sum
$\eta_{\a\b}+h_{\a\b}(X)$ of the fixed Minkowskian metric
$\eta_{\a\b}$ plus a small perturbation $h_{\a\b}(X)$. This is even
more inappropriate in the full quantum-gravity theory since there is
no obvious way of writing the DeWitt metric \eq{Def:DeW-metric} on
$\RS$ as a sum of this type.

    \item If a one-particle Hilbert space {\em can\/} be constructed,
what is the interpretation of states that are tensor products of the
vectors in such a space? In the case of particles, if $\ket{x_1}$ and
$\ket{x_2}$ are states corresponding to a particle localised at
points $x_1$ and $x_2$ respectively, then the tensor product
$\ket{x_1}\ket{x_2}$ describes a pair of particles, both of which
move in the {\em same\/} physical three-space. However, if
$\ket{g_1}$ and $\ket{g_2}$ are eigenstates of the
metric operator, it is not clear to what the product state
$\ket{g_1}\ket{g_2}$ refers. The simplest thing might be to say that
$g_1$ and $g_2$ are both metrics on the same space $\Si$, but this
has no obvious physical interpretation.
\ei

    2.\ A more meaningful interpretation is that $g_1$ and $g_2$ are
metrics on {\em different\/} copies of $\Si$ or, equivalently, a
single metric on the disjoint union of two copies of $\Si$. However
this also raises a number of problems:
\bi
    \item The transition from a state $\ket{g}$ to a state
$\ket{g_1}\ket{g_2}$ corresponds to a topology change in which $\Si$
bifurcates into two copies of itself. But this is unlikely to be
compatible with the Wheeler-DeWitt equation. Are we to look for some
non-linear interaction between the operators $\op\Psi[g]$ that
describes such a process; analogous perhaps to what is done in
string field theory? Any such term would signify a radical departure
from the usual field equations of general relativity.

    \item A normal Fock-space construction involves Bose statistics
but it is not clear what it means physically  to say that
$\ket{g_1,g_2}$ is {\em symmetric\/} in $g_1$ and $g_2$. The two
copies of $\Si$ are disjoint, and presumably no causal connection
can be made between them. So what is the operational significance of
a Bose structure? The use of Fermi statistics would be even more
bizarre!

    \item Once the original space $\Si$ has been allowed to bifurcate
into a pair of copies of itself it seems logical to extend the
topology change to include an {\em arbitrary\/} final three-manifold.
Thus the scope of the theory is increased enormously.

    \item It is difficult to see how the bifurcation of space helps
with the problem of time. Presumably the Wheeler-DeWitt equation will
continue to hold in each disconnected piece of the universe, and then
the problem of internal time reappears in each.
\ei

    3.\  If $\Psi[g]$ becomes a self-adjoint operator it should
correspond to some sort of observable. But what can that be, and
how could it be measured?

    The idea of third quantisation is intriguing and could lead to a
radical change in the way in which quantum gravity is perceived. But,
in the light of the comments above, it is difficult to see how it can
resolve the problem of time.


\subsection{The Semiclassical Approximation to Quantum Gravity}
\label{SSec:SemClass}
\subsubsection{The Early Ideas}
Studies of the semiclassical approach to quantum gravity date
back to the work of \citen{Mol62} which was based on the idea that a
consistent unification of general relativity and quantum theory
might not require quantising the gravitational field itself
but only the matter to which it couples. The suggested
implementation of this scheme was the system of equations
\beq
    G_{\a\b}(X,\g]=\bra{\psi}T_{\a\b}(X;\g,\op\f\,]\ket{\psi}
                                                        \label{2G=E(T)}
\eeq
and
\beq
    i\hbar{d\op\f\over dt}=\big[\,\op H[\g],\op\f\,\big]
                                                        \label{HE:G=E(T)}
\eeq
in which the source of the gravitational field is the expectation
value in  some state $\ket\psi$ of the energy-momentum tensor of the
quantised matter $\f$. The time label in the Heisenberg-picture
equation of motion \eq{HE:G=E(T)} must be related to the time
coordinate used in \eq{2G=E(T)}. If $\f$ is field, \eq{HE:G=E(T)}
would probably be replaced with a set of relativistically-covariant
field equations. The hope is that the pair of equations
\eqs{2G=E(T)}{HE:G=E(T)} is {\em exact\/} and comprises a consistent,
complete  solution to the problem of quantum gravity.

    The early 1980s saw a renewal of interest in this approach,
although the  results were not conclusive (see \citen{Kib81} for a
survey of the situation  at that time, and \citen{Duf81} for a
general criticism). One problem is  that if the matter is chosen to
be a quantised field, the right hand side of  \eq{2G=E(T)} can be
defined only after the quantum energy-momentum tensor  has been
regularised and renormalised \cite{RKK80}. This procedure has been
much studied in the general context of quantum field theory in a
curved  background spacetime and is widely agreed to be ambiguous
if the metric is non-stationary. Counter-terms arise that
involve higher powers of the Riemann curvature  and which have a
significant effect on the Einstein field equations. For  example,
there have been claims, and counter claims, that the system is
intrinsically unstable
\cite{HW78,Hor80,HW80,Hor81,Sue89a,Sue89b,Sim91,Ful90}.

    From the perspective of conventional quantum theory, the coupled
equations \eq{2G=E(T)} and \eq{HE:G=E(T)} are rather peculiar. In
particular,  different states $\ket{\psi_1}$ and $\ket{\psi_2}$ give
rise to different background spacetime geometries $\g_1$ and $\g_2$,
and so it is difficult to make much sense of the superposition
principle.  The theory is therefore difficult to interpret since none
of the standard quantum-mechanical rules are applicable. Another
question is the status of different states $\ket{\psi}$ and metrics
$\g$ that satisfy the coupled  equations \eq{2G=E(T)} and
\eq{HE:G=E(T)}. Is one particular state to be selected via, for
example, some quantum cosmological theory of the initial state? Or do
all possible solutions have some physical meaning?

    One might wonder if the whole idea of quantising everything but
the gravitational field is simply inconsistent---perhaps along the
lines of the famous argument in \citen{BR33} which showed that the
electromagnetic field {\em has\/} to be quantised if it is to couple
consistently to the current generated by quantised matter. However,
as Rosenfeld himself pointed out, there is no direct analogue for
gravity since the proof for electromagnetism involves taking to
infinity the ratio $e/m$ of the charge $e$ to the inertial mass $m$
of a test particle---a procedure that is impossible in the
gravitational case since the analogue of $e$ is the gravitational
mass whose ratio to the inertial mass is fixed by the equivalence
principle \cite{Ros63}. There have been several attempts since then
(\eg \citen{EH77}, \citen{PG81}) to clarify the situation but it is
still somewhat unclear.

\subsubsection{The WKB Approximation to Pure Quantum Gravity}
The semi-classical approach has reappeared in recent years in the guise of
a Born-Oppenheimer/WKB approximation to quantum gravity, and has been
particularly discussed in the context of the problem of time. The starting
point is no longer the equations \eqs{2G=E(T)}{HE:G=E(T)} describing a
purely classical spacetime metric coupled to quantum matter. Instead, one
begins with the full quantum-gravity theory in the form of the Wheeler-DeWitt
equation augmented to incorporate the matter degrees of freedom. A solution
$\Psi[g,\f]$ to this equation is then subject to a WKB-type expansion with
the aim of showing that the lowest-order term satisfies a functional {\em
Schr\"odinger\/} equation with respect to an internal time function that is
determined by the state function $\Psi$. Thus, to this order of
approximation, the second-order Wheeler-DeWitt equation is replaced by a
first-order Schr\"odinger equation, and hence by a system that can be
given a probabilistic interpretation using the associated inner product.
This is therefore a good example of a type-II scheme:
the physical interpretation appears only {\em after\/} a
time variable has been identified in a preliminary quantum theory.

    One aim of this approach is to provide a framework that
interpolates between quantum gravity proper and the, better
understood, subject of quantum field theory in a fixed background
spacetime. For example, there have been several discussions of how
the original semiclassical equations \eq{2G=E(T)} and \eq{HE:G=E(T)}
arise in this framework. However, since  these equations (or, rather,
their analogues) are now only {\em approximate\/}, some of the
problems discussed above disappear or, more precisely, appear in a
different light.

    As far as the general problem of time in quantum gravity is
concerned, the main idea can be summarised by saying that `time' is
only a  meaningful concept in a quantum state that has some
semi-classical component which can serve to define it. Thus time is
an {\em approximate\/}, semi-classical concept, and its definition
depends on the quantum {\em state\/} of the system.  In particular,
time would have no meaning in a quantum cosmology in which the
universe never `emerges' into a semi-classical region. Thus this
approach pays some deference to the general idea that time is part of
the  classical background assumed in the Copenhagen interpretation
of quantum theory.

    At a technical level, the starting point is a WKB technique for
obtaining an approximate solution to the Wheeler-DeWitt equation for pure
geometrodynamics (see \citen{SP89} for a comprehensive review, and
\citen{Kuc92a} for a recent, and very careful, discussion). This involves
looking for a solution in the form
\beq
    \Psi[g] = A[g]e^{iS[g]/\hbar\k^2}                   \label{Psi-WKB}
\eeq
where $S[g]$ is real, and where $A[g]$ is a positive, real function of
$g$ that is `slowly varying' in the sense that
\beq
    \hbar\k^2\left|{\de A[g]\over\de g_{ab}}\right| \ll
        \left|A[g]{\de S[g]\over\de g_{ab}}\right|.     \label{A<<S}
\eeq
Inserting \eqs{Psi-WKB}{A<<S} into the Wheeler-DeWitt equation shows that,
to lowest order in an expansion in powers of $\hbar\k^2\simeq (L_P)^2$
(where $L_P:=(G\hbar/c^3)^\half$ is the Planck length), the phase $S$
satisfies the Hamilton-Jacobi equation
\beq
    {\cal G}_{ab\,cd}(x,g]{\de S[g]\over\de g_{ab}(x)}
    {\de S[g]\over\de g_{cd}(x)}-\dg^\half(x)R(x,g]=0   \label{HJE}
\eeq
of classical general relativity (the supermomentum constraints
$\op\H_a(x)\Psi=0$ have the same implication as before). The amplitude factor
$A$ obeys the `conservation law'
\beq
    {\cal G}_{ab\,cd}(x,g]{\de\over\de g_{ab}(x)}
        \left(A^2[g]{\de S[g]\over\de g_{cd}(x)}\right) = 0.
                                                    \label{WKB-consv}
\eeq

    It should be noted that the precise form of \eq{HJE} and
\eq{WKB-consv} depends on the choice of operator ordering in the
Wheeler-DeWitt equation. I have used a simple ordering in \eq{WDE} in
which the functional derivatives stand to the right of the DeWitt
metric, but one might prefer, for example, a version in which the
kinetic energy term  is formally invariant under transformations of
coordinates on $\RS$. This will lead to minor changes in the
Hamilton-Jacobi equation \eq{HJE} and the conservation law
\eq{WKB-consv} (to illustrate this point see the analogous
equations in \citen{Kuc92a}).

\subsubsection{Semiclassical Quantum Gravity and the Problem of Time}
The recent surge of interest in the application of WKB methods to the
problem of time began with the important work of \citen{Ban85}. The basic idea
is a type of Born-Oppenheimer approach and has been developed further in a
number of papers; for example \citen{Har86}, \citen{Zeh86}, \citen{Zeh88},
\citen{Bro87}, \citen{BHW87}, \citen{BV89}, \citen{Eng89}, \citen{Hal87},
\citen{SP89}, \citen{Pad89a}, \citen{KS91} and \citen{Hal91a}.

    Banks considered a system of matter fields coupled to gravity for
which the Wheeler-DeWitt equation for the combined system can be
written as (\cf \eq{WDE})
\beq
    -\hbar^2\k^2{\cal G}_{ab\,cd}(x,g] {\de^2\Psi\over\de g_{ab}(x)\,
        \de g_{cd}(x)}[g] -\left({\dg^\half(x)\over\k^2}R(x,g] -
            \op{\cal H}_m(x,g]\right)\Psi[g]=0              \label{WDE-m}
\eeq
where ${\cal H}_m(x,g]$ is the Hamiltonian density for the matter. The next
step is to find solutions to \eq{WDE-m} of the special form
\beq
    \Psi[g,\f] = A[g]\,\Phi[g,\f]\,e^{iS[g]/\hbar\k^2}      \label{Psi-WKB-m}
\eeq
where $\f$ denotes the collection of matter variables.  The function
$\Phi[g,\f]$ is expanded as a power series
\footnote{As always, an expansion in a dimensioned constant needs to
be handled carefully. The physical expansion parameter will be a
dimensionless parameter constructed from $G$ and, for example, some
energy scale in the theory.}
in Newton's constant $G$
\beq
    \Phi[g,\f]=\psi[g,\f]+\sum_{n=1}^\infty G^n\psi_{(n)}[g,\f]
                                                        \label{Psi-expand}
\eeq
which, together with the ansatz \eq{Psi-WKB-m}, is inserted into the
Wheeler-DeWitt equation \eq{WDE-m}. Keeping just the lowest-order
terms in the expansion shows that the phase factor $S$ satisfies the
Hamilton-Jacobi equation \eq{HJE} as before. Furthermore, the
amplitude factor $A$ can be selected to satisfy the conservation
equation \eq{WKB-consv} (there is clearly some ambiguity in writing
the overall amplitude as a product $A[g]\Phi[g,\f]$). Finally, the
lowest-order term $\psi[g,\f]$ in \eq{Psi-expand} satisfies the {\em
first-order\/} functional differential equation
\beq
    -2i\hbar{\cal G}_{ab\,cd}(x,g]{\de S[g]\over\de g_{ab}(x)}
        {\de\psi[g,\f]\over\de g_{cd}(x)} +
            \big(\op{\cal H}_m(x,g]\psi\big)[g,\f] = 0.   \label{WDE-WKB}
\eeq

    The key step now is to find a functional $\T(x,g]$ such that
\beq
    2{\cal G}_{ab\,cd}(x,g]{\de S[g]\over\de g_{ab}(x)}
        {\de\T(x',g]\over\de g_{cd}(x)} = \de(x,x')         \label{WKB-T}
\eeq
which can serve as an intrinsic time functional for the system. This must
be augmented by a set $\s^R(x,g]$, $R=1\ldots5$, of functions on $\RS$
that are `comoving' along the flow lines generated by $\T$ in $\RS$ in the
sense that
\beq
    2{\cal G}_{ab\,cd}(x,g]{\de S[g]\over\de g_{ab}(x)}
        {\de\s^R(x',g]\over\de g_{cd}(x)} =  0.               \label{WKB-ZR}
\eeq
Then, using the functions $(\T,\s^R)$ as coordinates on $\RS$,
\eq{WDE-WKB} becomes the  functional Schr\"odinger equation
 \beq
    i\hbar{\de\psi[\T,\s,\f]\over\de\T(x)} =
        \big(\op{\cal H}_m(x;\T,\s]\psi\big)[\T,\s,\f],  \label{SE:WKB}
\eeq
which is the desired result.

    There are various things to note about this construction.
\be
\item The equation \eq{WKB-T} shows clearly how the definition of
the time function $\T(x,g]$ depends on the quantum state via the
choice of the solution $S$ to the Hamilton-Jacobi equation
\eq{HJE}. This approach to the problem of time in quantum gravity
can be traced back to the early ideas of \citen{DeW67a} and
\citen{Mis72}. It was developed further in \citen{LR79} and
\citen{Ban85}, and formed a central ingredient in the interpretation
of quantum cosmology given in \citen{Vil89}.

\item \citen{Pad90} and Greensite (1990, 1991a, 1991b)
\nocite{Gre90a,Gre91a,Gre91b} have suggested an extension of the idea above
in which time is defined with respect to the phase of {\em any\/} solution
of the Wheeler-DeWitt equation, not just one that is in WKB form. Their
construction can be regarded as a result of requiring quantum gravity to
satisfy the Ehrenfest principle (see also \citen{Vin92} and \citen{Squ91}).

\item Only the gravitational field is treated in this semiclassical
way. The matter fields are fully quantised, although the probability
associated with \eq{SE:WKB} is conserved only to the same order in
the expansion to which \eq{SE:WKB} is valid.

\item The scheme can be extended to include genuine quantum fluctuations
of the gravitational field itself (\ie fluctuations around the background
metrics associated with the solution to the Hamilton-Jacobi equation
\eq{HJE}).  Typical examples are the work by Halliwell and Hawking on
inhomogeneous perturbations of a homogeneous universe \cite{HH85,Hal91a},
and \citen{Vil89} who writes the spatial metric $g_{ab}(x)$ as the sum of a
classical background $g^{\rm class}_{ab}(x)$ and a quantum part $\op
h_{ab}(x)$.

\item If $S$ is a {\em real\/} solution to the Hamilton-Jacobi
equation \eq{HJE}, the function $\Psi[g]=A[g]\,e^{iS[g]/\hbar\k^2}$
oscillates rapidly along paths in superspace. However, imaginary $S$
solutions can also exist, and these produce an exponential type of
behaviour. This is often interpreted as indicating a Riemannian
rather than Lorentzian spacetime picture and has been much studied
in quantum cosmology with the idea that the universe tunnels from an
`imaginary-time' region \cite{Hal87,Hal90}.
\ee

\subsubsection{The Major Problems}
The WKB approach to the definition of time is interesting, but it
raises a number of difficult questions and problems. For example:
\be
\item In so far as one starts with the Wheeler-DeWitt equation, the
problems discussed earlier in \S\ref{SSSec:WDE} apply here too. These
include singular operator-products, factor ordering and, in particular,
the issue of whether or not the Wheeler-DeWitt equation is to be regarded
as a genuine {\em eigenvalue\/} equation for a set of self-adjoint
operators $\op\H_\perp(x)$, $x\in\Si$, defined on the Hilbert space
that carries the representation of the canonical operators $\big(\op
g_{ab}(x),\op p^{cd}(x)\big)$: \ie the question of the boundary conditions
in $\RS$ that are to be used in solving the Wheeler-DeWitt equation.

\item It is not obvious that there exist global solutions on
superspace to the defining equation \eq{WKB-T} for the internal
time. Indeed, the analysis in \citen{Haj86} of a minisuperspace
model suggests otherwise. This is related to the {\em global time
problem\/} that arises in the internal Schr\"odinger and
Klein-Gordon interpretations of time.

\item The internal time $\T$ defined by \eq{WKB-T} is a functional of the
metric $g_{ab}(x)$ only. However, as we argued earlier in the context
of the Klein-Gordon interpretation, an intrinsic function of this type
will not resolve the {\em spacetime problem\/} of constructing a
scheme that is independent of the initial choice of a reference foliation.

\item The simple WKB approximation breaks down at the turning points of
$S$, and a more careful treatment is needed near such regions.

\item The elegant first-order form of the Schr\"odinger equation \eq{WDE-WKB}
is lost at the next order in the WKB approximation. This
makes it difficult to assess the proper status of \eq{WDE-WKB} and
to understand how the physics of time changes as one gets nearer to
the Planck regime.

\item {The WKB ansatz \eq{Psi-WKB-m} is only one of a very large number of
possible types of solution to the Wheeler-DeWitt equation. Why should it
have such a preferred status?

    In particular, why should one not consider {\em superpositions\/}
of WKB solutions to the Wheeler-DeWitt equation? Indeed, in some of
the the original discussions it was assumed that the aim was to
construct a coherent superposition of such solutions to produce a
quantum state that approximates a single classical spacetime manifold
\cite{Ger69}. However, if the state vector is a sum
\beq
    \Psi[g] := \sum_j A_j[g]\,\Phi_j[g,\f]\,e^{iS_j[g]/\hbar\k^2}
                                                    \label{Sum-WKB}
\eeq
then each solution $S_j$ of the Hamilton-Jacobi equation will lead
to its {\em own\/} definition of time. A particularly relevant
example is a wave function of the form $e^{iS[g]}+e^{-iS[g]}$ which
is real, and is therefore a natural type of semiclassical solution
to the Hartle-Hawking ansatz for the wave function of the universe.
The two parts of this function are usually said to correspond to
expanding and contracting universes respectively (this can mean, for
example, their behaviour with respect to an extrinsic time variable
like $p^a{}_a(x)$), and it is very difficult to see what such a
quantum superposition could mean. The situation is reminiscent of
the Schr\"odinger-cat problem in ordinary quantum theory where a
single value for a macroscopic property has to be extracted from a
quantum state that is a linear superposition of eigenstates.}

\item The alternative is to keep just a single WKB function, but this also
raises several difficulties:
    \be
    \item As remarked earlier when discussing the Klein-Gordon inner
product \eq{SP:WDE}, the Wheeler-DeWitt equation is real, and
therefore naturally admits real solutions. On the other hand, the
time-dependent Schr\"odinger equation \eq{SE:WKB} is intrinsically
complex because of the $i=\sqrt{-1}$ in the left hand side. As
emphasised by \citen{BS88} and \citen{Bar90}, this means that any
attempt to derive the latter from the former will necessarily entail
the imposition by hand of some correlation between the real and
imaginary parts of $\Psi$. Keeping to just a single WKB function is
an example of this, essentially ad hoc, procedure.

    \item A quantum cosmologist might respond that this is precisely what
is to be expected in a theory that describes the quantum creation of the
universe via the medium of a unique solution to the Wheeler-DeWitt
equation, and the Vilenkin scheme does indeed produce a solution that is
naturally complex \cite{Vil89}. On the other hand, the Hartle-Hawking
ansatz leads to a {\em real\/} solution of the equation (although this
issue is clouded by ambiguities in deciding what the ansatz really means).

    \item With the aid of Wigner functions and other techniques, Halliwell
interprets a single WKB solution as describing a whole {\em family\/} of
classical spacetimes \cite{Hal87,Hal90}. In effect, these different
classical trajectories in superspace are labelled by the $\s^R(x,g]$
functions in \eq{SE:WKB}. But this again raises the `Schr\"odinger-cat'
problem in the guise of having to decide how any specific spacetime
arises.
    \ee

    \item We mentioned earlier that the WKB scheme has been extended to
include quantum fluctuations in the metric. The extreme example of such an
extension is to keep classical only those gravitational degrees of freedom
that can be used to define internal clocks and spatial reference frames.
This requires a split of the gravitational modes akin to that employed in
the internal Schr\"odinger interpretation, and similar difficulties are
encountered. In general there is a serious difficulty in deciding which
modes of the gravitational field are to remain classical and which are to
be subject to quantum fluctuations. There is no obvious physical
reason for making such a selection.

    \item If time is only a semi-classical concept, the notion of
probability---as, for example, in the usual interpretation associated
with a Schr\"odinger equation like \eq{SE:WKB}---is also likely to
be valid only in some approximate sense. At best, this leaves open
the question of how to interpolate between the Schr\"odinger equation
\eq{SE:WKB} and the starting, and as yet uninterpreted,
Wheeler-DeWitt equation \eq{WDE-m}; at worse, it throws doubt on the
utility of the entire quantum programme.
\ee

\subsection{Decoherence of WKB Solutions}
\subsubsection{The Main Idea in Conventional Quantum Theory}
Of the problems listed above, the two that are particularly awkward
at the conceptual level are:
\be
    \item A single $e^{iS[g]}$ corresponds to {\em many\/} classical
spacetimes.
    \item The lack of any {\em prima facie\/} reason for
excluding a combination $\sum_iA_i[g]\Phi_i[g,\f]e^{iS_i[g]}$ of WKB
solutions, each term of which gives rise to its {\em own\/}
intrinsic time function satisfying \eq{WKB-T}.
\ee
There have been a number of recent claims that these, and related,
problems can be solved by invoking the notion of {\em decoherence\/}. A
particularly useful general review is  \citen{Zur91}.

    The idea of decoherence has been developed as part of the general
investigation into the foundations of quantum theory that has been
growing steadily during the last decade. One reason for this activity
has been an increasing awareness of the inadequacy of the traditional
Copenhagen interpretation of quantum theory, especially the posited
dualism between the classical and quantum worlds, the emphasis on
measurement as a primary interpretative category, and the associated
invocation of `reduction of the state vector' as a descriptive
process that lies outside the deterministic evolution afforded by the
Schr\"odinger equation. Considerations of this type have been
enhanced by attempts to develop physical devices (\eg SQUIDS) that
are of macroscopic size but which are nevertheless expected to
exhibit genuine quantum properties.

    The other major motivation for the renewed interest in the
foundations of quantum theory is the many advances that have taken
place in cosmology, especially the realisation that the quantum
state of the very early universe could have been responsible for
the large-scale properties of the universe we see around us
today. But the universe itself is the ultimate closed system, and
there can be no external observer to make measurements. In
particular, the notion of state-vector reduction is an anathema to most
people who work in quantum cosmology.

    To see how the idea of decoherence arises, consider a
quantum-mechanical system $\cal S$ and an observable $A$ with
eigenstates
\footnote{The number of linearly-independent eigenstates $N$ can be
finite or infinite, but for simplicity I have assumed that the
spectrum of $\op A$ is discrete and non-degenerate.}
$\ket{a_1}_{\cal S},\ldots\ket{a_N}_{\cal S}$ corresponding to the
eigenvalues $a_1\ldots a_N$ of the associated self-adjoint operator
$\op A$.  Any (normalised) state $\ket{\psi}_{\cal S}$
in the Hilbert space of the system can be expanded as
\beq
    \ket{\psi}_{\cal S}=\sum_{i=1}^N \psi_i\ket{a_i}_{\cal S}
                                                    \label{psi-exp-ai}
\eeq
where the complex expansion-coefficients $\psi_i$ are given by
$\psi_i={}_{\cal S}\brac{a_i}{\psi}_{\cal S}$. The standard
interpretation is that if a measurement is made of $A$, then (i) the
result will necessarily be one of the eigenvalues of $\op A$; and (ii)
the probability of getting a particular value $a_i$ is $|\psi_i|^2$.
However, prior to the measurement, the state $\ket{\psi}_{\cal S}$
has no direct ontological interpretation vis-a-vis the observable
$A$ .

    Such a view of $\ket{\psi}_{\cal S}$ may be acceptable when
applied to sub-atomic systems, but it seems problematic if the system
concerned is Schr\"odinger's unfortunate cat or, even worse, if the
components in \eq{psi-exp-ai} represent the different terms in the
sum \eq{Sum-WKB} of WKB solutions to the Wheeler-DeWitt equation.
One might then prefer to interpret $\ket{\psi}_{\cal S}$ as
describing, for example, an ensemble of systems in which every
element {\em possesses\/} a value for $A$, and the fraction having
the value $a_i$ is $|\psi_i|^2$; \ie an essentially classical
probabilistic interpretation of the results of measuring $A$.
However, such a situation is described quantum mechanically by the
density matrix
 \beq
    \rho_{\rm mix}=\sum_{i=1}^N|\psi_i|^2\,
        \ket{a_i}_{\cal S}\,{}_{\cal S}\bra{a_i}    \label{rho-mix}
\eeq
whereas the density matrix associated with the pure state
$\ket{\psi}_{\cal S}$ is
\beq
    \rho_\psi:=\ket{\psi}_{\cal S}\,{}_{\cal S}\bra{\psi}=
            \sum_{i=1}^N\sum_{j=1}^N\psi_i\psi_j^*\,
        \ket{a_i}_{\cal S}\,{}_{\cal S}\bra{a_j}    \label{rho-psi}
\eeq
which differs from \eq{rho-mix} by off-diagonal terms. Of course, in
the conventional interpretation of quantum theory, after
the act of measurement the appropriate state {\em is\/} the mixed state
$\rho_{\rm mix}$, and the transformation
\beq
    \ket{\psi}_{\cal S}\,{}_{\cal S}\bra{\psi}
        \map \rho_{\rm mix}                         \label{pure->red}
\eeq
is what is meant by the `reduction of the state vector'. Such a
transformation can never arise from the effect of a unitary operator and
hence, for example, cannot be described as the outcome of applying the
Schr\"odinger equation to the combined system of object plus apparatus.
The `measurement problem' consists in reconciling this statement with the
need to regard the constituents of an actual piece of equipment as being
quantum mechanical.

    The goal of decoherence is to show that there are many situations in
which the replacement of the pure state $\ket{\psi}_{\cal S}$ by the mixed
state $\rho_{\rm mix}$ can be understood as a viable consequence of the
theory itself {\em without\/} the need to invoke measurement acts as a
primary concept. As emphasised by Zurek, no actual quantum system is
really isolated: there is always some environment $\cal E$ to which it
couples and which, presumably, can also be described quantum mechanically
\cite{Zur81,Zur82,Zur83,Zur86,Zur91}.
Suppose the environment starts in a state $\ket{\phi}_{\cal E}$ and with a
coupling between the system and environment such that the environment
states become correlated with those of the system, \ie $\ket{a_i}_{\cal
S}\ket{\phi}_{\cal E}$ evolves to $\ket{a_i}_{\cal S}\ket{\phi_i}_{\cal E}$
(thus the environment performs an `ideal measurement' of $A$).
Then, by the superposition principle, if the initial state of $\cal S$ is
the vector $\ket{\psi}_{\cal S}$ in \eq{psi-exp-ai}, the state for the
composite system ${\cal S}+{\cal E}$ will evolve as
\beq
    \left(\sum_{i=1}^N \psi_i\ket{a_i}_{\cal S}\right)\ket{\phi}_{\cal E}
    \map  \sum_{i=1}^N\psi_i\ket{a_i}_{\cal S}\ket{\phi_i}_{\cal E}.
                                            \label{psi+env-evol}
\eeq

    This state is thoroughly entangled. However, if one is interested
only in the properties of the system $\cal S$, the relevant state is
the reduced density-matrix $\rho_{\cal S}$ obtained by summing over
(\ie tracing out) the environment states. The result is
\beq
    \rho_{\cal S} =\sum_{i=1}^N\sum_{j=1}^N\psi_i\psi_j^*
        \ket{a_i}_{\cal S}\,{}_{\cal E}\brac{\phi_i}{\phi_j}_{\cal E}\,
            {}_{\cal E}\bra{a_j}    \label{rho-E}
\eeq
which, if the environment states are approximately orthogonal (\ie
${}_{\cal E}\brac{\phi_i}{\phi_j}_{\cal E}\simeq \de_{ij})$, becomes
\beq
    \rho_{\cal S}\simeq\sum_{i=1}^N|\psi_i|^2\,
        \ket{a_i}_{\cal S}\,{}_{\cal S}\bra{a_i}.
\eeq
Thus the system behaves `as if' a state reduction has take place to
the level of accuracy reflected in the $\simeq$ sign. This
is what is meant by decoherence.

    The mechanism works because the evolution of the traced-out
density matrix satisfies a master equation rather than being induced
by a unitary evolution \cite{CL83,JZ85}. In effect, the environment
`continuously measures' the system, and hence gives rise to a
continuous process of state reduction. This, it is claimed, is the
reason why Schr\"odinger's cat is in fact always seen to be either
dead or alive, never a superposition of the two. Numerical studies
have shown that the reduction of real physical systems can take
place very quickly. For example, the gas molecules surrounding a
piece of equipment in the laboratory will serve very well, and even
the background $3^0$K radiation is sufficient to produce the desired
effect in a very short time: a typical figure for a macroscopic
object is $10^{-23}s$ \cite{JZ85,Zur86,UZ89,Unr91}.

    Note that the actual collection of states into which the
initially pure state collapses (\ie whether they are eigenstates of
this or that particular observable of ${\cal S}$) is determined by
the coupling of the system to the environment. The essential
requirement is that the operator concerned should commute with the
interaction Hamiltonian describing the system-environment
coupling \cite {Zur81}.

\subsubsection{Applications to Quantum Cosmology}
For the purpose of this paper, the main interest in the above is the
possibility of applying these techniques to the semiclassical
solutions to the Wheeler-DeWitt equation discussed in the previous
section. However, some modification of these ideas is clearly
required if they are to be applied in the cosmological context: by
definition, the universe in its entirety is the one system which has
no environment. In practice, this is done by supposing that certain
modes of the gravitational or matter fields can serve as an
environment for the rest. For a brief, recent review see
\citen{Kie92}.

    The idea that classical spacetime might `emerge' in this way was
discussed in \citen{Joo86} and \citen{Joo87}. The first
quantum-cosmology calculation seems to have been that of \citen{Kie87} who
considered a scalar field coupled to gravity in a situation in which
the inhomogeneous modes serve to decohere the homogeneous modes of
both fields. Later papers by this author are \citen{Kie89a} and
\citen{Kie89b}; for a recent review see \citen{Kie92}. \citen{Hal89}
considered a homogeneous, DeSitter-space metric coupled to a scalar
field. The idea was that some of the inhomogeneous scalar modes can
act as the environment for the metric mode. Halliwell showed that
decoherence occurred for a single $e^{iS[g]}$ solution and also
that, in a Hartle-Hawking semiclassical solution of the form
$e^{iS[g]}+e^{-iS[g]}$, the two terms $e^{iS[g]}$ and $e^{-iS[g]}$
decohered by the same mechanism; see also \citen{Mor89}. Other
relevant work is \citen{Mel91}, who discusses decoherence in the
context of a Klein-Kaluza model, and \citen{Pad89b} who studied the
possibility of defining decoherence between different
three-geometries in $\RS$.

    The concept of decoherence is very interesting and could be of
considerable importance in the context of quantum cosmology and the
problem of time. However, several non-trivial problems arise that are
peculiar to the use of this concept in quantum gravity and which need
more study (see \citen{Kuc92a} for a further critique). In
particular:

    1.\ There does not seem to be any general way of deciding how to
separate the modes into those that are to be kept and those that are
to serve as an environment.

    2.\ In ordinary quantum theory, the transformation into the mixed
state \eq{rho-E} is only valid so long as the environment modes are
deliberately ignored: the `true' density matrix is the one associated
with the pure state \eq{psi+env-evol} and still contains the
off-diagonal interference terms. The transformation from pure to
mixed is therefore only `as if' and, although this may be more than
adequate for normal purposes, it is not really clear what is going on
in the case of quantum cosmology. There may be situations in which
information is genuinely lost: for example down an event horizon, as
in the suggestion by \citen{Haw82a} that the presence of a black hole
can transform a pure state into a mixed state, and some of the models
discussed in the literature involve a similar process using a
cosmological event horizon (\eg \citen{Hal87}). However, most of the
calculations that have been performed are not of this type, and their
meaning remains unclear.

    3.\ A rather serious technical problem arises when trying to
adapt the ideas of decoherence to quantum gravity. The process of
tracing-out modes requires a Hilbert space in which to take the
traces. But, as we have seen, the Hilbert space problem for the
Wheeler-DeWitt equation is still unsolved except, perhaps, in so far
as the semi-classical Schr\"odinger equation \eq{SE:WKB} comes
associated with a natural inner product. However, this equation
arises only {\em after\/} the decision has been made to select a
single $e^{iS[g]}$ solution, and so the corresponding inner product
cannot be used to perform operator tracing in the calculation of a
claimed decoherence effect. In practice, many people resort to the
simple inner product \eq{SP:naive} but, as we shall see in
\S\ref{SSec:NSI}, this brings problems of its own and is hard to
justify.

    4.\  This problem is related to various opaque features that
arise when decoherence is applied in discussions of the problem of
time. In particular, there is a tendency to show (or try to show)
that time itself is something that decoheres---in sharp
contradistinction to normal quantum theory where decoherence is a
{\em process\/} that happens {\em in\/} time. This is a natural
consequence of the fact that time is not an external parameter in
quantum gravity but rather something that must be constructed in some
internal way. However, it is by no means clear if, or in what sense,
time is represented by an operator in quantum gravity as is
suggested by the idea that it decoheres. Indeed, this is one of the
main distinctions between the internal Schr\"odinger interpretation
and those interpretations based on the Wheeler-DeWitt equation. This
difficult concept of decohering time is deeply connected with the
Hilbert space problem for the Wheeler-DeWitt equation, and is an area
in which the consistent histories interpretation (to be discussed in
\S\ref{SSec:CHI}) may have something to offer.


\section{TIMELESS INTERPRETATIONS OF QUANTUM GRAVITY}
\label{Sec:Timeless}
\subsection{The Na\"{\i}ve Schr\"odinger Interpretation}
\label{SSec:NSI}
The general philosophy behind all `timeless' interpretations of
quantum gravity is the belief that it should be possible to construct
a well-defined quantum formalism without the need to make a specific
identification of time at any stage. Approaches of this type
invariably employ some sort of `internal clock' to measure the
passage of time, but such a notion of time is understood to be purely
phenomenological, and hence of no fundamental conceptual or technical
significance. In particular, the choice of time plays no basic role
in the construction of the theory. It is fully accepted that such a
phenomenological time may only approximate the external time of
Newtonian physics and that as a consequence a Schr\"odinger
equation may arise at best as an approximate description of
dynamical evolution. However, it is affirmed that the theory
nonetheless admits a precise probabilistic interpretation with a
well-defined Hilbert space structure.

    The simplest example of such a scheme is the `na\"{\i}ve
\footnote{The appellation `na\"{\i}ve' was coined by \citen{UW89}.}
Schr\"odinger interpretation' whose central claim
is that quantum gravity should be approached by quantising before
constraining, and that the physically-correct inner product is
\eq{SP:naive}
 \beq
    \brac{\Psi}{\Phi}:=\int_{\RS}{\cal D}g\,\Psi^*[g]\,\Phi[g]
                                                    \label{2SP:naive}
\eeq
in which the measure ${\cal D}g$ is defined
\footnote{But recall my earlier caveats about the need to use
distributional metrics, the problem of constructing a proper measure,
\etc.}
on the space $\RS$ of Riemannian metrics on the three-manifold $\Si$.

    At a first glance, the use of the scalar product \eq{2SP:naive}
seems rather natural. After all, $L^2(\RS,{\cal D}g)$ is the Hilbert
space on which the canonical operators \eq{Def:op-g} and
\eq{Def:op-p} are formally self-adjoint. Indeed, by starting with
the  canonical commutation relations \eqs{CR:gg}{CR:gp} (or their
affine  generalisations \eqs{ACR:gg}{ACR:gp}), the spectral theory
associated with the abelian algebra generated by the commuting
operators $\op g_{ab}(x)$ means the (rigorous version of the) scalar
product \eq{2SP:naive} is bound to enter the theory somewhere. The
same Hilbert space is also used (sometimes implicitly) in
discussions of whether or not the constraint operators
$\op\H_\perp(x)$ and $\op\H_a(x)$ are self-adjoint.

    The scalar product \eq{2SP:naive} defines a large class of
square-integrable functions of $g_{ab}(x)$, but many of these are
deemed to be unphysical. More precisely, the constraints
\eqs{HaPsi=0}{HperpPsi=0}
\beqa
        \H_a(x;\op g,\op p\,)\Psi &=& 0         \label{2HaPsi=0}\\
     \H_\perp(x;\op g,\op p\,)\Psi &=& 0        \label{2HperpPsi=0}
\eeqa
serve to project out the `physical' subspace of the Hilbert space
$L^2(\RS,{\cal D}g)$ of such functions. Of course, \eq{2HperpPsi=0}
reproduces the Wheeler-DeWitt equation \eq{WDE}. However, and
unlike---for example---in the Klein-Gordon interpretation, the scalar
product \eq{2SP:naive} is assumed to have a direct physical meaning
with no specific reference to the Wheeler-DeWitt equation. This basic
interpretation of $\Psi[g]$ is that the probability of `finding' a
hypersurface in $\M$ on which the three-metric $g$ lies in the
measurable subset $B$ of $\RS$ is
\beq
    \Prob(g\in B;\Psi)=\int_B {\cal D}g\,|\Psi[g]|^2.   \label{Pr:ginB}
\eeq

    This interpretation has often been used in studies of quantum
cosmology, especially by Hawking and collaborators; for example
\citen{HH83}, \citen{Haw84a}, \citen{Haw84b}, \citen{HP86} and
\citen{HP88} (see also \citen{Cas88}). It has some attractive
properties, not least of which is its simplicity and the fact that,
unlike the Klein-Gordon pairing \eq{SP:WDE}, \eq{2SP:naive} defines a
genuine, positive-definite scalar product (modulo mathematical
problems in constructing a proper measure theory). This interpretation
also gives a clear `wave-packet' picture of how a classical spacetime
arises: one can say that the state $\Psi[g]$ is related to a specific
Lorentzian spacetime $\g$ if $\Psi[g]$ vanishes on almost every
metric $g$ except those that correspond to the restriction of $\g$ to
some spacelike hypersurface.

    Notwithstanding these advantages, the \naive Schr\"odinger
interpretation has some peculiar features that stem from the
`timeless' nature of the description. This is illustrated by the
example above of how to recover a classical spacetime: it is the
whole {\em spacetime\/} that is described by $\Psi$, not just the
configuration of the physical variables on a single time slice. In
general, a typical application of the \naive Schr\"odinger
interpretation is to pose questions of the type `What is the
probability of finding this or that universe?' rather than
questions dealing with this or that {\em evolution\/} of the same
universe.

    The issue becomes clearer if we think more carefully about what
is intended by the statement that \eq{Pr:ginB} is the probability of
`finding' $g$ in some subset of $\RS$. By analogy with normal quantum
theory, one would expect to talk about `measuring' the three-metric,
but it is hard to see what this means. As we have emphasised several
times, measurements are usually made at a single value of a time
parameter, and the results of time-ordered sequences of such
measurements provide the dynamical evolution of the system. But such
language is inappropriate here: the time parameter cannot be fixed
since, in effect, it is {\em part\/} of the metric $g_{ab}(x)$. We
might drop the use of measurement language and adopt a more realist
stance by saying that \eq{Pr:ginB} is the probability of $g$ {\em
being in\/} the subset $B$ of $\RS$, although, to make sense, this
probably needs to be augmented with some sort of many-worlds
interpretation of the quantum theory.

    However, the structure is still peculiar. This is partly because,
as it stands, the interpretation given above could apply to {\em
any\/} function $\Psi[g]$ of the $6\times\infty^3$ variables
$g_{ab}(x)$, $x\in\Si$, that are needed to specify a three-metric.
The supermomentum constraints $\op\H_a(x)\Psi=0$ remove
$3\times\infty^3$ variables
\footnote{The classical constraint functions $\H_a(x)$ generate the
$\DS$ action on $\RS$ and fibre it into orbits. Then, modulo
$\theta$-vacuum effects, the constraints $\op\H_a(x)\Psi=0$ can be
interpreted as saying that the theory is really defined on the
superspace $\RS/\DS$ of such orbits, in which case the inner product
\eq{2SP:naive} must be replaced with an integral over $\RS/\DS$.},
which leaves $3\times\infty^3$. However, to specify a physical
configuration of the gravitational field requires $2\times\infty^3$
variables, and so the functions $\Psi[g]$ depend on an extra
$1\times\infty^3$ variables which, of course, correspond to an
internal time function $\T(x,g]$. Thus the time variable is part of
the configuration space $\RS$ and, in effect, is represented by an
operator; that is why we get a timeless interpretation. For example,
in the simple minisuperspace model discussed in \S\ref{SSSec:WDE-RW},
the wave function $\psi(\Om,\f)$ is interpreted as the probability
amplitude for finding a matter configuration $\f$ {\em and\/} a
radius $a=\ln\Om$. Note that the imposition of the super-Hamiltonian
constraint $\op\H_\perp\Psi=0$ does not remove this extra
configuration variable (as it would if the constraint function
$\H_\perp$ was linear, rather than quadratic, in $p^{ab}(x)$) but
leads instead to the Wheeler-DeWitt equation on $\Psi[g]$.

    The analogue in ordinary wave mechanics would be to interpret
$|\psi(x,t)|^2$ as the probability density of `finding the particle
at point $x$ {\em and\/} time to be $t$', in which $t$ is regarded
as an eigenvalue of some time operator $\op T$. The Schr\"odinger
equation then seems to follow from imposing the constraint
\beq
    \big(\op p_T + H(t,\op x,\op p)\psi\big)(x,t)=0     \label{SE:consQM}
\eeq
on allowed state vectors with $\op p_T:=i\hbar\pa/\pa t$. However,
the conventional interpretation of such a state is that, for {\em
fixed\/} $t$, $|\psi(x,t)|^2$ is the probability distribution in $x$.
Thus the \naive Schr\"odinger interpretation of quantum gravity is
based on an idea that involves a significant change in the quantum
formalism. Note also that any solution to \eq{SE:consQM} will not be
square-integrable in $x$ and $t$, and hence, at best, it is
possible to talk about {\em relative\/} probabilities only. This arises
because the spectrum of the operator $\op p_T+\op H$ on
$L^2(\R^2,dx\,dt)$ is {\em continuous\/}, and hence its eigenstates
are not normalisable.

    A notable property of this construction is that the time operator
$\op T$ does not commute with the constraint operator $\op
p_T+H(t,\op x,\op p)$. In the quantum gravity case this is reflected
in the fact that the projection operator onto a measurable subset $B$
of $\RS$ does not commute with $\op\H_\perp(x)$ (since $[\,\op
g_{ab}(x'),\op\H_\perp(x)\,]\ne 0$). In this sense, the \naive
Schr\"odinger interpretation of probability is inconsistent with the
Wheeler-DeWitt equation unless one studiously avoids asking about the
state function {\em after\/} a hypersurface has been found with a
given three-metric. What is at stake here is the crucial question of
what is meant by an `observable' $\op A$, and what role the concept
plays in the construction of the physical Hilbert space. There is no
problem if this means only that $[\,\op  A,\op\H_a(x)\,]=0$: one
merely passes to the version of the formalism in which the states are
defined on $\RS/\DS$. The difficulty arises if it is required in
addition that $\op A$ commutes with the super-Hamiltonian operators
$\op\H_\perp(x)$. It does not help to impose the weaker condition
that $[\,\op A,\op\H_\perp(x)\,]=0$ only on physical states $\Psi$
that satisfy $\op\H_\perp(x)\Psi=0$: the operator $\op g_{ab}(x)$
maps such a state into one that is not annihilated by the
super-Hamiltonian operator, and this is incompatible even with the
weak condition.

    Another problem is that, analogously to the operator $\op
p_T+H(t,\op x,\op p)$ in \eq{SE:consQM}, the operators
$\op\H_\perp(x)$, $x\in\Si$, defined on the Hilbert space
$L^2(\RS,{\cal D}g)$ can be expected to have continuous spectra, and
so the solutions to the Wheeler-DeWitt equation are not
normalisable. Thus the inner product \eq{2SP:naive} does not induce
an inner product on the physical states. This is a serious difficulty
for the \naive Schr\"odinger interpretation and is one of the main
attractions of the Klein-Gordon programme in which the aim is to
define a scalar product only on {\em solutions\/} to the
Wheeler-DeWitt equation, not on a general function of $g_{ab}(x)$.

\subsection{The Conditional Probability Interpretation}
\label{SSec:CPI}
\subsubsection{The Main Ideas}
The conditional probability interpretation is a development of the
\naive Schr\"odinger approach that has been studied especially by
Page and Wooters
\cite{PW83,Haw84b,Woo84,Pag86a,Pag86b,Pag89,Pag91,PP92}; see also
\citen{Eng89}, \citen{Deu90}, \citen{Squ91} and \citen{CS92}. The
Hilbert space is the one used before, \ie the space of all
functionals $\Psi[g]$ that are square-integrable with respect to the
inner product \eq{2SP:naive}.  However, the interpretation is
different: $|\Psi[g]|^2$ is no longer regarded as the absolute
probability density of finding a three-metric $g$ but is instead
thought of as the probability of finding the $2\times\infty^3$
physical modes of $g$ {\em conditional\/} on the remaining
$1\times\infty^3$ variables---the internal-time part of $g$---being
equal to some specific function (I am assuming that the constraints
$\op\H_a(x)\Psi=0$ have already been solved). The claim or hope is that
this does not require any specific split of $g$ into physical parts
and an internal time function; indeed, the interpretation is supposed
to be correct for any such choice. Thus in the
minisuperspace model in \S\ref{SSSec:WDE-RW}, the
(suitably-normalised) wave-function $|\psi(\Om,\f)|^2$ can be
regarded equally as the probability density in $\f$ at fixed $\Om$
(\ie regarding $\Om$ as an internal metric-field time) or as the
probability density in $\Om=\ln a$ at fixed $\f$ (\ie regarding $\f$
as a matter field time).

    The original work of Page and Wooters was not aimed at quantum
cosmology alone but at the more general problem of the quantisation of any
closed system. They argued, {\em pace\/} Bohr, that in the normal
Schr\"odinger equation the time parameter $t$ is an external parameter,
and hence has no place in the quantum theory if the system is truly
closed. Instead, time must be measured with a physical clock that is part
of the system itself. Their interpretation of $|\psi(x,t)|^2$ is that it
gives the probability distribution in $x$ conditional on the value of the
internal clock being $t$. The normal Schr\"odinger time-dependent
equation is replaced by an eigenvalue equation
\beq
    \op H_{\rm tot}\psi=E\psi                       \label{Hpsi=Epsi}
\eeq
where $H_{\rm tot}$ is the Hamiltonian of the total system, which
includes the physical clock and its interaction with the rest of the
system. Whatever can be said about  `time development' has to be
extracted from this equation. This is done by studying the dependence
of conditional probabilities on the value of the internal-clock
variable on which the probabilities are conditioned.

    The relevance for quantum gravity of these ideas should be clear.
The eigenvalue equation \eq{Hpsi=Epsi} becomes the super-Hamiltonian
constraints \eq{2HperpPsi=0}, and the conditioning is on the internal
time functional $\T(x,g]$ being equal to a specific function
$\tau(x)$; \ie the internal clock is defined by a configuration of
the gravitational field. Of course, if matter fields are present,
they also can serve to define an internal time. Note that, as in the
\naive Schr\"odinger interpretation, the internal clock is
represented by a genuine operator on the Hilbert space of the total
system.

\subsubsection{Conditional Probabilities in Conventional Quantum Theory}
\label{SSSec:CondProb}
To understand the novel features of this idea is it useful to recall
briefly how the notion of conditional probability enters conventional
quantum theory; this will also be helpful in our discussion in
\S\ref{SSec:CHI} of the consistent histories interpretation.

    Let the (mixed) state of a quantum system at some time $t=0$ be
$\rho_0$. In the Schr\"odinger representation, the state $\rho_t$ at time
$t$ is related to the $t=0$ state by the unitary transformation
\beq
    \rho_t = U(t)\,\rho_0\,U(t)^{-1}                    \label{HP-rho}
\eeq
where $U(t):=\exp(-it\op H/\hbar)$. Therefore, if a measurement of an
observable $A$ is made at time $t_1$, the probability that the result will
lie in some subset $\a$ of the eigenvalue spectrum of the operator $\op A$
is
\beqa
\Prob(A\in\a,t_1;\rho_0)&=&\tr\big(P^A_\a\,\rho_{t_1}\big)
                                                        \nonumber\\
                             &=& \tr\big(P^A_\a(t_1)\,\rho_0\big)
                                                        \label{Pr:Aina}
\eeqa
where $P^A_\a(t_1)$ is the Heisenberg-picture operator defined by
\beq
    P^A_\a(t_1) := U(t_1)^{-1}\,P^A_\a\,U(t_1)
\eeq
(with the reference time chosen to be $t=0$) and $P^A_\a$ is the
operator that projects onto the subset $\a$; for example, if the
spectrum of $\op A$ is a set $a_1,\ldots a_N$ of non-degenerate discrete
eigenvalues, then
\beq
    P^A_\a:=\sum_{a_i\in\a}\ket{a_i}\bra{a_i}.
\eeq

    If the measurement of $A$ yields a result lying in $\a$, any further
predictions must be made using the density matrix
\beq
    \rho_\a:={P^A_\a(t_1)\,\rho_0\,P^A_\a(t_1)\over
            \tr\big(P^A_\a(t_1)\,\rho_0\big)}           \label{rho-a}
\eeq
and the transformation
\beq
    \rho_{t_1}\map\rho_\a = {P^A_\a(t_1)\,\rho_0\,P^A_\a(t_1)\over
            \tr\big(P^A_\a(t_1)\,\rho_0\big)}           \label{rho->rho-a}
\eeq
is the analogue for density matrices of the familiar reduction of the state
vector.

    Now let the system evolve until time $t_2$ when a measurement of
an observable $B$ is made. According to the discussion above, the
probability of finding $B$ in a range $\b$, given that (\ie
conditional on) $A$ was found to be in $\a$ at time $t_1$, is
\beq
\Prob(B\in\b,t_2\,|\,A\in\a,t_1;\rho_0)=\tr\big(P^B_\b(t_2)\rho_\a\big)
    ={\tr\big(P^B_\b(t_2)\,P^A_\a(t_1)\,\rho_0\,P^A_\a(t_1)\big)
        \over\tr\big(P^A_\a(t_1)\,\rho_0\big)}.      \label{Pr:BArho(t)}
\eeq

\subsubsection{The Timeless Extension}
The extension of the ideas above to the situation in which there is no
external time parameter proceeds as follows. The physical observables
in the theory are regarded as operators that commute with the total
Hamiltonian $H_{\rm tot}$. As a consequence, there is no difference
between the Heisenberg and the Schr\"odinger pictures of time evolution.
Indeed, there is no time evolution at all in the sense of a change with
respect to any external parameter $t$; in particular, the density matrix
$\rho$ of the system satisfies $[\,H_{\rm tot},\rho\,]=0$: a truly
`frozen' formalism.  Nevertheless, it is assumed that much of the
framework of conventional quantum mechanics is still applicable. In
particular, it is deemed meaningful to talk about the conditional
probability of finding $B$ in the range $\b$, given that $A$ lies in $\a$,
and to assert that, if the state of the system is $\rho$, the value of
this quantity is
\beq
\Prob(B\in\b\,|\,A\in\a;\rho)={\tr\big(P^B_\b\,P^A_\a\,\rho\,P^A_\a\big)
                \over\tr\big(P^A_\a\,\rho\big)}.    \label{Pr:BArho}
\eeq
The extension to the quantum gravity situation is obvious and, again,
there is no external time parameter.

    The suggested form \eq{Pr:BArho} should be compared carefully
with the expression \eq{Pr:BArho(t)} of conventional quantum theory.
There are no $t$-labels in \eq{Pr:BArho} and therefore, in
particular, no sense in which the quantities $B$ and $A$ are
time-ordered (as they are in \eq{Pr:BArho(t)}, with $t_2>t_1>t_0$).
Furthermore, and unlike the case in conventional quantum theory, the
expression \eq{Pr:BArho} is not obtained via any process of state
reduction. Rather it is simply {\em postulated\/} as one of the
fundamental interpretative rules of the theory. Correspondingly, the
concept of `measurement' is only a secondary one: like time, it is
not something that comes from `outside' but is instead only a way of
talking about a particular type of interaction between certain
sub-elements of the closed system. As a consequence we are confronted almost
inevitably with a many-worlds interpretation of the
theory; indeed, supporters of the conditional probability
interpretation of quantum gravity are almost always strong advocates
of a post-Everett view of quantum mechanics.

    Thus we have a timeless picture of quantum theory. This does not
mean that the notion of time-evolution is devoid of any content, but
the challenge is to recover it in some way from the
conditional probability expression.  This is done as follows. Let $T$
be a quantity that we wish to use as an internal clock to measure the
change in another quantity $A$. Then we study the probability
$\Prob(A\in\a|T=\tau;\rho)$ and see how it varies with $\tau$. This
is the dynamical evolution in the theory.

    The conditional probability interpretation is certainly
attractive. It captures nicely the idea that the passage of time
should be identified with correlations inside the system rather than
reflecting changes with respect to an external parameter. In this
respect it seems well-suited for application to the problem of time
in quantum gravity, and it is certainly an improvement on the \naive
Schr\"odinger interpretation. However, this new interpretation gives
rise to various problems of its own, and these deserve careful
consideration.

    1.\ The central difficulty is that the probabilistic rule
\eq{Pr:BArho} constitutes a significant departure from conventional
quantum theory, and the resulting structure may not be
self-consistent. In particular, we need clear guidelines for deciding
when a variable has the property that it is appropriate to condition
on its values. For example, if applied to conventional quantum theory
where there {\em is\/} an external time parameter $t$, the formalism
makes sense only if $T$ is a `good' clock in the sense discussed in
\S\ref{Sec:Problem}. If $T$ is a bad clock, and hence can take on the
same value $\tau$ at two different values of $t$, the probability of
$A$ conditional on $T=\tau$ is not well-defined. For a closed system,
this raises the general question of how we know whether or not a
particular quantity affords a consistent choice for an internal time
variable.

    2.\ Another feature of the conditional probability interpretation
is the peculiar lack of any sense of history. That is, there is no
way of directly comparing things at different times. All statements
are of the form `the probability of $A$ is this, when $B$ is that'
and, in that sense, always refer to the single `now' at which the
statement is made.  Page defends the situation by citing the general
philosophical position that statements about the past are really
counter-factual claims that certain consistency conditions would be
met if {\em present\/} records are examined. However, not everyone is
convinced by this argument; in particular see the critical analysis in
\citen{Kuc92a} and the discussion between Page and Kucha\v{r}
reported in \citen{PP92}

    3.\  There is also a potential problem with the constraint
$\op{H}_{\rm tot}\psi=E\psi$ or, in the gravitational case, the
super-Hamiltonian constraint $\op\H_\perp(x)\Psi=0$. As emphasised
in \citen{Kuc92a}, if an internal clock is to function as such, it
{\em cannot\/} commute with the constraint operator, and---in that
sense---it is not a physical observable. In the gravitational
context, this means that any derivation of \eq{Pr:BArho} based on a
state reduction with an internal time $\T(x,g]$ taking the value
$\tau(x)$,
 \beq
    \rho\map P^\T_{\tau}\,\rho\,P^\T_{\tau}/\tr(P^\T_{\tau}\,\rho)
\eeq
would be incompatible with the Wheeler-DeWitt equation for density
matrices (which is $[\,\H_\perp(x),\rho\,]=0$) since the reduced
$\rho$ violates this condition. In so far as \eq{Pr:BArho} is simply
{\em postulated\/}, this may not be a problem---indeed, expressions
of this type are used frequently in the many-worlds interpretation
of standard quantum theory without invoking the idea of a
collapse---but, together with the absence of any time labels, it
does show the extent to which the conditional probability
interpretation deviates from conventional quantum ideas.

    4.\ Finally, we still have the {\em spacetime\/} problem since in the
form above the programme uses an internal time $\T(x,g]$ which, as has
been mentioned several times, cannot lead to a local scalar function on
the spacetime manifold $\M$. However, this problem might be addressed by
applying the conditional probability interpretation to a system that
includes the type of matter reference fluids discussed in
\S\ref{SSec:RefFluid}.

\subsection{The Consistent Histories Interpretation}
\label{SSec:CHI}
\subsubsection{Preamble}
In some respects, the consistent histories approach to the problem of
time can be regarded as a development of the conditional probability
interpretation, with the particular advantage of enabling questions about
the history of the system to be addressed directly. In particular, it
takes account of the fact that decoherence is really a {\em process\/}
that develops in time and that, for example, a system that has decohered
could in principle recohere at some later time. The scheme culminates in
a suggestion that gravity should be quantised with something like a
functional integral over spacetime geometries.

    In itself, this does not seem so novel: formal quantisation
schemes of  this type have been considered for many years and
inevitably founder on intractable technical problems. Nor is it
obvious how such an approach solves the problem of time, or thows any
light on the related question of constructing the Hilbert space of
states. For example, one way of defining a real-time  functional
integral is to start with the canonical Hilbert space quantum theory
and then define the integral as the limit of a time-sliced
approximation to some matrix element of the unitary evolution
operator (\ie using the Trotter product formula). But this is no help
in a situation in which the Hilbert space structure is one of the
things we are trying to discover. The rival euclidean approach to
quantum gravity does not help much either. For example, in the
Hartle-Hawking ansatz the central object is the functional integral
over all euclidean-signature metrics $\g$ on a four-manifold $\M$
with a single  three-boundary $\Si$  \beq
        \Psi[g]:=\int{\cal D}\g\,e^{-iS_E[\g]/\hbar},
\eeq
where $S_E$ is the euclidean action, and where $\g$ restricted to $\Si$ is
the given three-metric $g$ \cite{HH83}. The function $\Psi[g]$ can be shown
formally  to satisfy the Wheeler-Dewitt equation \cite{Hal88,Hal91b}, and
this particular state is then regarded as the `wave-function of the
universe'. But this does not help with the problem of time since we are
simply faced once more with the difficult question of how to interpret
solutions of the Wheeler-DeWitt equation.

    However, there is far more to the idea of consistent histories
than simply a call to return to spacetime functional integrals. It
stems from what is in fact a radical revision of the formalism of
quantum theory in general. I must admit, when I first came across
the idea I was not very enthusiastic. But my feelings have undergone
a major change recently and I am inclined now
\footnote{This enantiadromia was entirely the result of gentle
pressure from Jim Hartle encouraging me to read the original papers
and to think about them carefully. I most grateful to him for his
efforts in this  direction!}
to rate it as one of the most significant developments in quantum
theory during the last 25 years.

    The consistent histories interpretation is a thorough-going
post-Everett scheme in the sense that measurement is not a
fundamental category; instead the theory itself prescribes when it is
meaningful to say that a measurement has taken place. In particular,
there is no external reduction of the state vector. In the original
formalism, `time' appears in the standard way as an external
parameter. However, the relegation of measurement to an internal
property gives rise to the hope that time may be treated likewise;
indeed, for our present purposes, this is one of the main attractions
of the approach.

\subsubsection{Consistent Histories in Conventional Quantum Theory}
Let us start by summarising the idea in the context of
non-gravitational quantum physics. The seminal papers are
\citen{Gri84}, Omn{\`e}s (1988a, 1988b, 1988c, 1989,
1990)\nocite{Omn88a,Omn88b,Omn88c,Omn89,Omn90}, \citen{Gel87} and
Gell-Mann \&Hartle (1990a, 1990b, 1990c)\nocite{GH90a,GH90b,GH90c}.
Comprehensive recent reviews of both the gravitational and the
non-gravitational case are \citen{Har91a} and \citen{Omn92}; see also
Alberich (1990, 1991, 1992),\nocite{Alb90,Alb91,Alb92} \citen{Ble91},
\citen{DH92} and \citen{Hal92}.

    Consider first the description in conventional quantum theory of
the process of making a series of measurements separated in time.
Each measurement can be regarded as asking a set of questions---the
answer to which is either yes or no---and each such question is
represented by a hermitian projection operator; typically the
projection onto a subset of the spectrum of some operator $\op A$ (so
that the question is `does the value of $A$ lie in the given
subset?'). Let $Q$ denote the set of all questions pertaining to a
particular measurement. Then if $q$ is one such question, the
associated projection operator will be written $P^Q_q$. We want the
set of questions to be mutually exclusive (\ie the answer to at most
one question is `yes') and exhaustive (\ie the answer to at least
one question is `yes'), which means the projection operators must
satisfy
\beq
        P^Q_q\,P^Q_{q'} = \de_{q\,q'}\,P^Q_q        \label{PqPq'=}
\eeq
and
\beq
        \sum_{q\in Q}P^Q_q = 1.                     \label{sum-Pq=}
\eeq

    Now consider making a series of measurements at times $t_1<
t_2\ldots< t_N$ with corresponding sets $Q_1,Q_2,\ldots, Q_N$ of yes-no
questions. The quantity of interest is the absolute probability
$\Prob(q_N\,t_N,q_{N-1}\,t_{N-1},\ldots,q_1\,t_1;\rho_0)$
of obtaining `yes' to questions $q_1\in Q_1$ at time $t_1$, $q_2\in Q_2$
at time $t_2$, $\ldots$, $q_N\in Q_n$ at time $t_N$, given that the state
at time $0\le t_1$ was the density matrix $\rho_0$. The discussion of
conditional probabilities in \S\ref{SSec:CPI} leading to \eq{Pr:BArho}
can be extended to show that
\beqa
\lefteqn{\Prob(q_N\,t_N,q_{N-1}\,t_{N-1},\ldots q_1\,t_1;\rho_0)=}
                                                    \nonumber\\
& & \tr\big(P^{Q_N}_{q_N}(t_N)\,P^{Q_{N-1}}_{q_{N-1}}(t_{N-1})\ldots
    P^{Q_1}_{q_1}(t_1)\,\rho_0\,P^{Q_1}_{q_1}(t_1)\,\ldots
    P^{Q_{N-1}}_{q_{N-1}}(t_{N-1})\,P^{Q_N}_{q_N}(t_N)\big)\ \ \ \
                                                    \label{Pr:1-N}
\eeqa
where the projection operators are in the Heisenberg picture as defined
by \eq{HP-rho}.

    It must be emphasised that \eq{Pr:1-N} is derived using
conventional ideas of sequences of measurements and associated
state-vector reductions like \eq{rho->rho-a}. However, the intention
of the consistent histories interpretation of quantum theory is to
sidestep this language completely by talking directly about the
probability of a history; the idea of `measurement' is then regarded
as a secondary concept that can be described using the history
language applied to the entire system (\ie including what used to be
regarded as an observer). For this reason, following the example of
John Bell, I shall talk about a hermitian operator $\op A$
representing a `beable' rather than an `observable'. Used in this
way, a `history' means any sequence of projection operators
\beq
    P^{Q_N}_{q_N}(t_N)\,P^{Q_{N-1}}_{q_{N-1}}(t_{N-1})\ldots
        P^{Q_1}_{q_1}(t_1)                          \label{Def:hist}
\eeq
satisfying the conditions \eqs{PqPq'=}{sum-Pq=}. Note that this is a
considerable generalisation of the notion of history as used in a
standard path integral where it usually means a path in the
configuration space of the system. A history of this particular type
can regarded as a limit of a sequence of histories \eq{Def:hist} of a
special type in which the projection operators project onto
vanishingly small regions of the configuration space, and the
separation between time points tends to zero.

    The desire to assign probabilities to histories is initially
frustrated by the fact that this is precisely what {\em cannot\/} be
done in conventional quantum theory. All that is possible there is to
give a probability {\em amplitude\/} for a history, but then the
passage to the probability itself introduces interference terms
between different histories. This is seen most clearly in the Feynman
path-integral approach. The amplitudes for paths $a(t)$, $b(t)$ in
configuration space are $A[a]:=e^{iS[a]/\hbar}$ and
$A[b]:=e^{iS[b]/\hbar}$ respectively, where $S[a]$ denotes the
classical action evaluated on the path $a$. But then, generally
speaking, $|\,A[a]+A[b]\,|^2\ne|\,A[a]\,|^2+|\,A[b]\,|^2$ because of
the interference term $|\,A[a]\,A[b]\,|$. The classic example is the
two-slit experiment.

    The central idea in the consistent histories approach is that,
although generic histories cannot be assigned probabilities, this may be
possible for certain special families of histories: the so-called
`consistent'' families.  The key technical ingredient is the {\em
decoherence functional\/} $D(h',h)$ which is a function of pairs of
histories $h',h$ associated to the same collection of questions
$Q_1,Q_2,\ldots,Q_N$. If
\beq
    h':=P^{Q_N}_{{q'}_N}(t_N)\,P^{Q_{N-1}}_{{q'}_{N-1}}(t_{N-1})\ldots
        P^{Q_1}_{{q'}_1}(t_1)                          \label{Def:h'}
\eeq
and
\beq
    h:=P^{Q_N}_{q_N}(t_N)\,P^{Q_{N-1}}_{q_{N-1}}(t_{N-1})\ldots
        P^{Q_1}_{q_1}(t_1)                          \label{Def:h}
\eeq
then $D(h',h)$ is defined by
\beq
D(h',h):=
 \tr\big(P^{Q_N}_{{q'}_N}(t_N)\,P^{Q_{N-1}}_{{q'}_{N-1}}(t_{N-1})\ldots
        P^{Q_1}_{{q'}_1}(t_1)\,\rho_0\,P^{Q_1}_{q_1}(t_1)\,\ldots
            P^{Q_{N-1}}_{q_{N-1}}(t_{N-1})\,P^{Q_N}_{q_N}(t_N)\big)
                                                    \label{Def:Dhh'}
\eeq
which provides a good measure of the size of the interference terms
between the two histories.
\footnote{This is most easily seen by considering the special case where
$\rho$ is a pure state $\ket{\psi}\bra{\psi}$.}
The family of histories is said to be {\em consistent\/} if $D(h',h)=0$
for all pairs $h,h'$ for which $h\ne h'$. If this is so, we {\em
assign\/} the probability to $h$ given by \eq{Pr:1-N}. Thus, in this
approach to quantum theory, the fundamental interpretative rule is
\beq
    D(h',h) = \de_{h'\,h}\,
        \Prob(q_N\,t_N,q_{N-1}\,t_{N-1},\ldots q_1\,t_1;\rho_0)
                                                \label{Dh'h=deh'h}
\eeq
where $\Prob(q_N\,t_N,q_{N-1}\,t_{N-1},\ldots q_1\,t_1;\rho_0)$ is {\em
defined\/} by \eq{Pr:1-N}; but note again that this probability is
assigned to the history {\em only\/} if the consistency conditions
\eq{Dh'h=deh'h} are satisfied
\footnote{In his original paper, Griffiths showed that it is sufficient
if the {\em real\/} part of the off-diagonal parts of $D(h',h)$ vanish.}
for all histories in the family under consideration. It is
straightforward to show that probabilities arrived at in this way obey
all the basic rules of classical probability theory. It must be
emphasised that the decoherence functional is {\em computed\/} using the
mathematical techniques of standard quantum theory: it is only the
probability {\em interpretation\/} that is new.

    The main task is to find families of consistent histories. In
practice, one may decide that exact consistency is not needed: it may be
sufficient if \eq{Dh'h=deh'h} is {\em approximately\/} true
\footnote{This raises the intriguing notion of `approximate
probabilities'.},
although the degree of approximation that is deemed appropriate will
depend on the physical situation involved. However, even with this
weakened requirement most families of histories will not satisfy the
consistency condition. The most discriminating sets of projection
operators $\{P^Q_q\,|\,q\in Q\}$ are those in which each operator $P^Q_q$
projects onto a one-dimensional range. This would happen if
$\{P^Q_q\,|\,q\in Q\}$ is the set of spectral projection operators for a
complete set of commuting `beables' with discrete eigenvalue spectra.
Families of histories associated with collections $Q_1,Q_2,\ldots,Q_N$ of
sets $Q_i$ of questions of this type are least likely to be consistent.
To gain consistency starting with such a family it will be necessary to
{\em coarse-grain\/} the histories---another important concept in the
general programme.

    To coarse-grain a set $Q$ of questions means to partition $Q$ into
subsets of less precise questions. If $\bar Q$ denotes the new set of
questions, and if $\bar q$ is one of the partitions, then the projection
operator corresponding to the new question $\bar q$ (`do any of the
questions in the set $\{q\in\bar q\subset Q\}$ have the answer `yes'?')
is
\beq
            P^{\bar Q}_{\bar q} = \sum_{q\in\bar q}P^Q_q,
\eeq
and the associated decoherence functional is
\beq
    D(\bar h',{\bar h}) =
        \sum_{h'\in{\bar h}'}\sum_{h\in\bar h}D(h',h).  \label{D-CG}
\eeq
The idea is that with an appropriate coarse-graining this new set of
histories may be consistent. One extreme act of coarse-graining is to
choose a single partition for one of the questions, at time $t_j$ say.
This results in the trivial question whose answer is always `yes' and is
represented by the unit operator; in effect that particular time $t_j$
is removed from the sequence. Of course, there is a converse in which a
family of histories can be `fine-grained' by inserting a set of questions
at a time intermediate between a consecutive pair in the original family.

    The conventional interpretation of quantum physics can be recovered
using the idea of a `quasi-classical domain'. A quantum theory is said
to have a quasi-classical domain if there exists a consistent family of
histories with the property that the values of certain, sufficiently
coarse-grained `beables' are correlated in time in a way that reproduces
the equations of some piece of classical physics. Such variables could
include the coarse-grained features of actual pieces of measuring
equipment, with the histories involved describing, for example, the
production of persistent records: a property that has frequently been
seen as the signature of a successful `measurement'.

    It must be emphasised that consistency is a property of a
complete {\em family\/} of histories. Many different such families
may exist, giving different perspectives on the picture of reality
portrayed by quantum theory. Properties like complementarity arise
from the existence of families that are mutually incompatible. In a
situation like this, a `many-worlds' (or, `many-histories')
interpretation of quantum theory seems inevitable. But note that the
`many histories' involved come not only from the different histories
associated with a fixed collection of questions $Q_1,Q_2,\ldots
Q_N$; the collections of questions themselves are also variable---any
collection leading to a consistent family of histories is
admissible.

\subsubsection{The Application to Quantum Gravity}
The consistency condition \eq{Dh'h=deh'h}
depends on the state $\rho_0$; in particular, this is true of the
existence of quasi-classical domains. As emphasised by Gell-Mann and
Hartle, this implies that the manifest existence in our current world of
a quasi-classical domain depends ultimately on the initial state $\rho_0$
that existed shortly after the `initial' big-bang. From this perspective,
the classical features of our present-day world must be seen as a
contingent property of the big-bang: they could have been otherwise.
Indeed, for all we know, the quantum theory of our universe may admit
other consistent families of histories with no quasi-classical domains at
all; a valid concept in the context of a post-Everett interpretation of
quantum theory.

    Many discussions of this type can be carried out within the framework
of a non-quantised background metric. However, problems arise when we
come to quantum gravity itself. The very concept of a `history' rests on
the notion of a time parameter and, as we have seen, this is an elusive
entity. Gell-Mann and Hartle address this problem by proposing an
extension of the formalism in which the notion of `history' becomes a
primary one with no {\em a priori\/} reference to sequences of
questions or beables ordered in any external time. The basic ingredients are:
\be
    \item families of mathematical objects called `histories';
    \item a notion of `coarse-graining' whereby families of histories are
partitioned into exclusive and exhaustive sub-families;
    \item a `decoherence functional' $D(h',h)$  defined on pairs of
histories.
\ee
The decoherence functional must have the following properties:
\bi
    \item {\em Hermiticity\/}: $D(h',h)=D^*(h,h')$
    \item {\em Positivity\/}: $D(h,h)\ge 0$
    \item {\em Normalisation\/}: $\sum_{h',h}D(h',h)=1$
    \item {\em The principle of superposition\/}:
$D({\bar h}',\bar h):=\sum_{h'\in{\bar h}'}\sum_{h\in\bar h}D(h',h)$
where ${\bar h}'$ and $\bar h$ are coarse-grained histories.
\ei
A particular family of histories is said to be {\em consistent\/} if
$D(h',h)=0$ unless $h'=h$, and then a probability $\Prob(h)$ is assigned
to each member of such a family by the rule
\beq
            D(h',h)=\de_{h'\,h}\,\Prob(h).          \label{Prob(h)}
\eeq
As before, the strict equality might be replaced with an approximate
equality where the approximation reflects the physical situation to
which the formalism is applied.

    These rules constitute the entire theory. In particular, there is
no {\em prima facie\/} Hilbert space structure, although a
`phenomenological' one may `emerge' in some domains of the theory.
However, this absence of the standard mathematical formalism can
cause problems when trying to implement the scheme. For example,
finite or countably-infinite sums are used in \eqs{PqPq'=}{sum-Pq=}
because the underlying Hilbert space is assumed to be separable. But
it is not obvious that sums are sufficient in the absence of any such
structure. Some of the collections of histories could well be
non-countably infinite, which suggests that integrals are more
appropriate, and this is likely to produce major technical problems.
It also places in doubt the notion of a `most-discriminative' set
of histories from which all others can be obtained by
coarse-graining. This happens already in the Hilbert space theory for
an operator with a continuous spectrum, but the spectral theory for
such operators enables one to avoid using integrals and to keep to
the well-defined sums in \eqs{PqPq'=}{sum-Pq=}. As we shall see,
this problem is relevant to the application of these ideas in quantum
gravity.

    Several attempts have been made to apply this generalised
formalism to general relativity. The first involves an extension of
the conventional sum-over-histories formalism to situations in
which, although there are paths in the configuration space, the
theory is invariant under reparametrisations of these paths; in this
sense the paths are a generalised form of a `history'. Theories of
this type were studied in some depth by Teitelboim (1982, 1983a,
1983b, 1983c, 1983d),\nocite{Tei82,Tei83a,Tei83b,Tei83c,Tei83d} but
their development in the context of the consistent histories
formalism has been mainly at the hands of Hartle who has emphasised
the significance of the existence of many types of
spacetime-oriented coarse-graining that have no analogue in a
conventional Hamiltonian quantum theory \cite{Har88a,Har91b}; in
particular, there may be no notion of a state being associated with a
{\em spacelike\/} hypersurface of spacetime.  \citen{Har88b} has
also shown in a simple model how the notion of time, and conventional
Hamiltonian quantum mechanics, can emerge from the formalism as a
reading on a physical clock---the same general philosophy that
underlies the conditional probability interpretation discussed
earlier. These path-integral constructions are the subject of a
careful critique in \citen{Kuc92a}.

    More recently, \citen{Har91a} has proposed that the
generalised consistent histories interpretation be extended to
quantum gravity by defining a (most-discriminative) `history' to be
a Lorentzian metric $\g$ on the spacetime manifold $\M$ plus a
specification of the values of a set of spacetime fields $\f$. The
decoherence functional is then defined as
\beq
D(h',h):=\int_{h'}{\cal D}\g'\,{\cal D}\f'\int_h{\cal D}\g\,{\cal D}\f\,
            e^{i(S[\g',\f']-S[\g,\f])/\hbar}     \label{D-grav}
\eeq
where $S[\g,\f]$ is the classical action, and where the integral is
over the constituents of the coarse-grained histories $h$ and $h'$.
The motivation for this expression is that the analogous object in
the path-integral version of normal Hamiltonian quantum mechanics is
the correct choice to reproduce the conventional theory. Note that
\eq{D-grav} apparently contains no reference to a state $\rho$.
However, if desired, this can be thought of as boundary conditions
that could be imposed on the spacetime geometries and matter field
configurations appearing in the integrals. In particular, quantum
cosmological considerations are coded into the behaviour of these
fields near the big-bang region.

    The development of the theory now proceeds as discussed earlier.
Thus one seeks consistent families of such histories from which the
probabilistic interpretation can be extracted. In particular, the
problem of time reduces to studying the classical correlations
between the various variables, including actual physical clocks, in
a quasi-classical domain. Hence the view taken of `time' is
essentially the same as in the conditional probability
interpretation, but the structural framework of the theory is better
defined.

\subsubsection{Problems With the Formalism}
The consistent histories approach has many attractive features, but also
some difficult problems and challenges that need to be taken seriously.

    1.\ The expression \eq{D-grav} illustrates the problem mentioned
earlier about the need to use integrals rather than sums. Functional
integrals can provide valuable heuristic insights into the structure
of a would-be quantum theory, but they are rarely well-defined
mathematically. On the contrary, in the case of general
relativity the theory is known to be perturbatively
non-renormalisable, and hence the chances of making proper
mathematical sense of \eq{D-grav} are not high. One might adopt a
semi-classical approximation (\eg \citen{Kie91}, \citen{Har91a}) but
this is not terribly satisfactory given the lack of the
proper theory that is supposedly being approximated.
In particular, in the absence of a non-perturbative evaluation, the
functional integral \eq{D-grav} is at best a low-energy
phenomenological description that must be cut-off at energies where
the effects of the more basic theory (superstrings?) become
significant. This may well be the best way of justifying the use of the
WKB approach (via a saddle-point approximation to the functional
integral), but it leaves unanswered the question of what happens at
the Planck length and, in particular, the problem of time at that
scale.

    2.\ Even at a formal level, there is a problem attached to
\eq{D-grav} that arises from the presumed $\DM$ invariance of the
theory.  This could be implemented by requiring the elements being
integrated over to be $\DM$-invariant equivalence classes of fields,
but it is notoriously difficult to construct the $\DM$-invariant
measures needed to facilitate this process. The conventional,
heuristic approach is to choose a gauge, define the gauge-fixed
functional integrals in the standard way, and then to show that the
theory is independent of the choice of gauge. However, in the case of
gravity, fixing a gauge means making a choice of internal time \etc,
and then we must confront once more all the problems discussed  in
earlier sections. In other words, to construct the decoherence
functional it may be necessary first to solve the problem of time, and
so we are in danger of going round in circles. Thus further study is
needed into the possibility of performing a functional integral like
\eq{D-grav} {\em without\/} having to invoke a conventional
Hilbert-space formalism.
\footnote{Of course, this does not rule out the possibility that the
functional integral may be defined using {\em some\/} Hilbert
space, but one that is not that of the conventional Hamiltonian formalism.}
Indeed, if something like \eq{D-grav} {\em could\/} be defined properly, it
would be consistent with the general Gell-Mann-Hartle philosophy to
expect the conventional, Hilbert-space structure to emerge only in
some coarse-grained limit of the theory.

    3.\ A major challenge is to find what type of coarse-graining is
needed to produce a consistent family of histories using the spacetime
fields $\g$ and $\f$. At the very least, we need families that are
consistent up to the approximations that may be inherent in pretending
that \eq{D-grav} is a fundamental expression rather than a
phenomenological reflection of a more basic theory.

    4.\  This issue is connected to one of the major questions of
quantum cosmology: `What types of consistent families of histories
give rise to a quasi-classical domain, and how is this related to
the conditions in the early universe?' (\eg \citen{GH92}). A
related issue is the extent to which the {\em only\/} relevant
Planck-length era is that of the very early universe. More
precisely, what physics would we find at the Planck scale if we
could probe it here and now? The question is whether spacetime has
some type of foam structure, and if so how this affects, or is
reflected in, the consistent histories approach to quantum gravity.

    Let me conclude by reaffirming my belief that the Gell-Mann-Hartle
axioms constitute a significant generalisation of quantum theory.
Their suggested implementation via the decoherence functional
\eq{D-grav} represents a rather conservative approach to quantum
gravity and runs into the difficulties mentioned above. But one can
imagine more radical attempts involving, for example, some notion of
generalised causal sets. The entire scheme certainly deserves very
serious further study.

\subsection{The Frozen Formalism: Evolving Constants of Motion}
Rovelli has advocated recently an interesting approach to the
problem of time that shares the central philosophy of the other
`timeless' schemes discussed earlier (Rovelli 1990, 1991a, 1991b,
1991c)\nocite{Rov90,Rov91a,Rov91b,Rov91c}. Thus the main claim is
that it is possible to construct a coherent quantum gravity scheme---
including a probabilistic interpretation---without making any
specific identification of time, which will rather emerge as a
phenomenological concept associated with  physical clocks and
internal time variables.

    Rovelli's starting point is his affirmation that, in the canonical
version of classical general relativity, an {\em observable\/} is any
functional $A[\,g,p\,]$ of the canonical variables
$\big(g_{ab}(x),p^{cd}(x)\big)$ whose Poisson bracket (computed using the
basic relations \eqs{PB:gg}{PB:gp}) with all the constraint functions
vanishes:
\beqa
        \{A,\H_a(x)\}&=&0                       \label{PB:AHa=0}\\
        \{A,\H_\perp(x)\}&=&0.                  \label{PB:AHperp=0}
\eeqa
Properly speaking, it is probably more correct to require these
Poisson brackets to vanish only {\em weakly\/} (as in the right hand
side of \eq{Def:obs}), but this point is not addressed in the
original papers and I shall not go into it here (it is not of any
great significance).

    Since the Hamiltonian \eq{Def:H[N]} for the canonical theory is
$H[\,\Np,\N\,]:=\int_\Si d^3x\,(\N\H_\perp+N^a\H_a)$, these conditions
imply that
\beq
            {dA\over dt}\big(g(t),p(t)\big)=0.
\eeq
Thus, as emphasised in \S\ref{SSSec:ConsAlg}, an observable is
automatically a constant of motion with respect to evolution along
the foliation associated with any choice of lapse function $\Np$ and
shift vector $\N$. This is the `frozen formalism' of classical,
canonical general relativity.

    There are two different approaches to the construction of the
quantum theory of this system. The first uses the group-theoretical
scheme advocated in\citen{Ish84} with the aim of finding a
self-adjoint operator representation of the classical
Poisson-bracket algebra of all observables (or, perhaps, some
selected subset of them) obeying \eqs{PB:AHa=0}{PB:AHperp=0}. The
feasibility of adopting such an approach lies in the observation that
if $A[\,g,p\,]$ and $B[\,g,p\,]$ are a pair of functions which
satisfy  \eqs{PB:AHa=0}{PB:AHperp=0} then the Jacobi identity implies
that $\{A,B\}$ also satisfies these conditions. Thus the set of all
observables is closed under the Poisson bracket operation. If the
resulting algebra is a genuine Lie algebra, a self-adjoint operator
representation can be found by looking for unitary representations
of the associated Lie group. Note that, by constructing
the physical Hilbert space in this way one arrives at a
probabilistic interpretation without making any specific
identification of time.

    Unfortunately, sets of observables that generate a true Lie
algebra are rather rare, and the algebra seems more likely to be one
in which the coefficient of the Poisson bracket of two generators is
a non-trivial function of the canonical variables. It is difficult to
find self-adjoint representations of algebras of this type because of
awkward problems involving the ordering of the generators and their
$q$-number coefficients. Certainly, no one has succeeded in
constructing a proper quantum gravity scheme in this way, although
this is partly due to the difficulty in finding {\em classical\/}
functions that satisfy \eqs{PB:AHa=0}{PB:AHperp=0}. An interesting
model calculation is given in \citen{Rov90} (but note the criticism
in \citen{Haj91} and the response of \citen{Rov91c}). For a
comprehensive analysis of schemes of this type applied to
finite-dimensional examples see \citen{Tat92}.

    An alternative approach is to start with the scheme employed in
\S\ref{Sec:TAfter} which is based on an operator representation of the
canonical commutation relations \eqs{CR:gg}{CR:gp} (or their affine
generalisation \eqs{ACR:gg}{ACR:gp}) on some Hilbert space $\H$. The
physical state space $\H_{\rm phys}$ is deemed to be all vectors in $\cal
H$ that satisfy the operator constraints \eqs{HaPsi=0}{HperpPsi=0}
\beqa
        \H_a(x;\op g,\op p\,]\Psi&=&0               \label{3HaPsi=0}\\
        \H_\perp(x;\op g,\op p\,]\Psi&=&0,          \label{3HperpPsi=0}
\eeqa
and a physical observable is then defined to be any operator $A[\,\op
g,\op p\,]$ that satisfies the operator analogue of
\eqs{PB:AHa=0}{PB:AHperp=0}
\beqa
        [\,\op A,\op\H_\a(x)\,]&=&0                 \label{CR:AHa=0}\\
        {[}\,\op A,\op\H_\perp(x)\,]&=&0.           \label{CR:AHperp=0}
\eeqa
These equations are compatible with \eqs{3HaPsi=0}{3HperpPsi=0} in the
sense that any operator satisfying them maps the physical subspace
$\H_{\rm phys}$ into itself. A weaker version of
\eqs{CR:AHa=0}{CR:AHperp=0} is to require the commutators to
vanish only on the physical subspace.

    The next step is to place a suitable scalar product on $\H_{\rm
phys}$. As emphasised earlier, this cannot be done simply by
regarding $\H_{\rm phys}$ as a subspace of the original Hilbert space
$\H$: the continuous nature of the spectra of the constraint
operators means that the vectors in $\H_{\rm phys}$ all have an
infinite $\H$-norm. It is by no means clear how to set about finding
the correct scalar product but presumably a minimal requirement is
that the physical observables satisfying \eqs{CR:AHa=0}{CR:AHperp=0}
should be self-adjoint in the new Hilbert space structure.

    This issue raises the general question of how physical
observables are actually to be constructed (this is also very
relevant for the first approach). Rovelli claims that a particularly
important class is formed by the so-called `evolving constants of
motion', which serve also to introduce some notion of dynamical
evolution. The basic idea is best explained in a simple model with a
single super-Hamiltonian constraint $H(q,p)$ defined on a
finite-dimensional phase space $\cal S$. A classical physical
observable is then defined to be any function $A(q,p)$ such that
$\{A,H\}=0$ (or, perhaps, $\{A,H\}\approx 0$). The next step is to
introduce some internal time function $\T(q,p)$ with the property
that, for any $t\in\R$, the hypersurface
\beq
        {\cal S}_t:=\{(q,p)\in{\cal S}|\T(q,p)=t\}
\eeq
intersects each dynamical trajectory (on the constraint surface)
generated by $H$ once and only once (of course, there may be global
obstructions to finding such a function). Note that this requirement
means that $\{\T,H\}\ne0$, and hence the internal time function $\T$
is {\em not\/} a physical observable in the sense above.

    The key idea is to associate with each function $F$ on $\cal S$ a
one-parameter family of observables (\ie constants of motion) $F_t$,
$t\in\R$, defined by the two conditions
\beqa
                \{F_t,H\}&=&0                           \label{Def:Ft1}\\
                F_t|_{{\cal S}_t} &=& F|_{{\cal S}_t}   \label{Def:Ft2}
\eeqa
\ie the observable $F_t$ is equal to $F$ on the subspace ${\cal S}_t$
of the phase space $\cal S$. Dynamical evolution with respect to the
internal time is then described by saying how the {\em family\/} of
observables $F_t$, $t\in\R$, depends on $t$. This is therefore a
classical analogue of the Heisenberg picture of time development in
quantum theory.

    A direct Poisson-bracket calculation shows that
\beq
        \{\T,H\}\,{dF_t(q,p)\over dt} = \{F,H\}.        \label{dFt/dt}
\eeq
Note that if $\T$ is a `perfect Hamiltonian clock' then, by definition,
we have
\beq
        H=p_\T+h
\eeq
where the clock Hamiltonian is $p_\T$---the conjugate to the internal time
function, so that $\{\T,H\}=1$---and the Hamiltonian $h$ describing the
rest of the system is independent of $p_\T$. It follows that
$\{F,h\}=\{F_t,h\}$, so that \eq{dFt/dt} becomes
\beq
    {dF_t(q,p)\over dt} = \{F_t,h\},
\eeq
which is the usual equation of motion. Thus \eq{dFt/dt} can be viewed
as a {\em bona fide\/} generalisation of conventional mechanics to
the situation where the only time variable is an internal one.

    Rovelli's suggestion is that these evolving constants of motion
should form the basis for a quantisation of the system. Thus, in the
group-theoretical approach, the key algebra to be represented is the
Poisson-bracket algebra generated by the classical quantities $F_t$,
$t\in\R$. The main problem here will be to decide whether or not the
set of all such objects forms a genuine Lie algebra. If it does, a
unitary representation of the associated Lie group will yield the
desired quantum observables. If---as seems more likely---it forms
only a function algebra (\ie with $q$-number coefficients), it will
be necessary to think again about how to find self-adjoint operator
representations.

    In the alternative, constraint-quantisation approach one needs
operator equivalents of the defining equations
\eqs{Def:Ft1}{Def:Ft2}. The hope is that the inner product on the
physical states $\H_{\rm phys}$ can then be determined by the
requirement that all operators $\op F_t$ are self-adjoint. A number
of severe technical problems arise in this version of the programme
and are articulated in \citen{Kuc92a}. For example:
\bi
\item The operator form of \eq{Def:Ft1} is ill-defined and
ambiguous. Neither is it clear that, even if they could be defined
properly, the conditions \eqs{Def:Ft1}{Def:Ft2} are sufficient to
yield a unique operator $\op F_t$ from a given operator $\op F$.
This particular problem can be avoided by starting with the
classical versions $F_t$, which {\em are\/} well-defined, and then
trying to make them into operators. But severe operator-ordering
problems will inevitably enter at this point and are likely to be
intractable. This is because the classical object $F_t$ can be
obtained only by {\em solving\/} the classical equations of motion,
and hence it is likely to be, at best, an implicit function of the
starting function $F$.

\item The global time problem means that no globally-defined internal
time function exists. In this circumstance there is a good case
for arguing that the associated operators $\op F_t$ should {\em
not\/} be self-adjoint. This is the basis of the objection to
Rovelli's procedure in \citen{Haj91}.

\item It seems most unlikely that a single Hilbert space can be used
for all possible choices of an internal time function $\T$. Thus the
multiple choice and Hilbert space problems appear once more.
\ei
These are real difficulties and need to be taken seriously.
However, they are no worse than those that arise in any of the other
approaches to the problem of time and Rovelli's scheme deserves
serious attention, not least because it emphasises once again the
importance of the still-debated question of what is to be regarded
as an observable in a quantum theory of gravity.


\section{CONCLUSIONS}
We have discussed three main ways of approaching the central
question of how time should be introduced into a quantum theory of
gravity. In theories of type I, time is defined internally at a
classical level: a procedure that is associated with the removal of
all redundant variables before quantisation and which culminates in
the production of a standard Schr\"odinger time-evolution equation
for the physical modes of the gravitational and matter fields. This
approach is relatively uncontroversial at a conceptual level but it runs
into severe technical problems including obstructions to global
existence, and local non-uniqueness. It also seems rather {\em ad
hoc\/} and it is aesthetically unattractive.

    In approaches of type II, all the canonical variables are
quantised and the constraints are imposed at the quantum level {\em
\`a la\/} Dirac as constraints on allowed state vectors.
Unfortunately, there is no universally-agreed way of interpreting the
ensuing Wheeler-DeWitt equation; certainly none of the ideas produced
so far is satisfactory. However, it must be emphasised that there
is no real justification for extending the Dirac approach to
constraint generators that are {\em quadratic\/} functions of the
momentum variables. Therefore, although it may be heretical to suggest
it, the Wheeler-DeWitt equation---elegant though it be---may be
completely the wrong way of formulating a quantum theory of gravity.

    Approaches of type III differ from types I and II in ascribing to
the concept of `time' only a secondary, phenomenological status: a
move that is inevitably associated with some change in the quantum
formalism itself. Techniques of this sort are particularly
well-suited for handling the deep philosophical issues that arise in
quantum cosmology when quantum theory is applied to the universe as a
whole. To my mind, the consistent-histories approach is the most
far-reaching in its implications, but it needs further development,
especially in the direction of finding a more adventurous definition
of what is meant by a `history' in the context of quantum gravity.

    Let me emphasise once more that most of the problems of time in
quantum gravity are {\em not\/} associated with the existence of
ultraviolet divergences in the weak-field perturbative quantisation;
in particular, many interpretative difficulties arise already in
infinity-free, minisuperspace models. Therefore, I feel it is correct
to say that the problems encountered in unravelling the concept of
time in quantum gravity are grounded in a fundamental inconsistency
between the basic conceptual frameworks of quantum theory and general
relativity. In responding to this situation the main task is to
decide  whether `time' should preserve the basic role it plays in
classical general relativity---something that is most naturally
achieved  by incorporating it into the quantum formalism by the
application of a quantization algorithm to the classical theory---or
if it is a concept that should emerge phenomenologicaly from a
theoretical framework based on something very different from
`quantising' classical general relativity.

    If the former is true, which suggests a type I approach to
the problem, the best bet could be some `natural' choice of internal
time dictated by the technical requirements of mathematical
consistency in a quantisation scheme; for example the programme
currently being pursued by Abhay Ashtekar and collaborators.

    If the latter is true, two key questions arise: (i) what is this
new framework?, and (ii) how, if at all, does it relate to the
existing approaches to quantum gravity, especially the semi-classical
scheme? In particular, how does the framework yield conventional
quantum theory and our normal ideas of space and time in their
appropriate domains?

    The most widely-studied scheme of this sort is superstring
theory but, in its current manifestation, this is not well-suited for
addressing these basic questions. The idea of strings moving in a
spacetime already presupposes a great deal about the structure of
space and time; and the quantisation techniques employed presuppose
most of structure of standard quantum theory, particularly at a
conceptual level. It may well be that a new, non-perturbative approach
to superstring theory will involve a radical reappraisal of the ideas
of space, time and quantum theory; but this remains a task for the
future. Perhaps the answer is to find a superstring version of
Ashtekar's formalism (or an Ashtekarisation of superstring theory),
and with the conceptual aspects of quantum theory being handled by
a consistent-histories formalism. A nice challenge for the next few
years!

\bigskip\noindent
{\bf Acknowledgements}

\smallskip\noindent
I would like to reiterate my remarks in the preamble concerning my
great indebtedness to Karel Kucha\v{r} for sharing his ideas with me.
I have also enjoyed recent fruitful discussions and correspondence on
the problem of time with Julian Barbour, Jim Hartle and Ranjeet Tate.
Finally, I would like to thank the organisers of the Advanced Study
Institute for their kindness and friendship during the course of a
very pleasant meeting.

\end{document}